\documentclass{aa}  
\usepackage{stfloats}
\usepackage{placeins}
\usepackage{graphicx}
\usepackage{txfonts}
\usepackage{comment}
\usepackage{soul}
\usepackage{amsmath}
\usepackage[colorlinks=true,linkcolor=blue,filecolor=blue,urlcolor=blue,citecolor=blue,breaklinks=true]{hyperref}
\usepackage[flushleft]{threeparttable}
\usepackage{lastpage}
\usepackage{float}
\usepackage{multirow}
\usepackage{makecell}
\usepackage{xcolor}
\usepackage{cleveref}
\usepackage{appendix}
\usepackage{multicol}
\usepackage[utf8]{inputenc}
\usepackage{float}

\begin{document} 

\title{A Pixel-by-Pixel Path to Population III Discovery with JWST}

    \author{
   Patricia Iglesias-Navarro\inst{1,2}
   \and
   Thomas Harvey\inst{3} \and Marc Huertas-Company\inst{1,2} \and Christopher C. Lovell\inst{4,5} \and  Johan H. Knapen\inst{1,2} \and Christopher Conselice\inst{3} \and Brant Robertson\inst{6} \and Andrew J. Bunker\inst{7} \and Stéphane Charlot\inst{8} \and Natalia C. Villanueva\inst{9} \and Hannah \"Ubler\inst{10} \and Zhiyuan Ji\inst{11} \and Kevin Hainline\inst{11} \and Christina C. Williams\inst{12}
   } 

   \institute{
   Instituto de Astrof\'isica de Canarias (IAC), C/ V\'ia L\'actea s/n, 38205 La Laguna, Tenerife, Spain
   \and Departamento de Astrof\'isica, Universidad de La Laguna, 38200 La Laguna, Tenerife, Spain
   \and Jodrell Bank Centre for Astrophysics, University of Manchester, Oxford Road, Manchester M13 9PL, UK
   \and Kavli Institute for Cosmology, Madingley Road, Cambridge CB3 0HA, UK
   \and Institute of Astronomy, Madingley Road, Cambridge CB3 0HA, UK
   \and Department of Astronomy and Astrophysics, University of California, Santa Cruz, 1156 High Street, Santa Cruz, CA 95064, USA
   \and Department of Physics, University of Oxford, Denys Wilkinson Building, Keble Road, Oxford OX1 3RH, UK
   \and Sorbonne Universit\'e, CNRS, UMR 7095, Institut d'Astrophysique de Paris, 98 bis bd Arago, 75014 Paris, France
   \and Department of Astronomy, The University of Texas at Austin, Austin, TX, USA
\and Max-Planck-Institut f\"ur extraterrestrische Physik (MPE), Gie{\ss}enbachstra{\ss}e 1, 85748 Garching, Germany
   \and Steward Observatory, University of Arizona, 933 N. Cherry Avenue, Tucson, AZ 85721, USA
   \and NSF National Optical-Infrared Astronomy Research Laboratory, 950 North Cherry Avenue, Tucson, AZ 85719, USA
   }

   \date{}

   \abstract
   {The identification of the first generation of metal-free stars, known as Population III (Pop~III), remains a primary goal of modern observational astronomy. While JWST has discovered an abundance of UV-bright galaxies at $z > 10$, distinguishing primordial stellar populations from early metal-enriched systems is a significant challenge. We present an end-to-end framework that combines physically motivated forward modelling from Yggdrasil primordial models with simulation-based inference (SBI) to test Pop~III detectability in JWST-like observations, from isolated sources to realistic overlap with enriched (Pop~II) hosts. Our analysis spans several Pop~III initial mass function (IMF) assumptions, nebular configurations, and Lyman-$\alpha$ transmission scenarios, while mocking the noise properties and filter coverage of the JWST Advanced Deep Extragalactic Survey (JADES).

We find that unresolved or integrated analyses are strongly limited by host-galaxy contamination, whereas spatially resolved, pixel-based model comparison substantially improves recoverability. In our resolved experiments, detectability is highest for young and massive Pop~III clumps in nearly-quenched hosts at larger projected separations from their centres, reaching $\sim 90\%$ recovery in favourable configurations, while older and centrally embedded clumps are rarely recovered. Explainability analysis (SHAP) confirms that distance, redshift, mass ratio and host star formation rate are the dominant drivers of detection, with secondary dependence on the age of the Pop~III clump and the morphology of the host galaxy. Applying the framework to a literature candidate yields spatially differentiated behaviour: a compact blue companion is preferentially described by Pop~III-like models, while the host is better explained by fiducial Pop~I/II models. Our pipeline provides practical criteria for future searches and motivates imaging-first, spectroscopy-assisted strategies for identifying primordial stellar populations in JWST data.
}

   \keywords{galaxies: evolution -- galaxies: high-redshift -- galaxies: formation}

   \maketitle

\section{Introduction}


Population~III (Pop~III) stars are expected to be the first metal-free stellar generation and a key driver of early chemical enrichment, radiative feedback, and the timing of reionization, making their detection central to a physical picture of the first galaxies \citep{1999ApJ...527L...5B,2002A&A...382...28S,2003A&A...397..527S, 2004ARA&A..42...79B,2005ARA&A..43..531B,2010MNRAS.403...45S,2010ApJ...716..510G,Glover2025}. In practice, however, their signatures are likely short-lived and rapidly diluted by subsequent star formation and enrichment, so observational constraints still rely on indirect diagnostics and rare candidate systems \citep{2014MNRAS.442.2963K,2023MNRAS.524..351K,2024ApJ...973L..12V}.

The James Webb Space Telescope (JWST) era has transformed this landscape by uncovering multiple high-redshift systems with unusually hard ionizing spectra and weak metal-line signatures. Recent analyses have reported increasingly compelling Pop~III-like candidates, including lensed, ultra-faint sources at $z\sim6$--$7$ and He\,II emitters near massive high-z galaxies, while also emphasizing that alternative channels—Active Galactic Nuclei (AGN) activity, direct-collapse black hole ionization, etc.—can reproduce parts of the same observable space \citep{Cameron2023,Fujimoto2025,Visbal2025, Reumert2026,Maiolino2026,bler2026}. This combination of progress and ambiguity motivates a framework that can quantify when primordial interpretations are favoured over enriched-population solutions or other alternative explanations.

Early searches for Pop~III stars have relied on photometric and spectroscopic signatures expected for metal-free stellar populations, including very blue continua, strong hydrogen and He\,II emission, weak or absent metal lines, and colour selections in NIRCam and MIRI bands \citep{Nakajima2022,Visbal2023,Trussler2023}. These strategies have already motivated the identification of several promising candidates, including the GN-z11 He\,II clump, RXJ2129-z8HeII and LAP1-B \citep{Maiolino2024,Maiolino2026,bler2026,2024ApJ...967L..42W,Nakajima2025}.

However, a core limitation of many current approaches is their dependence on integrated photometry or integrated line measurements, where host-galaxy light can wash out the contrast of a Pop~III component and increase degeneracies with very low-metallicity Pop~II stars. This confusion is exacerbated by varying stellar age and dust attenuation, as well as nebular physics \citep{Nakajima2022,Trussler2023}. Indeed, some of the most promising candidates lie in the vicinity of enriched hosts \citep{2023A&A...677A..88B,Reumert2026,Maiolino2026,bler2026}, highlighting the need to model confusion between a Pop~III-dominated system and its local environment incorporating realistic morphology, point-spread function (PSF) convolution, and survey-dependent noise.

Addressing this blending between compact primordial signals and their local environments requires a spatially resolved approach that can disentangle stellar populations at the pixel level. However, the massive volume of high-resolution data provided by JWST necessitates a shift away from traditional spectral energy distribution (SED) fitting techniques, which remain computationally intensive and often prohibitively slow for large-scale pixel-level studies. 

Conventional Bayesian methods---critical for capturing the complex degeneracies and uncertainties in stellar properties---can demand up to 10--100 CPU hours per source \citep{Tacchella2022}, creating a significant bottleneck for deep-field surveys. To address this, simulation-based inference \citep[SBI;][]{Cranmer2020} has emerged as a transformative framework: models are trained on large suites of forward-modelled synthetic observations with known physical parameters, learning complex posterior distributions directly from simulated data. By using neural density estimators, SBI enables amortised inference: once trained, the model can estimate full posterior distributions for new observations almost instantaneously \citep{Alsing2019, Hahn2022, Iglesias-Navarro2024}. This approach has been successfully applied to both integrated and spatially-resolved properties \citep{Khullar2022, Wang2023, iglesias-navarro25, synference}, offering a scalable solution that maintains a Bayesian treatment of uncertainties even in the presence of the low signal-to-noise data typical of the high-redshift Universe.

To this end, we present a simulation and inference framework that combines physically motivated forward models with SBI to compare Pop~III and fiducial Pop~I/II hypotheses directly at the pixel level. Our models include different initial mass function (IMF) families, nebular covering fractions, and Lyman-$\alpha$ escape fractions, and are evaluated in observational setups matched to JWST imaging. This allows us to map the detectability surface in terms of mass, age, redshift, and geometric separation from the host. The key novelty is the transition from isolated/integrated tests to resolved contamination experiments, where we inject Pop~III clumps into host galaxies and evaluate recovery under controlled overlap conditions. 

The rest of the paper is organized as follows: Sect.~\ref{sec:inference} presents the forward modelling and SBI setup; Sect.~\ref{sec:isolated} quantifies recovery and model identification for isolated Pop~III sources; Sect.~\ref{sec:contamination} explores integrated and spatially resolved contamination by enriched host galaxies; Sect.~\ref{sec:blueberry} applies the framework to a specific high-redshift candidate; and Sects.~\ref{sec:limitations} and \ref{sec:prospects} discuss limitations, observational prospects, and future extensions.


\section{Simulation-based inference}
\label{sec:inference}

Our aim is to characterise the relationship between the galaxy properties, denoted by \(\theta\), and the observations, \(\{X_i\}\). In traditional Bayesian inference, this relationship is encoded in the likelihood function \(P(X\mid\theta)\), which describes the probability of the data given a set of parameters and is used, through Bayes' theorem, to derive the posterior distribution. In practice, however, evaluating this likelihood becomes difficult in high-dimensional parameter spaces with strong degeneracies. To overcome this limitation we use SBI, in which a forward model generates synthetic observations from physical parameters, producing paired samples \((\theta, X')\). Comparing these simulated data \(X'\) with the actual observations \(X\) then allows us to infer the posterior distribution of the parameters. \citep{Hahn2022,iglesias-navarro25,synference}

This comparison is carried out with a backward, or inverse, model. Early likelihood-free approaches such as approximate Bayesian computation estimate the posterior by simulating many realisations and retaining the parameter values that best match the observations \citep[see][for a review]{marin2011}. More recent methods use neural networks to approximate complex posterior distributions, for example with normalising flows (NFs) \citep{durkan2019}. SBI therefore avoids an explicit likelihood calculation and offers substantial flexibility in the modelling assumptions, but it also relies on high-fidelity simulations. As in any inference problem, the quality of the inferred parameters depends on the quality of the forward model, and on modelling choices such as the priors, wavelength coverage, and resolution, all of which can significantly affect the recovered stellar-population properties \citep{pacifi2023}. Developing a reliable forward model is therefore essential for robust parameter inference.

\subsection{Forward model}

\subsubsection{Population III models}
\label{sect:pop3}

We model primordial stellar populations using the Yggdrasil simple stellar populations (SSPs) framework \citep{Zackrisson2011}, based on the metal-free templates of \cite{Schaerer2002}. We adopt two zero-metallicity Pop~III families. The first is Pop~III.2, a zero-metallicity population with a moderately top-heavy IMF (log-normal with characteristic mass $M_c=10~M_{\odot}$, dispersion $\sigma=1$ and wings extending from $1-500 M_{\odot}$) from \cite{raiter2010}. The second is a Pop~III model with a Kroupa IMF \citep{kroupa2001} in the range $0.1$--$100\,\rm{M}_{\odot}$, implemented as a rescaled metal-free SSP. In all Pop~III runs, we assume an instantaneous burst star formation history (SFH). This choice is motivated by the rapid nature of primordial star formation in low-mass haloes, where strong internal feedback is predicted to quench further star formation on timescales shorter than the dynamical time of the system \citep[e.g.,][]{2004ARA&A..42...79B, 2010ApJ...716..510G}. Furthermore, a purely Pop~III phase is physically constrained to be short-lived: the first massive stars enrich the surrounding gas in $\lesssim 3$~Myr, driving the system toward a "self-polluted" or metal-enriched state. Consequently, an extended SFH built entirely from pristine Pop~III stars would lack self-consistency, whereas an SSP provides a more realistic approximation for the brief metal-free episode we aim to detect.

We skip the analysis based on Pop~III.1, following the terminology of \cite{Zackrisson2011}, with an extremely top-heavy Salpeter IMF in the stellar-mass interval $50$--$500\,M_{\odot}$, because, although these populations are intrinsically brighter, they are largely photometrically indistinguishable from Pop~III.2 in broad-band data and are theoretically less likely to be found in the redshift range $z<15$ \citep{Zackrisson2011,Trussler2023,Rusta2026}. Pop~III.1 stars are expected to form preferentially in the very first low-mass minihaloes, so by the epochs considered here they should be substantially rarer than Pop~III.2-like systems and therefore less relevant for the observational scenarios explored in this work.

The Yggdrasil templates incorporate both nebular lines and continuum, which were self-consistently calculated using \texttt{Cloudy} photoionisation models \citep{Ferland2017}. These templates parametrise the nebular covering fraction, $f_{\mathrm{cov}}$, to represent the fraction of ionising photons reprocessed by the surrounding gas. This enables a controlled transition between nearly pure stellar spectra and strongly nebular-dominated Pop~III SEDs. In this work, we adopt these pre-computed Yggdrasil grids, where the associated Lyman-continuum escape fraction is defined as
\begin{equation}
f_{\mathrm{esc}} = 1 - f_{\mathrm{cov}}.
\end{equation}

We explore two limiting cases: $f_{\mathrm{cov}} = 1.0$ (maximal nebular contribution; $f_{\mathrm{esc}} = 0.0$), and $f_{\mathrm{cov}} = 0.0$ (stars-only SED; $f_{\mathrm{esc}} = 1.0$). At $z \gtrsim 5$, even when the nebula produces strong Lyman-$\alpha$ emission (high $f_{\mathrm{cov}}$), resonant scattering by neutral hydrogen in the IGM efficiently attenuates transmission of Lyman-$\alpha$ photons, strongly suppressing the observed line flux \citep{Zackrisson2011}, and making $f_{\mathrm{esc,Ly\alpha}} \sim 0$. We leave for future analysis hybrid cases, e.g., $f_{\mathrm{cov}} = 0.5$ (mixed stellar+nebular; $f_{\mathrm{esc}} = 0.5$).

\begin{figure}[htbp]
\centering
\includegraphics[width=\columnwidth]{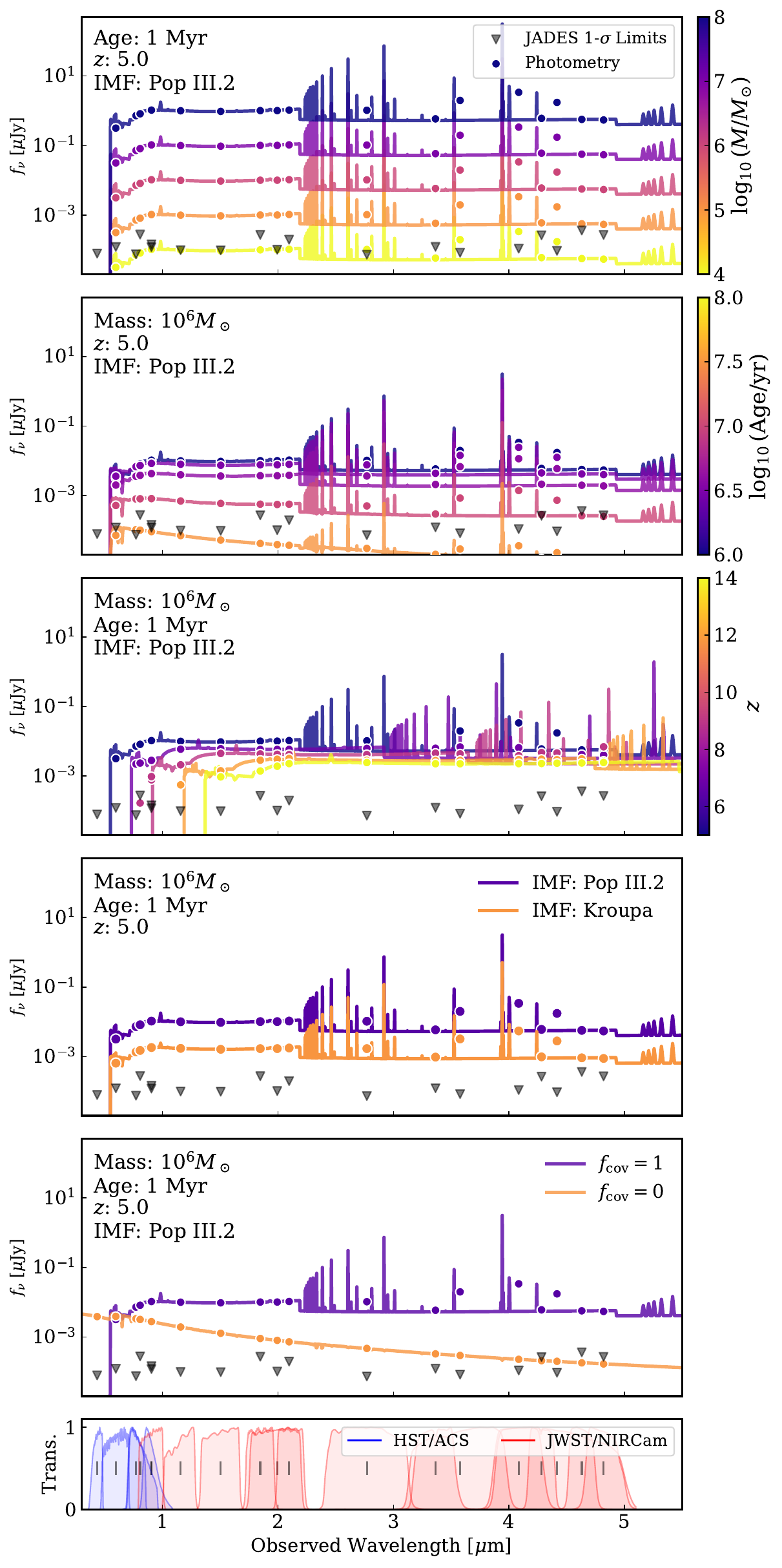}
\caption{Comparison of Pop~III SEDs showing the impact of varying one parameter at a time: stellar mass (top), age, redshift, IMF, and nebular covering fraction $f_{\rm cov}$ (bottom), while keeping the remaining parameters fixed in each panel. The sequence highlights the expected luminosity scaling with mass, SED reddening and UV-slope evolution with age, wavelength shifting with redshift, continuum and ionising-photon differences between IMF assumptions, and the effect of nebular emission: $f_{\rm cov}=1$ corresponds to a fully covered nebula with no Lyman-$\alpha$ transmission, while $f_{\rm cov}=0$ corresponds to a purely stellar SED with no nebular contribution. On the bottom we show the ACS+NIRCam filter response curves used in this work, while in the main panels we show the respective photometry values for each SED and average sensitivity values (as grey triangles) per pixel from the JADES deep survey.}
\label{fig:sed_parameter_variations}
\end{figure}

The Pop~III parameter grids are listed in Table~\ref{tab:pop3_grids}. Some examples of the SEDs when varying those ingredients are shown in Fig.~\ref{fig:sed_parameter_variations}.

\begin{table}
\caption{Pop~III forward-model ingredients and fixed assumptions.}
\label{tab:pop3_config}
\centering
\footnotesize
\begin{tabular}{p{3.2cm}p{4.8cm}}
\hline
Component & Adopted options \\
\hline
SSP library & Yggdrasil metal-free SSPs based on \cite{Schaerer2002} and \cite{Zackrisson2011} \\
IMF option 1 & Pop~III.2
A zero-metallicity population with a moderately top-heavy IMF (log-normal with characteristic mass $M_c=10 M_{\odot}$, dispersion $\sigma=1$ and wings extending from $1-500 M_{\odot}$) from \cite{raiter2010}\\
IMF option 2 & Pop~III (Kroupa): Kroupa IMF, $0.1$--$100\,M_{\odot}$ \\
SFH & Instantaneous burst (single-age SSP) \\
$f_{\mathrm{cov}}$ & $1.0$, $0.0$ \\
Lyman-$\alpha$ transmission & $0.0$, $1.0$ \\
\hline
\end{tabular}
\end{table}

\begin{table*}
\caption{Parameter grids for Pop~III simulations.}
\label{tab:pop3_grids}
\centering
\small
\begin{tabular}{lccc}
\hline
IMF & Redshift ($z$) & $\log_{10}(M/M_{\odot})$ & $\log_{10}(\text{Age/yr})$ \\
\hline
Pop~III.2 (Log-normal with $M_c = 10 \, M_\odot$) & $5.0 - 14.0$ & $4.0 - 8.0$ & $ 6.0, 6.5, 6.7, 7.0, 7.5, 7.7, 8.0$ \\
\rule{0pt}{4ex} 
Pop~III (Kroupa) & $5.0 - 14.0$ & $4.0 - 8.0$ &  Same as above, plus $4.00, 5.04, 5.32, \dots, 6.56$ (37 steps) \\
\hline
\end{tabular}
\end{table*}

The expected stellar mass of individual Pop~III star-forming events remains poorly constrained, depending on halo mass, radiative feedback, and the local environment \citep{2004ARA&A..42...79B,Liu2024}. While theoretical and observational studies suggest typical clusters range from $\sim10^3\,M_{\odot}$ up to $10^7\,M_{\odot}$ in environmentally boosted cases \citep{2024ApJ...973L..12V,Visbal2025,Nakajima2025,venditti2026}, we adopt the mass range of $10^4$--$10^8\,M_{\odot}$, covering the cluster masses most likely to be detectable. The high-mass end ($>10^7\,M_{\odot}$) is included for theoretical completeness, though detections at these masses would probably represent unresolved mixtures of Pop~III and Pop~II populations \citep{Trussler2023,Jeong2026}, while the low-mass end would require high gravitational magnification ($\mu$) in order to be detected. Furthermore, the observable window for these systems is narrow; after $\approx 3$\,Myr ($10^{6.5}$\,yr), the death of massive O-type stars abruptly curtails the ionizing flux and nebular continuum, leaving a surviving low-mass population that is intrinsically too faint for non-lensed JWST observations \citep{Trussler2023}.

\subsubsection{Fiducial models}
\label{sec:fiducial}

Our fiducial model represents enriched stellar populations (Pop~I/II) and follows the baseline non-primordial framework adopted in \cite{iglesias-navarro25}.

We first generated a dataset of simulated composite stellar populations (CSPs; interpretable as galaxies or galaxy pixels), with their corresponding ACS 
and NIRCam photometry and associated uncertainties, given their physical parameters $\theta$. We used a stellar population synthesis (SPS) approach: starting from a library of empirical stellar spectra and assuming an IMF and a set of isochrones, we constructed SEDs for SSPs spanning a range of ages and metallicities. These SSPs were then combined according to prescribed SFHs, assuming a constant metallicity and applying a dust attenuation law. For more detailed reviews of this methodology, we refer to \cite{Conroy_2013} and \cite{iyer2025spectralenergydistributionsgalaxies}.

We used the pipeline provided by \texttt{FSPS} \citep{Conroy2009, Conroy_2010}, a stellar-population synthesis code for generating model SEDs, together with \texttt{Dense Basis} \citep{Iyer2017,Iyer19}, a framework for constructing flexible non-parametric star-formation histories and composite stellar populations. We employed MIST isochrones \citep{Choi_2016}, the MILES stellar spectral library \citep{vazdekis2010}, the DL07 dust emission library \citep{draine2007}, and a Kroupa IMF \citep{kroupa2001}. We adopted a Calzetti dust attenuation law \citep{Calzetti2000}, coupling the birth-cloud attenuation to that from older stars. We generated both nebular continuum and line emission using \texttt{Cloudy} \citep{Ferland2013, Ferland2017}, setting the gas-phase metallicity equal to the stellar metallicity and assuming it remains constant over the galaxy lifetime.

We used non-parametric SFHs with the \texttt{Dense Basis} module \texttt{GP-SFH} \citep{Iyer2017,Iyer19}, allowing for complex behaviour such as rejuvenation events, bursts, and sudden quenching without imposing a fixed functional form. Our adopted prior assumes that the sSFR (i.e. the SFR normalised by stellar mass) across three equally spaced time bins follows a Dirichlet distribution with concentration parameter \(\alpha=1\). This SFH prior was studied in detail by \cite{Leja2019}. We also included the asymptotic giant branch (AGB) circumstellar dust model from \cite{Villaume2015} and accounted for stellar mass loss in the total mass budget.

The priors for the parameters controlling these components are listed in Table~\ref{tab:prior_fiducial}. Rather than matching the observed distributions, we kept the priors as uniform as possible, since inference with strongly unbalanced priors is more susceptible to systematic effects and biases \citep{hahn2023}. We set a lower stellar-mass prior limit of \(\log_{10}(M_{*}/M_{\odot})=4.0\). At these very low stellar masses per pixel, SSP models become unreliable because of statistical fluctuations in the stellar population, particularly those caused by stochastic IMF sampling \citep{cervino2003,martin-navarro26a,martin-navarro26b}. In these regimes, inferred stellar population properties such as age and metallicity are highly uncertain. In practice, however, this lower limit is required to model the faint pixels expected for most Pop~III candidates. We note that the dust absorption and emission models also assume energy balance, an approximation that is likely valid globally but not necessarily within each individual pixel.

\begin{table}
\caption{Priors for the parameters of the simulation using the Dirichlet SFH model.}
\label{tab:prior_fiducial}
\centering
\footnotesize
\begin{tabular}{p{3.2cm}p{3.4cm}}
\hline
Parameter & Prior distribution \\
\hline
$\log_{10} (M_{*}^{\rm{formed}}/\rm{M}_{\odot})^{\dagger}$ & U$(4.0,12.0)$ \\
$\log_{10}$(sSFR) & U$(-11.5,-7.5)$ \\
$t_{25\%,50\%,75\%}$$^{\dagger\dagger}$ & Dir$(\alpha=1.0,N=3)$ \\
$[M/\mathrm{H}]$ & U$(-2.3,0.4)$ \\
$A_{V}$ & U$(0.0,4.0)$ \\
$\log(U)$ & U$(-4.0,-1.0)$ \\
$z$ & U$(0.0,14.0)$ \\
\hline
\multicolumn{2}{c}{{Fixed parameters}} \\
\hline
$\gamma_{\rm{dust}}$ & $0.01$ \\
$U_{\rm{min}}$ & $1.0$ \\
$Q_{\rm{PAH}}$ & $2.0$ \\
\hline
\end{tabular}
\tablefoot{Fixed parameters include dust emission, nebular continuum, and emission information. \textbf{U}: Uniform and \textbf{Dir}: Dirichlet.}
\begin{tablenotes}
\footnotesize
\item{ $\dagger$ Formed stellar mass (integral of SFH). We also inferred the surviving stellar mass.}
\item{ $\dagger\dagger$ Cosmic times at which 25\%, 50\%, and 75\% of total stellar mass was formed, normalised to the age of the Universe.}
\end{tablenotes}
\end{table}

\subsubsection{Mock observations}
\label{sect:mocking}

The output of the forward model consists of fluxes in the 19 filters listed in Table~\ref{filters}, including ACS filters and NIRCam wide and medium-band filters. To derive the mock observational properties, we used the publicly available JWST and HST imaging from the JADES \citep{Eisenstein2026,2023ApJS..269...16R}, JEMS \citep{williams2023}, and  FRESCO \citep{Oesch2023} surveys. Table~\ref{filters} summarizes the filters, instruments, and average $5\sigma$ depths per pixel. 

To simulate realistic photometry, we modelled the flux uncertainties as a function of flux in each filter. We used the JADES segmentation map to identify all pixels classified as galaxies and, for each filter, binned the pixel fluxes into four flux intervals. Within each bin, we collected the flux uncertainties ($\sigma_i$) from the real observations and fitted their distribution with a truncated Gaussian ($x>0$) for each of those four flux bins. These fitted Gaussians capture the empirical uncertainty--flux relation, $f(\mu_{\sigma,i},\sigma_{\sigma,i})$, and are used to add noise to the simulated fluxes. For each simulated SED, we sample the noise standard deviation, $\sigma$, from the appropriate truncated Gaussian according to its flux in each filter. We then sample the corresponding noise realization from a Gaussian with mean zero and standard deviation $\sigma$.

This approach captures a broad range of uncertainties consistent with the observational data. Following the procedure described in \cite{Hahn2022}, the network inputs include both the noisy fluxes and their associated standard deviations, enabling the inference model to produce noise-aware posteriors.

We transformed both fluxes and standard deviations to AB magnitudes to avoid numerical instabilities during training caused by the wide dynamic range of possible flux values. We also accounted for the $1\sigma$ depth limit in each filter, $m_i^b$, using the values obtained by \cite{Hainline2024} for the JADES mosaics. We repeated the calculation of the $1\sigma$ depth limits in the regions where the galaxies are selected and find negligible differences. To amortize this process, we therefore used a mean depth limit independent of the specific region in which each galaxy is located. If the simulated magnitude of a CSP in a given filter, $m_i$, is fainter than the corresponding depth limit ($m_i>m_{\sigma_i}^b$), we set it to $m_i^{\prime}=99$ and set the input uncertainty to $m_{\sigma_i}^b$. Although $m_i^{\prime}=99$ is not physically meaningful (the formal value would be infinite), it is a convenient numerical convention that avoids convergence problems and clearly identifies dropouts, i.e. filters in which the source is not detected. These dropouts typically occur blueward of the Lyman break for high-redshift galaxies, although non-detections can also occur in redder filters, especially for low-S/N photometry.

\begin{table}[htbp]
\centering
\caption{Filters, cameras, and surveys used, along with the 5$\sigma$ depth limits per pixel.}
\footnotesize
\begin{tabular}{cccc}
\hline
Filter & Instrument & Survey & 5$\sigma$ [nJy] \\
\hline
F435W & ACS & HUDF & 0.394 \\
F606W & ACS & HUDF & 0.611 \\
F775W & ACS & HUDF & 0.376 \\
F814W & ACS & HUDF & 1.388 \\
F850LP & ACS & HUDF & 0.724 \\
F090W & NIRCam & JADES & 0.601 \\
F115W & NIRCam & JADES & 0.496 \\
F150W & NIRCam & JADES & 0.489 \\
F182M & NIRCam & JEMS+FRESCO & 1.361 \\
F200W & NIRCam & JADES & 0.509 \\
F210M & NIRCam & JEMS+FRESCO & 0.987 \\
F277W & NIRCam & JADES & 0.367 \\
F335M & NIRCam & JEMS & 0.616 \\
F356W & NIRCam & JADES & 0.416 \\
F410M & NIRCam & JEMS & 0.547 \\
F430M & NIRCam & JEMS & 1.327 \\
F444W & NIRCam & JADES+FRESCO & 0.472 \\
F460M & NIRCam & JEMS & 1.813 \\
F480M & NIRCam & JEMS & 1.351 \\
\hline
\end{tabular}
\label{filters}
\end{table}

To quantify how host-galaxy light affects Pop~III identification, in Sect.~\ref{sect:contamination} we build controlled composite systems with \texttt{synthesizer} \citep{Lovell2025Synthesizer,Roper2026}, an open-source framework for generating morphologically and spectrophotometrically self-consistent mock observations. In our case, its role is not simply to produce individual SEDs, but to construct galaxy realizations in which spatial structure and pixel-level spectral properties are modelled consistently. 

The host galaxy stellar populations are modelled using the FSPS v3.2 templates \citep{Conroy2009} based on the MILES stellar library \citep{vazdekis2010} and MIST isochrones \citep{Choi_2016}, adopting a Kroupa-like IMF and a delayed exponential SFH. To constrain the parameter space dimensionality, we adopt fixed metallicity ($Z=0.001$) and negligible dust attenuation—justified at high-$z$, where young ages and limited enrichment histories restrict both metal content and dust production \citep{trump2023, curti2025, isobe2026}. Nebular emission for the Pop~II host is self-consistently included by coupling the incident radiation with \texttt{Cloudy} grids via the incident and nebular emission models within \texttt{synthesizer}, while the Pop~III component follows the modelling described in Sect.~\ref{sect:pop3}. 

For self-consistent mock observations, we apply SPH kernel smoothing to the particles. We use a $0.1$\,kpc smoothing length for host stars to mimic diffuse light from high-redshift dwarf galaxies ($R_e \approx 0.2$--$0.8$\,kpc), whereas a smaller $0.05$\,kpc length preserves the compact, point-like appearance of Pop~III clusters within minihalo cores. The resulting simulations are projected onto the JWST/NIRCam and HST/ACS filter sets. Finally, we apply the realistic noise model described at the beginning of Sect.~\ref{sect:mocking}

To ensure the reliability of the inferred properties for the  candidate of \cite{Reumert2026}, located in the PRIMER UDS \citep{dunlop_primer,austin_inprep}, we repeat the empirical noise modelling and training process specifically for this field. This step is critical as the observational noise fingerprint---including the average $\sigma$ and the depth of the imaging---varies significantly between the ultra-deep JADES footprints and the wider, shallower PRIMER survey. We then re-characterize the uncertainty distributions by binning the local pixel-level data as described above. The filter configurations are adjusted to match the specific available photometry, a combined set of 12 filters, incorporating HST/ACS (F606W, F814W) and NIRCam wide-bands (F090W through F444W). By training survey-specific SBI models, we guarantee that the resulting posteriors are conditioned on the exact instrumental noise and depth limits ($m_{\sigma_i}^b$).

\subsection{Inverse model}

We first infer the physical properties of the observed sources from their SEDs using neural posterior estimation (NPE), implemented with normalising flows. Specifically, we adopt a masked autoregressive flow with 15 transforms and 500 hidden features to model the posterior distribution $P(\theta \mid D)$ directly. We implement this approach with \texttt{SBIPIX} \citep{iglesias-navarro25} and the \texttt{SBI} \citep{Tejero-Cantero2020} Python packages. By learning the inverse mapping from simulated observables to physical parameters, NPE captures complex parameter degeneracies without requiring the evaluation of an explicit tractable likelihood function.

The redshift of each source is assumed to be known from spectroscopic or photometric measurements and is passed directly as an additional conditioning input to the network, appended to the observed per-pixel photometry and uncertainties. In practice, redshift is often inferred from photometry using photo-$z$ codes; our framework naturally accommodates redshift uncertainty as an extra input, allowing the network to learn how this propagates into the inferred physical parameters. When spectroscopic redshifts are available, they can be used instead. Importantly, since redshift is determined once per galaxy rather than per-pixel, this two-step inference strategy is computationally efficient and physically motivated. By conditioning on redshift, the network can account for the strong redshift dependence of the SED without requiring marginalization over a high-dimensional redshift dimension, freeing capacity to resolve parameter degeneracies in stellar population properties.

We then perform the fitting procedure on a pixel-by-pixel basis in order to account for spatial variations in the stellar populations. Because our approach is likelihood-free, observational constraints such as non-detections and upper limits are incorporated directly during the forward-modelling stage used to generate the training set. The prior distributions for the parameters $\theta$ are defined as described in Table~\ref{tab:pop3_grids} and Table~\ref{tab:prior_fiducial}, and are chosen to be as uniform as possible.

Because NPE does not inherently provide the marginal likelihood required for formal Bayesian model selection, we assess model performance through posterior predictive checks (PPCs). We generate synthetic mock observations, $F_{\mathrm{mock}}$, by drawing parameter sets from the inferred posterior and passing them back through the forward model. To quantify the agreement between the observed SEDs and these posterior predictions, we compute the $\chi^2$ statistic,

\begin{equation}
    \chi^2 = \sum_{i=1}^{N_{\mathrm{bands}}} \left( \frac{F_{\mathrm{obs},i} - F_{\mathrm{mock},i}}{\sigma_i} \right)^2,
\end{equation}

where $F_{\mathrm{obs},i}$ and $F_{\mathrm{mock},i}$ are the observed and mock fluxes in band $i$, respectively, and $\sigma_i$ is the total observational uncertainty. To handle non-detections in the PPC $\chi^2$ consistently, we adopt a soft upper-limit penalty. For photometric bands in which the source is undetected, we set the observed flux to zero ($F_{\text{obs}} = 0.0$) and the corresponding uncertainty to the $1\sigma$ detection limit for those photometric bands ($\sigma = \sigma_{\text{limit}}$), specified in Table~\ref{filters}. The $\chi^2$ contribution from these bands then becomes:

\begin{equation}
    \chi^2_{\text{non-det}} = \left( \frac{F_{\text{mock}}}{\sigma_{\text{limit}}} \right)^2.
\end{equation}

This formulation acts as a half-Gaussian penalty: it strongly penalises posterior samples that predict significant emission in dropout bands ($F_{\text{mock}} \gg \sigma_{\text{limit}}$), while remaining permissive of models that correctly predict faint, undetectable fluxes ($F_{\text{mock}} \lesssim \sigma_{\text{limit}}$). In this way, the physical information carried by the non-detections --- such as the location of the Lyman break --- is preserved during model comparison.

We use this PPC $\chi^2$ distribution to compare the fits obtained using our Pop~III model directly with the fiducial model. This comparison allows us to determine whether a framework incorporating primordial stellar populations systematically reproduces the observed photometry better than standard enrichment scenarios. At the same time, this statistic should be interpreted as a predictive goodness-of-fit comparison rather than as a formal model-selection quantity. The fiducial model spans a broader and more weakly constrained parameter space than the Pop~III model, so differences in effective flexibility, prior volume, and number of free parameters can in principle introduce systematic biases in $\Delta\chi^2$. In practice, this means that a Pop~III preference is a conservative result: because the Pop~III family is more restrictive, cases in which it still outperforms the fiducial model are likely to be cleaner and therefore more scientifically compelling, even if the PPC-based comparison does not fully account for model complexity.

In summary, our approach infers, for each model family, the range of physical properties that can plausibly reproduce the observations under the measured uncertainties, and then re-simulates the corresponding observables from posterior samples. We subsequently compare the posterior predictive performance of the Pop~III and fiducial models on a pixel-by-pixel basis. Carrying out the same analysis with conventional MCMC methods would be computationally expensive, because it would require repeated posterior sampling for two distinct models in every pixel, together with an explicit choice for the likelihood function. The main advantage of that approach would be direct access to the Bayesian evidence, enabling a more formal model comparison. In future work, this limitation could be addressed with neural ratio estimation (NRE) or evidence networks \citep{2024MLS&T...5a5008J}, as discussed below, or by combining MCMC with NPE-informed proposal distributions.

The primary metric for model preference is defined as the logarithmic difference in the goodness-of-fit derived from the PPC:

\begin{equation}
\label{eq:model_comparison_eq}
    \Delta \log_{10} \chi^2 = \log_{10}(\chi^2_{\mathrm{fiducial}}) - \log_{10}(\chi^2_{\mathrm{PopIII}}).
\end{equation}

A positive $\Delta \log_{10} \chi^2$ indicates a statistical preference for the Pop~III model. Because the fiducial model is more flexible, Pop~III preference only occurs when its distinctive 
spectral features are genuinely present, as demonstrated in the following sections. To interpret the complex, non-linear dependencies that determine when a Pop~III signature is statistically favoured over a standard stellar population, we use SHapley Additive exPlanations \citep[SHAP;][]{Lundberg2017SHAP,Lundberg2020TreeSHAP}.  We train a random forest regressor with 100 decision trees as a surrogate model. We use the Python package \texttt{SHAP} \citep{Lundberg2017SHAP}. This surrogate is trained to predict $\Delta \log_{10} \chi^2$ directly from the ground-truth physical parameters of our simulated catalogue.

By applying SHAP to the regressor, we decompose the predicted $\Delta \log_{10} \chi^2$ into additive marginal contributions from each physical feature. This allows us to rank the importance of individual parameters in driving the detectability of Pop~III sources. It also helps us identify approximate physical thresholds --- such as a critical mass ratio or a maximum age beyond which diagnostic emission lines fade --- where the preference for the Pop~III model collapses and the signal becomes indistinguishable from a fiducial Pop~I/II source.

As the detectability predictions depend sensitively on the treatment of nebular emission and the Lyman-$\alpha$ escape/transmission assumptions, the SBI training must be re-run for each model configuration to enable a consistent and quantitative benchmark across scenarios. The same requirement applies to IMF assumptions, since changing them modifies the ionising-photon budget and continuum shape, and therefore the resulting detectability landscape.

\section{Analysis of isolated Pop~III sources}
\label{sec:isolated}

To establish the baseline performance of our SBI pipeline, we first evaluate parameter recovery for a control sample of isolated Pop~III sources. These mock targets are generated assuming the Pop~III.2 IMF, maximal nebular continuum and line emission (100\%), and full Lyman-$\alpha$ obscuration ($f_{\mathrm{esc, Ly}\alpha}=0.0$). To replicate realistic observational conditions, we inject pixel-level empirical noise as described in Sect.~\ref{sect:mocking}. We then run inference on each source photometry, adopting a detection threshold of $S/N \geq 1.0$ in the first band redward of the observed Lyman-$\alpha$ break. This band typically contains the highest Pop~III stellar flux, so the cut guarantees a detection in at least one band, which is necessary (but, as shown below, not sufficient) to distinguish Pop~III from the fiducial model. The expected S/N as a function of stellar mass and redshift for this configuration is shown in Appendix~\ref{fig:salpeter_snr_1e6} and~\ref{fig:salpeter_snr_1e7}. Before turning to Pop~III identification performance, we present global recovery trends to validate the accuracy and stability of the inferred posteriors under these baseline conditions, together with the PPC results.

\subsection{Recovering properties of isolated Pop~III sources}

As discussed in Sect.~\ref{sec:fiducial}, we train a fiducial model on standard Pop~I/Pop~II stellar population models using the MILES library. Rather than adopting a narrowly defined scenario, we use a broad prior space and a non-parametric SFH family, which in a pixel-based context provides sufficient flexibility to describe most Pop~I/Pop~II. We extend this model to $z=14$ and train on $1{,}000{,}000$ simulations, obtaining $R^2$ values\footnote{$R^2$, the coefficient of determination, measures the fraction of variance in the target variable explained by the model; values closer to 1 indicate better predictive performance.} on a 1,000-simulation test set comparable to those reported in \cite{iglesias-navarro25}, ranging from $R^2=0.98$ for the stellar mass to $R^2=0.39$ for the metallicity.

On the other hand, we train the NPE model on Pop~III simulations (Pop~III.2 IMF, $f_{\mathrm{cov}}=1.0$, $f_{\mathrm{esc,Ly\alpha}}=0$), constructing a continuous mock catalogue of $100{,}000$ synthetic observations based on pre-computed Yggdrasil stellar population models. We uniformly sample the physical parameter space in redshift ($z$), stellar mass ($\log_{10}(M_*/M_{\odot})$), and stellar age. To map these continuous physical parameters to observational photometry, we apply a 2D regular-grid logarithmic interpolation across the mass and age axes, using nearest-neighbour matching for the discrete redshift grid ($\Delta z = 0.1$). During this sampling phase, we enforce a strict cosmological prior to ensure that no sampled stellar age exceeds the age of the Universe at its corresponding redshift. We then convert the resulting integrated fluxes in the 19 bands to magnitudes, and apply the mocking procedure to build the final training set.

 Once trained, we evaluate the model performance for recovering Pop~III mass and age from multi-band photometry and associated uncertainties, with redshift included as an explicit conditioning variable as specified above. While this approach requires \textit{a priori} redshift constraints---either from spectroscopy or from external photometric-redshift codes---it provides the most direct and computationally tractable framework to isolate and assess the network's ability to identify Pop~III signatures without compounding the inference problem with redshift-property degeneracies.

\begin{figure}[htbp]
\centering
\includegraphics[width=\columnwidth]{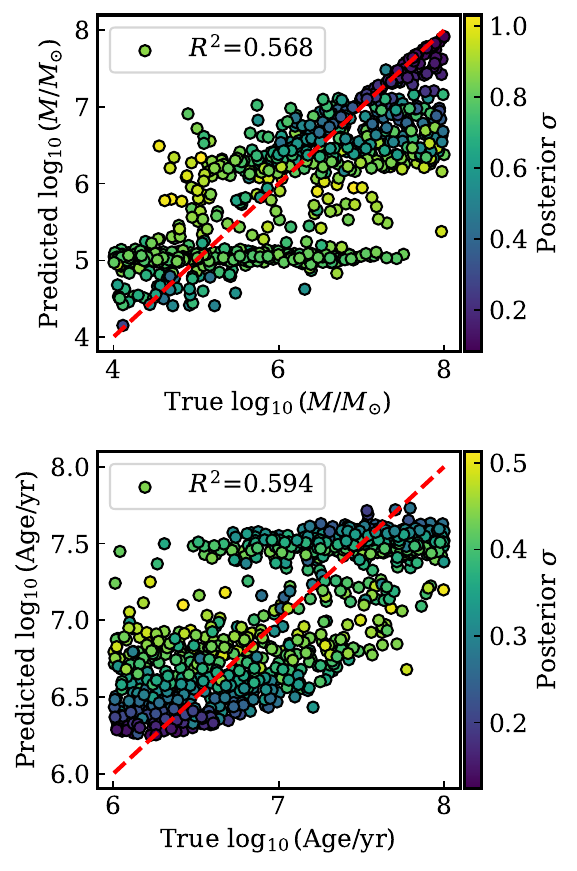}
\caption{Performance diagnostics for the Pop~III model with Ly$\alpha$ transmission fixed to 0.0 and nebular covering fraction fixed to 1.0. The top panel shows stellar mass and the bottom panel stellar age. The points represent the posterior medians for $1000$ simulated Pop~III sources, color-coded with the standard deviations and plotted against the true values. The one-to-one relation is shown with a red dashed line.}
\label{fig:r2}
\end{figure}

We obtain posterior distributions for the stellar mass and age for the full test set.
The $R^2$ values, computed with the median of the posterior distributions, are $0.57$ and $0.59$ respectively, as shown in Fig.~\ref{fig:r2}. 

\begin{figure}[htbp]
\centering
\includegraphics[width=\columnwidth]{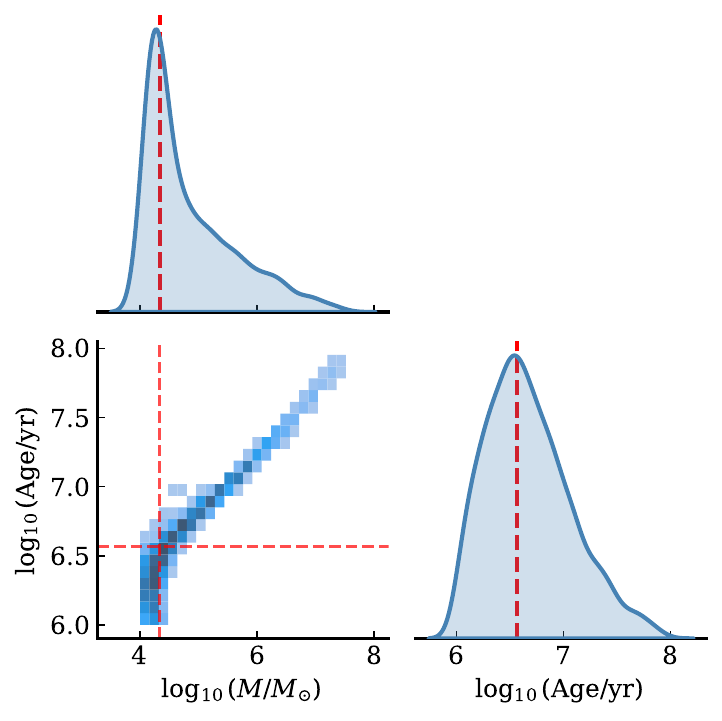}
\caption{Corner plot showing the posterior distributions of stellar mass and age for one simulated source fitted with the Pop~III model, with Ly$\alpha$ transmission fixed to 0.0 and nebular covering fraction fixed to 1.0. The true parameter values are indicated by the red dashed lines. The two parameters are strongly degenerate over several orders of magnitude.}
\label{fig:example_posts}
\end{figure}

The modest performance in the mass-recovery diagram is closely linked to the mass--age degeneracy, which is evident both in Fig.~\ref{fig:sed_parameter_variations} and in Fig.~\ref{fig:example_posts}. In particular, these two properties are strongly correlated: older and more massive populations can produce photometry similar to that of younger, lower-mass populations. For example, as seen in Fig.~\ref{fig:example_posts}, a Pop~III source with $M_* \sim 10^4\,M_\odot$ and $\log_{10}(\mathrm{Age/yr}) \lesssim 6.5$ can produce a broadband SED comparable to one with $M_* \sim 10^5\,M_\odot$ and $\log_{10}(\mathrm{Age/yr}) \approx 7.0$, making the two configurations difficult to distinguish from photometry alone.

Even so, the age posterior is constrained at a level comparable to that of the mass posterior, and it consistently recovers two distinct age regimes. This behaviour is physically driven by the lifetimes of massive Pop~III O stars. During the first $\sim$3~Myr after a burst ($\log_{10}(\mathrm{Age/yr}) \approx 6.5$), the SED is dominated by very hot, massive stars and by the strong nebular continuum and emission lines they power.

After $\sim$3~Myr, as the most massive O stars reach the end of their main-sequence lifetimes and explode as supernovae, the ionizing-photon production drops sharply, the nebular emission fades, and the continuum shape changes substantially. This transition becomes even more pronounced at later times (e.g., $>10$~Myr, $\log_{10}(\mathrm{Age/yr}) > 7.0$).

Because our main objective is model comparison rather than the most precise possible recovery of individual physical parameters, the key requirement is not simply accurate point estimates, but well-calibrated, noise-aware posterior distributions for both competing models. This is what allows a fair comparison between Pop~III and fiducial interpretations under realistic observational uncertainties. We verify this calibration in Appendix~\ref{sec:SBC} using simulation-based calibration, specifically performing Tests of Accuracy with Random Points (TARP) tests \citep{lemos2023sampling}, for both models.

Once trained, the amortised NPE models are extremely fast to evaluate. Drawing 500 posterior samples per pixel takes ${\sim}15$--$20\,\mathrm{ms}$ for the Pop~III model and ${\sim}120\,\mathrm{ms}$ for the fiducial model on an Apple M3 Max (14-core CPU).

We also train the model for the additional configurations listed in Table~\ref{tab:pop3_config}. First, adopting a Kroupa IMF yields $R^2$ values of 0.75 and 0.16 for mass and age, respectively. In this case, mass recovery improves and becomes more continuous compared with the top-heavier IMF, as shown in Fig.~{\ref{fig:r2_kroupa}}. Then, keeping the Pop~III.2 IMF fixed, we change the nebular coverage fraction to 0 (as shown in Fig.~\ref{fig:r2_fcov0}), obtaining similar trends and $R^2$ scores of 0.61/0.25 for mass and age respectively.

\subsection{Identifying isolated Pop~III}

\begin{figure}[htbp]
\centering
\includegraphics[width=\columnwidth]{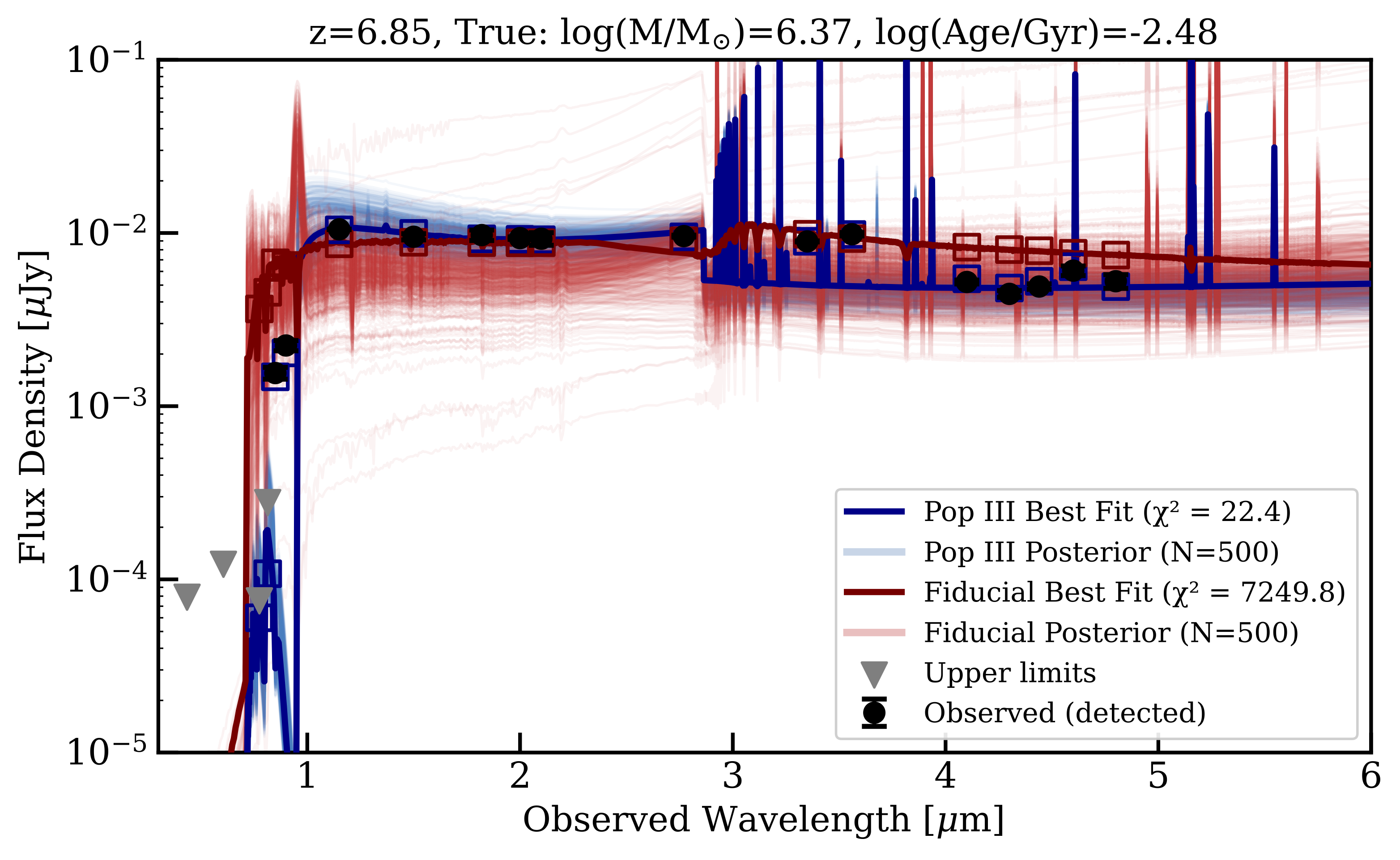}
\caption{Posterior predictive checks for one simulation using the Pop~III configuration (Pop~III.2 IMF with maximal nebular contribution and Lyman-$\alpha$ escape fraction of 0.0). Black dots with error bars show detected photometry, downward grey triangles mark upper limits, blue lines indicate the spectra corresponding to the Pop~III model posterior samples, with a darker blue line for the best-fit and blue squares for its corresponding photometry. The same elements in red indicate the SEDs from the posterior samples and the best fit from the fiducial model.}
\label{fig:ppc}
\end{figure}

As a visual reference for this analysis, Fig.~\ref{fig:ppc} shows a representative PPC example from our full set of isolated-source tests. For each source, we generate 500 SED realizations by sampling the Pop~III posterior (Pop~III.2 IMF, $f_{\mathrm{cov}}=1.0$, $f_{\mathrm{Ly}\alpha}=0.0$), and we display the minimum-$\chi^2$ realization as the best-fit Pop~III model. The same is done with the fiducial model. In this example the Pop~III model is preferred ($\chi^2=22.4$) over the fiducial fit ($\chi^2=7249$).

\begin{figure*}[htbp]
\centering
\includegraphics[width=0.98\textwidth]{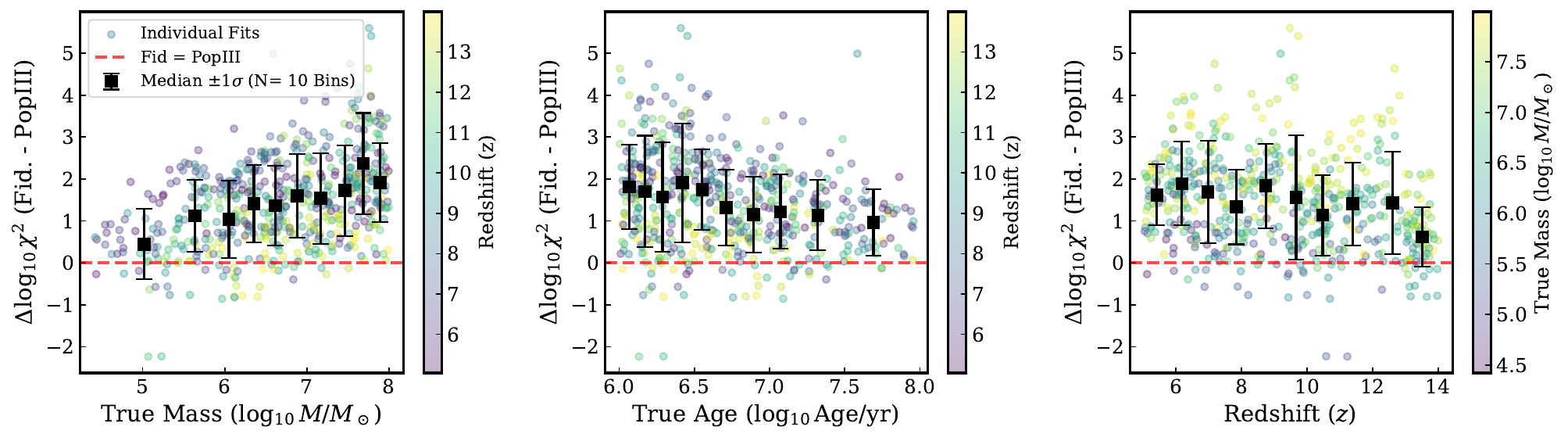}
\caption{Difference in goodness-of-fit, $\Delta\log_{10}\chi^2 = \log_{10}(\chi^2_{\mathrm{fid}})-\log_{10}(\chi^2_{\mathrm{PopIII}})$, for isolated Pop~III mocks (Pop~III.2 IMF, $f_{\mathrm{esc}}^{\mathrm{Ly}\alpha}=0$, $f_{\mathrm{cov}}=1.0$) as a function of true mass (left), true age (middle), and redshift (right). Individual simulations are shown as circles. Colors encode redshift in the left and middle panels, and true mass in the right panel. Black squares with error bars indicate the median and $1\sigma$ interval in bins of the x-axis variable, and the red dashed line marks $\Delta\log_{10}\chi^2=0$ (equal fit quality). Positive values indicate a preference for the Pop~III model over the fiducial model.}
\label{fig:delta_logchi2_vs_physical_properties}
\end{figure*}

For each simulated source, we summarize the relative model preference with Eq.~\eqref{eq:model_comparison_eq},
so that positive values favour the Pop~III interpretation. Figure~\ref{fig:delta_logchi2_vs_physical_properties} shows that most simulations lie at $\Delta\log_{10}\chi^2>0$, indicating successful recovery of the primordial model in the isolated-source regime, even though the fiducial model's higher flexibility would nominally favour it. The preference becomes stronger toward higher stellar masses, younger ages, and lower redshift, and weakens gradually toward older, fainter, and higher-$z$ systems, where fewer bands are detected and the spectral signatures become less distinctive. The small population of outliers with $\Delta\log_{10}\chi^2<0$ is concentrated mainly in the low-mass and older-source part of parameter space, where the fiducial model better reproduces the broadband SED. At low masses and higher redshifts, fewer bands reach the detection threshold, reducing the number of constraining data points and therefore the discriminatory power between the two models. In principle, MIRI \citep{2015PASP..127..584R} imaging could extend the wavelength coverage and provide additional constraints on the high-$z$ regime; however, the sensitivity depths currently achievable are likely insufficient to detect faint Pop~III sources \citep{Trussler2023}.

To interpret these trends beyond pairwise projections, we apply a two-step explainability pipeline to the same isolated-simulation catalogue. First, we train a random forest regressor as a surrogate model to predict $\Delta\log_{10}\chi^2$ from the three physical parameters. This choice is motivated by the strongly non-linear parameter couplings in the problem (e.g., mass and redshift jointly setting effective detectability). The surrogate reaches $R^2=0.72$ for the model with Pop~III.2 IMF, $f_{\mathrm{esc}}^{\mathrm{Ly}\alpha}=0$, and $f_{\mathrm{cov}}=1$, indicating that it captures a substantial fraction of the variance while preserving enough complexity for feature-attribution analysis. Second, we compute SHAP values for each simulation to quantify the marginal contribution of each parameter to the model preference score.

\begin{figure}[htbp]
\centering
\includegraphics[width=\columnwidth]{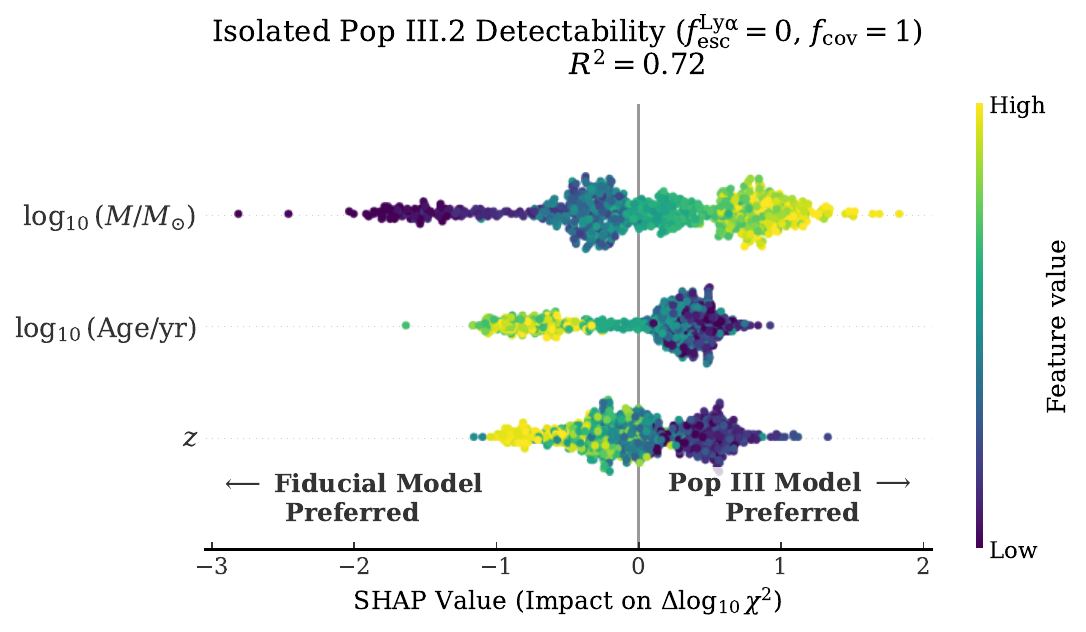}
\caption{SHAP beeswarm analysis (Pop~III.2 IMF, $f_{\mathrm{esc}}^{\mathrm{Ly}\alpha}=0$, $f_{\mathrm{cov}}=1$). Positive values indicate a preference for the Pop~III model over the fiducial model. Features are ranked from top to bottom by decreasing importance; the colour encodes the feature value from low (purple) to high (yellow), with the parameter ranges shown in Fig.~\ref{fig:delta_logchi2_vs_physical_properties}. The surrogate model reaches $R^2=0.72$.}
\label{fig:shap_single_clump}
\end{figure}

 In Fig.~\ref{fig:shap_single_clump}, the x-axis gives the SHAP impact on $\Delta\log_{10}\chi^2$ (positive values push toward Pop~III preference; negative values toward fiducial preference), the y-axis ranks features by global importance, and the point color encodes the feature value. The ranking identifies $\log_{10}(M/M_{\odot})$, $z$, and $\log_{10}(\mathrm{age}/\mathrm{yr})$ as the dominant drivers. High masses (red points in the top row) shift predictions toward positive SHAP values, whereas low masses shift toward negative values. For redshift, lower-$z$ systems preferentially contribute positive SHAP values, while higher-$z$ systems tend to reduce Pop~III preference. For age, younger bursts drive positive SHAP contributions and older bursts drive negative ones. Together, these diagnostics confirm (as expected) that isolated Pop~III recovery is most robust for massive, young, and comparatively lower-redshift systems in the explored parameter space.

The physical drivers identified by the SHAP analysis can be understood directly in terms of the underlying spectral features accessible to JWST photometry. At the masses and ages where Pop~III recovery is most robust ($M_* \gtrsim 10^5\,M_\odot$, age $\lesssim 3$ Myr), the SED is dominated by nebular continuum and strong emission lines --- principally He\,II $\lambda1640$, Ly$\alpha$, and the two-photon continuum --- which together produce a distinctive blue spectral slope with a hard ionising continuum that is difficult to replicate with enriched stellar populations at the same mass. As the burst ages beyond $\sim3$--$5$ Myr, the most massive O stars die off, the ionising-photon rate drops sharply, and this spectral contrast collapses: the SED transitions toward a fainter, redder stellar continuum that is nearly indistinguishable from a young low-metallicity Pop~II source in broadband photometry. This explains the strong age dependence recovered by the SHAP analysis. The redshift dependence arises because the position of the Lyman-$\alpha$ break relative to the NIRCam bandpass determines which filters sample the nebular-dominated rest-frame UV: at lower redshifts ($z \lesssim 7$), multiple blue filters constrain the continuum slope and line excess simultaneously, whereas at $z \gtrsim 9$ the rest-frame optical diagnostics (He\,II $\lambda4686$, H$\beta$) shift beyond the NIRCam long-wavelength channel into the MIRI regime, where current survey depths are insufficient to add constraining power. These spectral considerations directly inform the practical target selection criteria derived from our framework.

For the other model configurations, the trends in Fig.~\ref{fig:delta_logchi2_vs_physical_properties} remain broadly similar, as does the SHAP analysis, shown in Fig.~\ref{fig:shap_single_clump_ly1} and \ref{fig:shap_single_clump_kroupa}. As a control test, we also apply the same model-comparison procedure to Pop~I/II galaxies, using both simulated enriched galaxies (Appendix~\ref{sec:app_control_galaxies}) and real JADES DR5 sources (Appendix~\ref{sec:app_jades_control}). In neither case do we find Pop~III preference. This confirms that the isolated-source preference is not produced generically by the Pop~III model, but instead appears when the input SED is genuinely Pop~III-like.

\section{Contamination from a host galaxy}
\label{sec:contamination}
\label{sect:contamination}

In practice, detecting the very first Pop~III stars in complete isolation is expected to be extremely difficult; their short lifetimes, low intrinsic luminosities, and rapid local metal enrichment make genuinely pristine systems both transient and observationally rare, even at JWST depths. An expected and more common scenario is that a galaxy has already formed stars and later accretes pristine gas, triggering a Pop~III episode embedded in pre-existing host light, where the primordial signal is diluted by older stellar populations and nebular emission from the host \citep{venditti2026}.

To quantify how host-galaxy light affects Pop~III identification, we build controlled composite systems with \texttt{synthesizer}\footnote{\url{https://github.com/synthesizer-project/grid-generation}} \citep{Lovell2025Synthesizer,Roper2026} in two stages. First, we combine a standard Pop~II component with a $\tau$-delayed star-formation history and a superimposed Pop~III burst, and compute the total unresolved photometry by integrating both components in the same flux units and filter set. This experiment isolates purely spectral confusion (without morphology) and tests whether the Pop~III signal can still be recovered when diluted by realistic Pop~II emission. Second, we move to spatially structured mocks by embedding a Pop~III clump in a S\'ersic host galaxy generated from the same $\tau$-delayed Pop~II model, exploring different separations, contrast ratios, and Pop~III/host configurations. Comparing inference performance between the integrated and resolved setups allows us to determine when unresolved analyses fail and to motivate a spatially resolved strategy as the key route to robust Pop~III recovery in realistic galaxies.

\subsection{Integrated Pop~III--Pop~II Overlap}

For this first blending experiment we ignore morphology and focus on composite SFHs in integrated photometry. We generate mock systems as the sum of (i) a Pop~II host described by a $\tau$-delayed SFH and a Kroupa IMF and (ii) a superimposed Pop~III burst, with a Pop~III.2 IMF, $f_{\mathrm{cov}}=1.0$, and $f_{\mathrm{esc,Ly\alpha}}=0.0$, both modelled with \texttt{synthesizer} and combined before adding observational noise. We adopt the simplified configuration shown in Table~\ref{tab:parameter_grid_int} to restrict the size of the multidimensional parameter space.

\begin{table}
\caption{Parameter grid for the integrated Pop~III--Pop~II overlap simulations.}
\label{tab:parameter_grid_int}
\centering
\footnotesize
\begin{tabular}{p{3.5cm}p{4.4cm}}
\hline
Component & Adopted values \\
\hline
Redshift & $z=5.0,\ 8.0$ \\
Host stellar mass & $M_{\mathrm{host}}=10^6\,M_{\odot}$ \\
Host maximum age & $t_{\mathrm{host,max}}=500\,\mathrm{Myr}$ \\
Host SFH timescale & $\tau=10,\ 100,\ 500\,\mathrm{Myr}$ (delayed exponential) \\
Pop~III clump mass & $M_{\mathrm{PopIII}}=10^4,\ 10^5,\ 10^6\,M_{\odot}$ \\
Pop~III age & $t_{\mathrm{PopIII}}=1,\ 2,\ 4\,\mathrm{Myr}$ \\
Pop~III metallicity & $Z=0.0$\\
Host galaxy metallicity & $Z=0.001$ \\
\hline
\end{tabular}
\end{table}

This grid is designed to span the transition from clearly detectable, UV-bright Pop~III bursts to regimes where the Pop~III contribution is strongly diluted by the underlying Pop~II continuum. 
We adopt a single host stellar mass ($M_{\mathrm{host}}=10^6\,M_{\odot}$) providing a regime that favours the Pop~III detectability, with Pop~III stellar masses between 0.01 and 1.0 times the host stellar mass. While a host metallicity of $Z=0.001$ provides a clean photometric contrast to validate our spatial decomposition framework, we note that this represents a relatively evolved ambient environment \citep{trump2023, curti2025, isobe2026}. The most stringent test for distinguishing primordial populations occurs when contrasting pure Pop~III stars against extremely metal-poor Pop~II populations ($Z \sim 10^{-4}$--$10^{-5}$ or lower). At such extremely low metallicities, the stellar continuum and ionizing spectra of Pop~II stars closely mimic those of metal-free populations, rendering any standard colour selection techniques degenerate. We adopt this moderately enriched host as an initial baseline case to evaluate morphological dilution, leaving the degeneracies associated with more pristine hosts for a dedicated future analysis. Similarly, for the host galaxy component, we exclusively model extended, $\tau$-delayed SFHs, ensuring a physically motivated stellar population that contrasts sharply with the instantaneous burst assumed for the primordial clump.

\begin{figure*}[htbp]
\centering
\includegraphics[width=\textwidth]{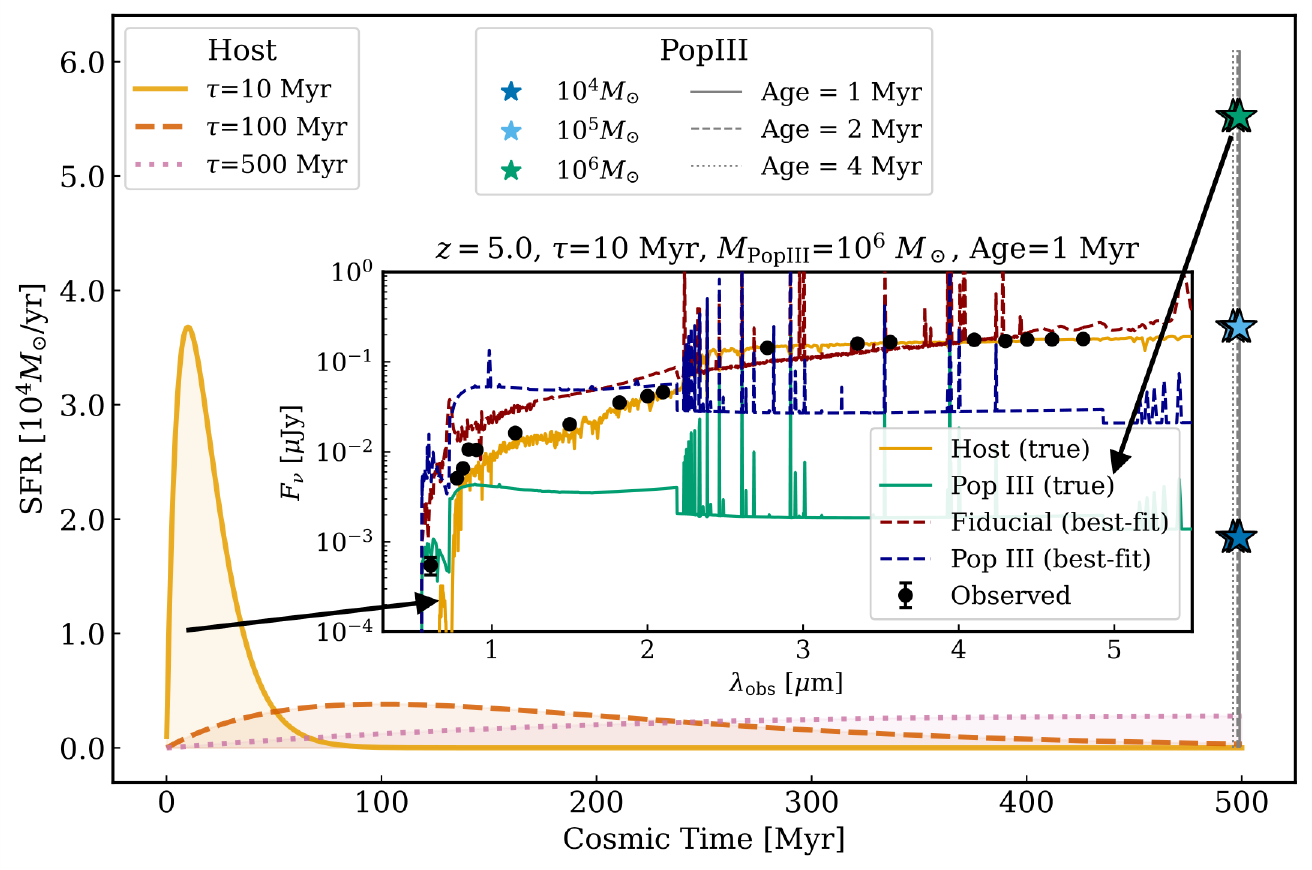}
\caption{Overlapping integrated setup \& posterior predictive checks ($z=5$): Main plot displays host-galaxy SFHs for delayed-exponential timescales $\tau = \{10, 100, 500\}$~Myr (yellow/orange/pink curves), with the x-axis indicating cosmic time and the y-axis showing SFRs. Vertical lines and star symbols mark the Pop~III instantaneous burst ages ($1, 2, 4$~Myr) and masses ($10^4, 10^5, 10^6\,M_{\odot}$) explored. The vertical placement of star symbols is arbitrary, purely for visualization. \textit{Inset:} Integrated SEDs and posterior predictive check for a representative 
configuration ($\tau_{\mathrm{host}}=10$~Myr, $t_{\mathrm{Pop~III}}=1$~Myr, 
$M_{\mathrm{Pop~III}}=10^6\,M_{\odot}$). The host SED is shown in yellow and the Pop~III 
component in green. Mock integrated photometry (black circles with errorbars; errors smaller than 
marker size) is overlaid. Best-fit SEDs for the fiducial (dark red dashed line) and 
Pop~III (blue dashed line) models are shown. The Pop~III model systematically underpredicts 
red-ward fluxes ($\lambda_{\mathrm{obs}} \gtrsim 2.5\,\mu\mathrm{m}$), whereas the fiducial 
model successfully recovers the observed continuum with lower $\chi^2$, providing a better fit to the observed data.}
\label{fig:sfhs}
\end{figure*}

The goal of this subsection is to quantify recovery in the integrated case alone, establishing a baseline for the resolved analysis presented next. An illustrative summary of the SFH configurations for $z=5$ is shown in Fig.~\ref{fig:sfhs}. It also includes an example SED decomposition (host, Pop~III burst, and total observed SED) for $z=5.0$, $\tau_{\mathrm{host}}=10$ Myr, $t_{\mathrm{PopIII}}=1$ Myr, and $M_{\mathrm{PopIII}}=10^{6}\,M_{\odot}$. In this example, the bluest filters (observed wavelengths below about $2\,\mu\mathrm{m}$) are primarily driven by the Pop~III slope and flux, whereas at longer wavelengths the total SED remains dominated by the host continuum, to which long-lived low-mass Pop~II stars contribute most of the light.

\begin{figure}[htbp]
\centering
\includegraphics[width=\columnwidth]{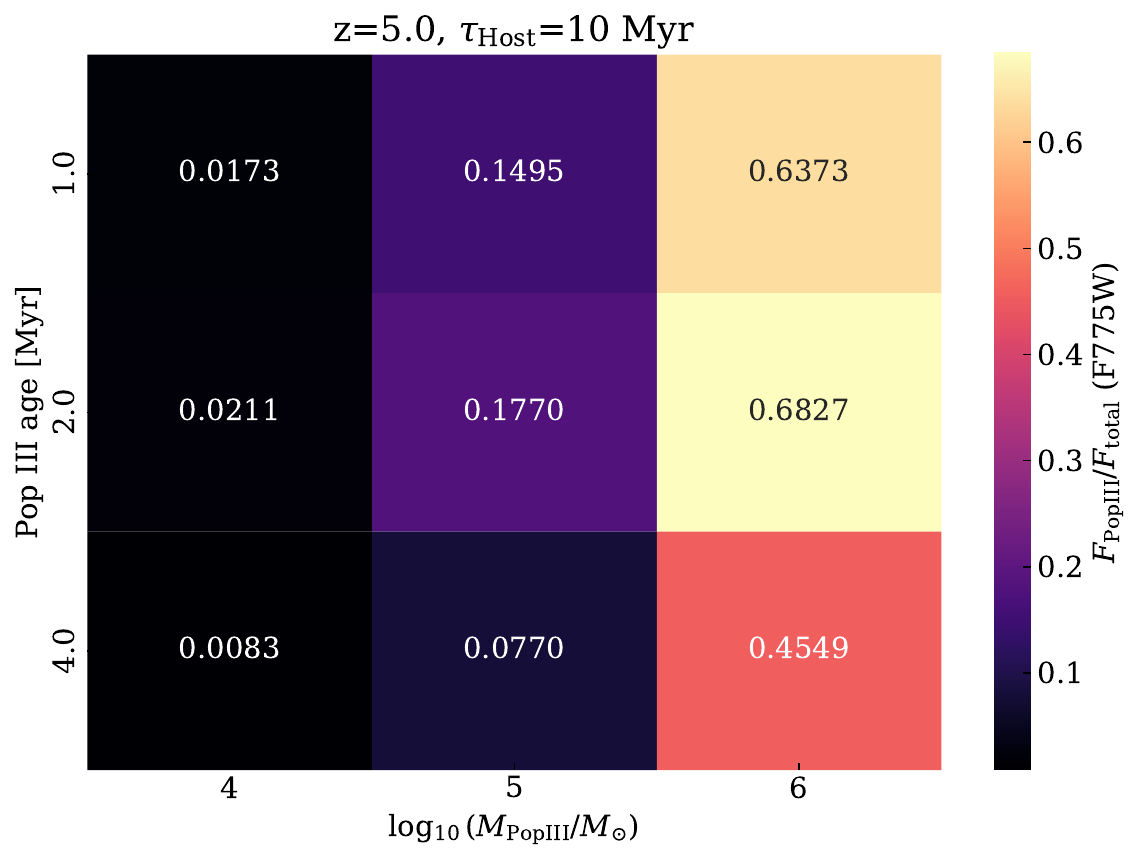}
\caption{Pop~III-to-total flux-ratio heatmap for the integrated SFH-overlap experiment at $z=5$ and host SFH timescale $\tau=10$ Myr. The flux ratio is measured in the filter immediately redward of the Lyman-$\alpha$ line (F775W at $z=5$).}
\label{fig:flux_contribution}
\end{figure}

To explore the most favourable integrated configuration in more detail, Fig.~\ref{fig:flux_contribution} shows the Pop~III-to-total flux ratio at $z=5$ for the rapidly declining host SFH ($\tau=10\,\mathrm{Myr}$). The ratio is measured in F775W, the filter immediately redward of Ly$\alpha$ at this redshift, where the Pop~III contrast is expected to be strongest. Even in this favourable case, the Pop~III contribution remains strongly dependent on both burst age and mass: young, massive bursts provide the largest fraction of the observed flux, whereas older or lower-mass bursts are rapidly diluted by the host continuum. In particular, Pop~III components older than a few Myr contribute little to the integrated light almost independently of mass, while $10^4\,M_{\odot}$ bursts remain difficult to detect in unresolved photometry.

\begin{figure}[htbp]
\centering
\includegraphics[width=\columnwidth]{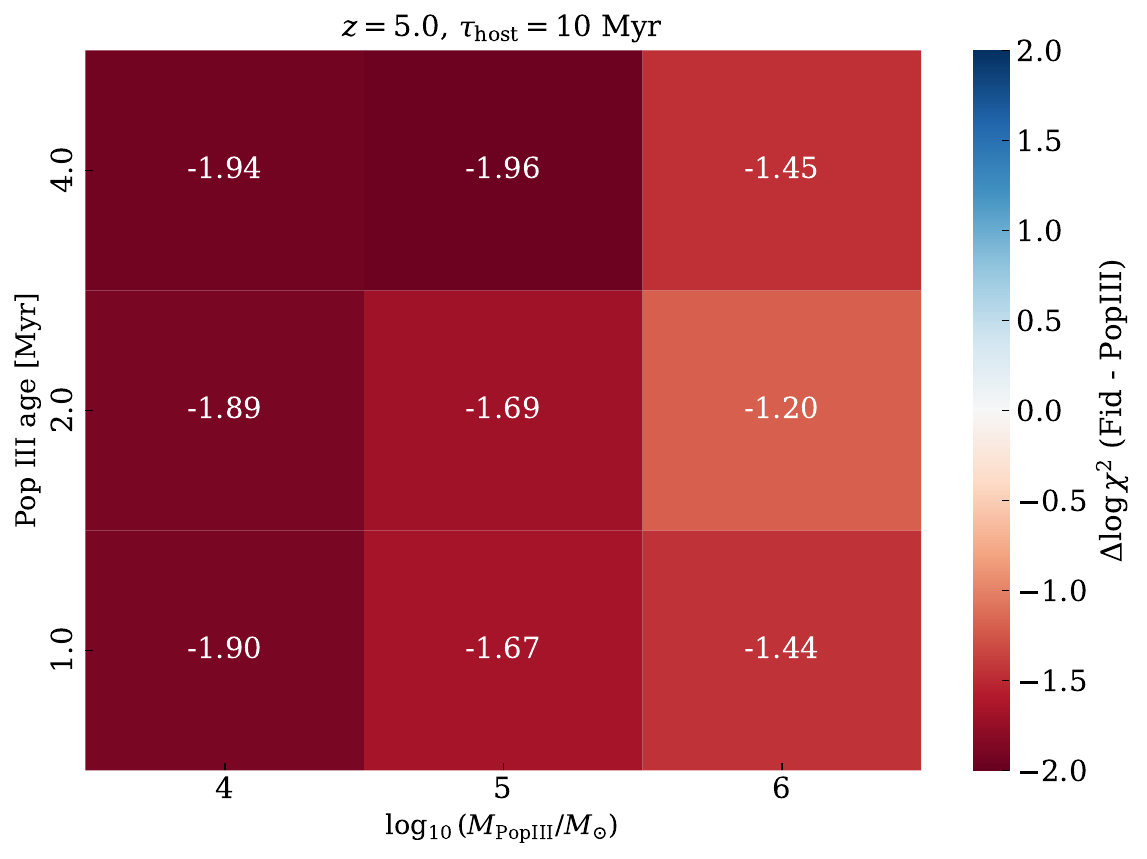}
\caption{Model-preference heatmap for the integrated SFH-overlap experiment at $z=5$ and host SFH timescale $\tau=10$ Myr. It shows $\Delta\log_{10}\chi^2=\log_{10}(\chi^2_{\mathrm{fid}})-\log_{10}(\chi^2_{\mathrm{PopIII}})$ as a function of Pop~III age (y-axis) and $\log_{10}(M_{\mathrm{PopIII}}/M_{\odot})$ (x-axis). All values are negative, indicating that the fiducial model is generally preferred in this unresolved integrated setup, even in this configuration where the Pop~III component makes its maximum flux contribution.}
\label{fig:heatmap_sfhs}
\end{figure}

We then fit these integrated SEDs with the two models: (i) the fiducial Pop~I/II model and (ii) the Yggdrasil Pop~III model with a Pop~III.2 IMF, $f_{\mathrm{cov}}=1.0$ (and $f_{\mathrm{esc,Ly\alpha}}=0.0$). For each fit, we draw posterior samples of the physical parameters and repeat the PPC procedure used in the isolated Pop~III analysis. In this integrated setup, the outcome depends not only on the relative flux contribution of each stellar component, but also on how well each model can reproduce the full combined SED shape. In practice, the higher flexibility of the fiducial model allows it to match most composite SEDs more effectively, typically with flatter continua, larger stellar masses, and younger ages that mimic the blue excess. 

Figure~\ref{fig:heatmap_sfhs} shows the resulting $\Delta\log_{10}\chi^2$ for $z=5$ and $\tau=10$ Myr. Importantly, even the configurations corresponding to the highest Pop~III blue flux fractions are most strongly favoured by the fiducial fit. This highlights the core challenge: a non-negligible Pop~III contribution does not necessarily imply Pop~III model preference when a flexible enriched-population model can reproduce the same broadband colours. 

In this configuration, the Pop~III-only fit systematically underpredicts the redder observed fluxes. We illustrate this with the PPC example in an inset in Fig.~\ref{fig:sfhs} ($z=5.0$, $\tau=10$ Myr, $t_{\mathrm{PopIII}}=1$ Myr, and $M_{\mathrm{PopIII}}=10^6M_{\odot}$): both models are comparable in the bluest bands, but beyond $\sim3\,\mu$m the fiducial model tracks the detections much better, while the Pop~III model remains too faint. Consistently, the fiducial fit yields better goodness-of-fit values  than the Pop~III model. The corresponding posteriors give $\log(M/M_{\odot})=8.53\pm0.19$ and $\log_{10}(\rm{Age}_{50\%}/\mathrm{yr})=6.94\pm1.32$ for the fiducial model, versus $\log(M/M_{\odot})=6.70\pm0.22$ and $\log_{10}(\mathrm{Age}/\mathrm{yr})=5.55\pm0.64$ for the Pop~III fit.

Furthermore, as redshift increases, key rest-frame optical information (where host Pop II red/older stars and continuum-shape differences have significant diagnostic power) shifts beyond NIRCam and toward the MIRI regime. With only the available high-S/N bands, the fit is driven by a narrower part of the SED, increasing degeneracy between Pop~III+host and fiducial solutions.

The SHAP analysis is also consistent with this behaviour, indicating that parameter changes modulate the strength of the model preference but do not reverse its sign. The surrogate model reaches $R^2=0.98$, with the following feature-importance ranking: host $\tau$, redshift, mass ratio, and Pop~III age. The improved surrogate performance relative to the isolated setup is driven by the presence of the host galaxy and its low-mass, red stellar populations. While the isolated scenario depends on subtle, highly degenerate spectral variations intrinsic to the Pop~III population, adding a host galaxy introduces larger-scale photometric changes. In particular, variations in host $\tau$ strongly modulate the global broadband spectral slope, while the mass ratio acts as an effective scale factor that controls the dilution of the primordial signal. Consequently, the regressor can predict the attenuation of Pop~III signatures in this toy model with high accuracy. Overall, we find no robust Pop~III detections in the unresolved integrated photometry tests performed: long-lived red Pop~II stars continue to contribute significantly at redder wavelengths, distorting the broadband shape away from a pure Pop~III SED and preventing a clean Pop~III-only fit.

\subsection{Resolved Pop~III--Pop~II Morphological Overlap}

Because Pop~III signatures may arise in compact clumps embedded in a chemically enriched host, a spatially resolved analysis provides a natural test of whether the primordial component can be isolated from the host continuum. To evaluate confusion in resolved observations, we build a controlled Pop~II+Pop~III image suite in which a compact Pop~III clump is injected into a smooth S\'ersic Pop~II host, then convolved with the instrumental PSF and perturbed with heteroscedastic noise as in the previous scenarios. We use the PSFs of the ACS and NIRCam filters derived for the JADES survey in \cite{synference}. 

\begin{figure*}[htbp]
\centering
\includegraphics[width=0.98\textwidth]{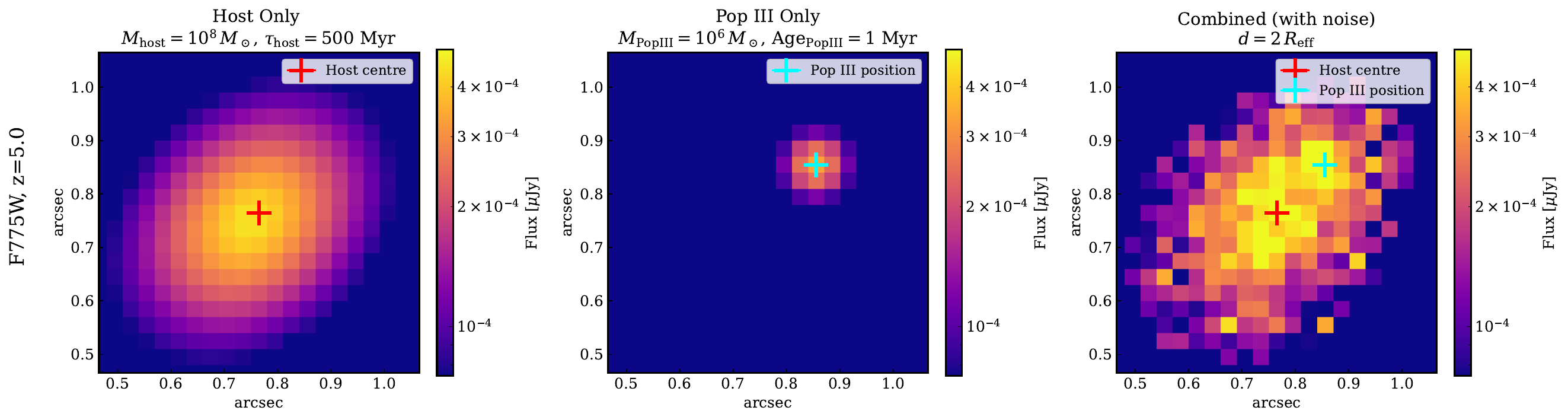}
\caption{Example of a resolved Pop~III--Pop~II morphological-overlap configuration at $z=5$. A compact Pop~III clump is injected into a smooth S\'ersic Pop~II host with fixed host mass ( $M_{\mathrm{host}}=10^{8}\,M_{\odot}$) and size ($0.2\arcsec$), then convolved with the instrumental PSF and perturbed with realistic pixel noise. This realization corresponds to $z=5.0$, $n=1.0$, $\tau=500\,\mathrm{Myr}$, $M_{\mathrm{PopIII}}=10^{6}\,M_{\odot}$, $\rm{Age}_{\mathrm{PopIII}}=1\,\mathrm{Myr}$, $d=2R_{\mathrm{eff}}$, and $\phi=0^{\circ}$. From left to right, we show the PSF-convolved noise-free host, the same for Pop~III clump, and the final PSF-convolved noisy combined image.}
\label{fig:sersic_examples}
\end{figure*}

The final resolved grid is summarized in Table~\ref{tab:resolved_grid}. Considering all combinations of the varied parameters gives $2\times3\times4\times3\times3\times4\times3=2592$ resolved simulations. At each configuration, we fit both the Pop~III and fiducial models and perform the PPCs at the pixel where the Pop~III clump is injected, so the comparison directly probes local recoverability in the highest-contrast region. This corresponds to 5184 model fits and, with 500 posterior samples drawn per fit, $2.592\times10^{6}$ posterior-predictive realizations. Thus, even this deliberately simple toy model already spans many configurations and effective degrees of freedom. A representative simulation is shown in Fig.~\ref{fig:sersic_examples}.

\begin{table}
\caption{Parameter grid for the resolved Pop~III--Pop~II morphological-overlap simulations.}
\label{tab:resolved_grid}
\centering
\footnotesize
\begin{tabular}{p{3.0cm}p{4.9cm}}
\hline
Component & Adopted values \\
\hline
Redshift & $z=5.0,\ 8.0$ \\
Host stellar mass & $M_{\mathrm{host}}=10^{8}\,M_{\odot}$ \\
Host effective radius & $R_{\mathrm{eff}}=0.2\arcsec$ \\
Host SFH timescale & $\tau=10,\ 100,\ 500\,\mathrm{Myr}$ \\
Host S\'ersic index & $n=0.5,\ 1.0,\ 2.0,\ 4.0$ \\
Pop~III stellar mass & $M_{\mathrm{PopIII}}=10^{4},\ 10^{5},\ 10^{6}\,M_{\odot}$ \\
Pop~III age & $t_{\mathrm{PopIII}}=1,\ 2,\ 4\,\mathrm{Myr}$ \\
Pop~III intrinsic size & $0.05\,\mathrm{kpc}$ smoothing length; sub-pixel before PSF convolution\\
Projected separation & $d/R_{\mathrm{eff}}=0,\ 1,\ 2,\ 3$ \\
Azimuthal angle & $\phi=0^{\circ},\ 45^{\circ},\ 90^{\circ}$ \\
\hline
\end{tabular}
\end{table}

\begin{figure*}[htbp]
\centering
\includegraphics[width=0.98\textwidth]{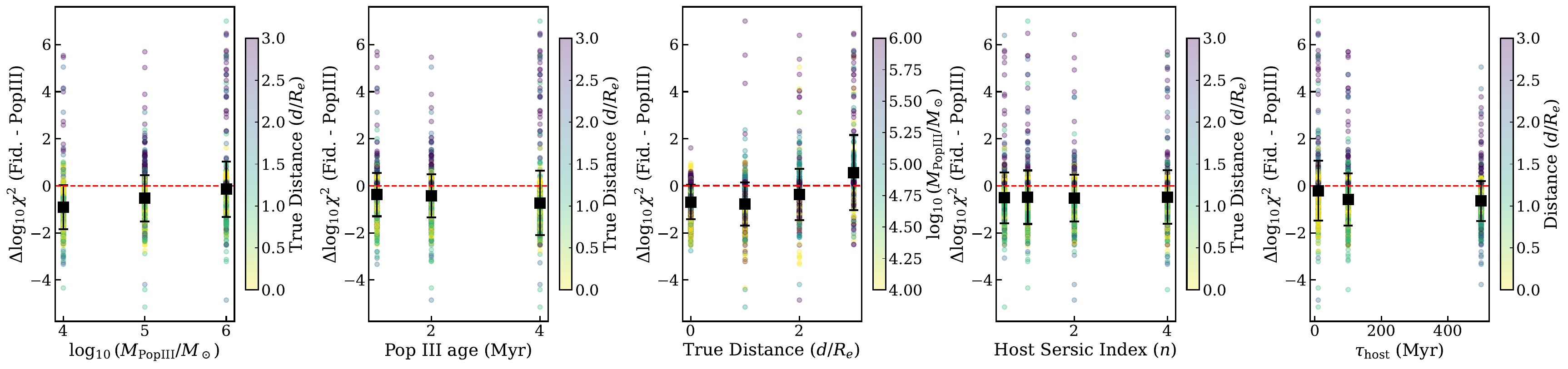}
\caption{Model-preference summary for the resolved Pop~III--Pop~II overlap experiment across the explored parameter grid. We set $z=5.0$ and $M_{\mathrm{host}}=10^{8}\,M_{\odot}$ for simplicity. Each panel shows $\Delta\log_{10}\chi^2=\log_{10}\chi^2_{\mathrm{fid}}-\log_{10}\chi^2_{\mathrm{PopIII}}$ as a function of $\log_{10}(M_{\mathrm{PopIII}}/M_\odot)$, Pop~III age, projected distance ($d/R_{\mathrm{e}}$), host Sérsic index, and host SFH timescale $\tau$. Individual simulations are shown as coloured points; black squares with error bars indicate the median and $1\sigma$ dispersion per bin. Positive values indicate a Pop~III-model preference, negative values indicate a fiducial-model preference. The dominant trend is a transition toward Pop~III preference at larger projected separations and higher Pop~III masses, especially for younger clumps, while blended central configurations remain mostly fiducial-dominated.}
\label{fig:model_preference_sersic}
\end{figure*}

The model preference across the resolved overlap grid is summarized in Fig.~\ref{fig:model_preference_sersic}. Overall, Pop~III-favoured regions concentrate in young, more massive Pop~III clumps, and spatially separated configurations, while host-dominated central cases are predominantly classified as fiducial.

\begin{figure}[htbp]
\centering
\includegraphics[width=0.98\columnwidth]{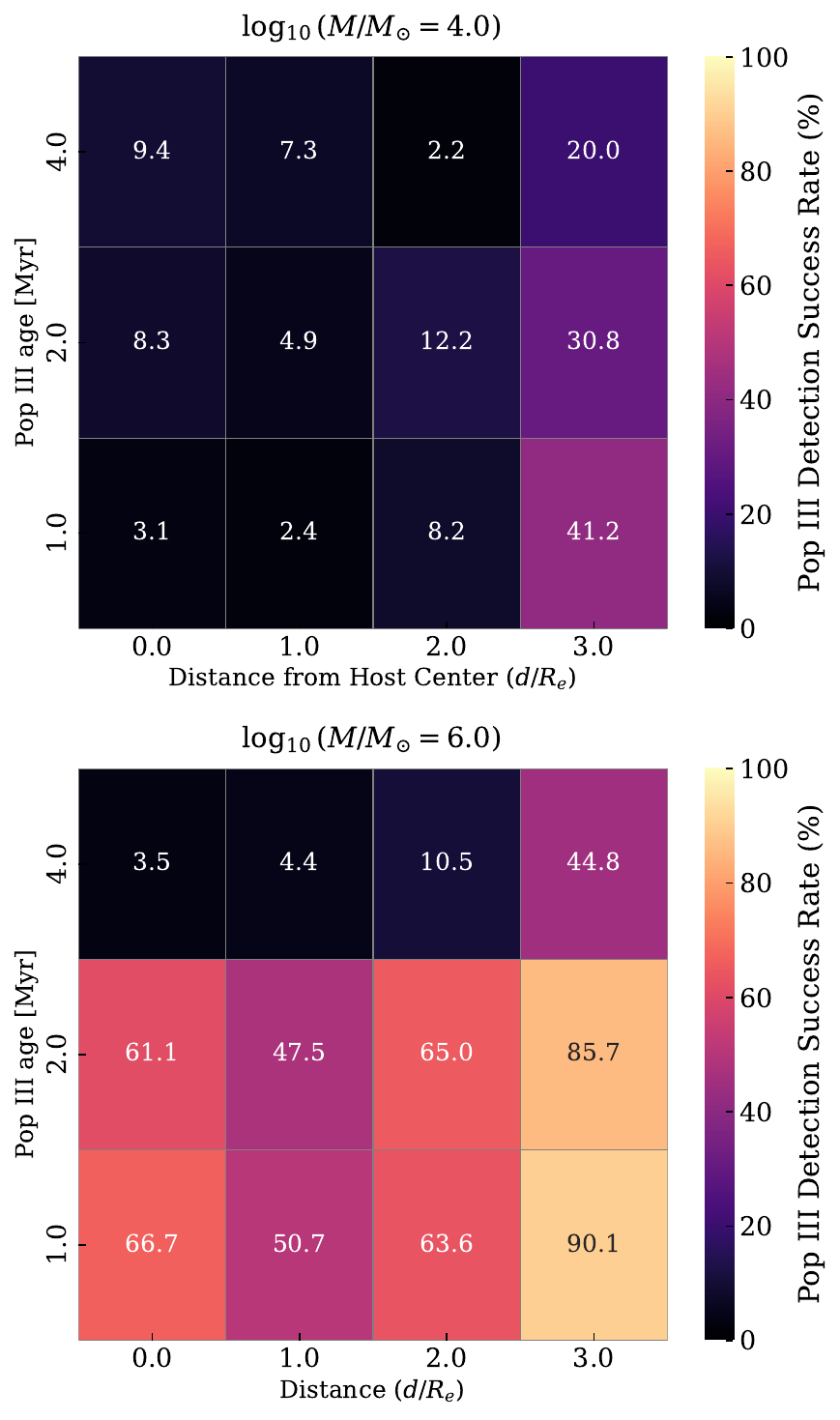}
\caption{Pop~III detection success rate (\%), defined as $\Delta \log_{10}\chi^2=\log_{10}\chi^2_{\mathrm{fid}}-\log_{10}\chi^2_{\mathrm{PopIII}}>0$,  in the resolved S\'ersic-overlap experiment with a host galaxy mass of  $M_{\mathrm{host}}=10^{8}\,M_{\odot}$ as a function of projected clump distance ($d/R_{\mathrm{e}}$) and Pop~III age, split by clump mass (top: low-mass, $\log_{10}(M/M_{\odot})=4$; bottom: high-mass, $\log_{10}(M/M_{\odot})=6$). Values in each cell indicate the fraction of simulations classified as Pop~III-favoured. Recovery is systematically higher for the high-mass clumps and increases toward larger projected separations, with the best performance at $d/R_{\mathrm{e}}=3$ and young ages, with $90\%$ of detections, while low-mass clumps remain difficult to recover across most of the parameter space.}
\label{fig:heatmap_mass_comparison_age_distance}
\end{figure}

As a complementary view to the $\Delta\log\chi^{2}$ map (Fig.~\ref{fig:model_preference_sersic}), Fig.~\ref{fig:heatmap_mass_comparison_age_distance} presents the empirical Pop~III success fractions in the same morphological setup (defined as $\Delta \log_{10}\chi^2=\log_{10}\chi^2_{\mathrm{fid}}-\log_{10}\chi^2_{\mathrm{PopIII}}>0$), as a function of the physical properties of the Pop~III and distance to the centre of the host. It confirms that detectability rises sharply with projected separation and that high-mass clumps are recovered far more often than low-mass clumps, while older bursts (4 Myr) show systematically lower success than younger cases at fixed distance. These trends are consistent with the model-preference diagnostics and reinforce that geometry, clump mass, and age jointly control resolved Pop~III recoverability. Pop~III recovery rises with increasing clump--host separation and is maximized for younger and more massive clumps, while centrally embedded and older/lower-mass configurations remain strongly confusion-limited. 

\begin{figure}[htbp]
\centering
\includegraphics[width=0.98\columnwidth]{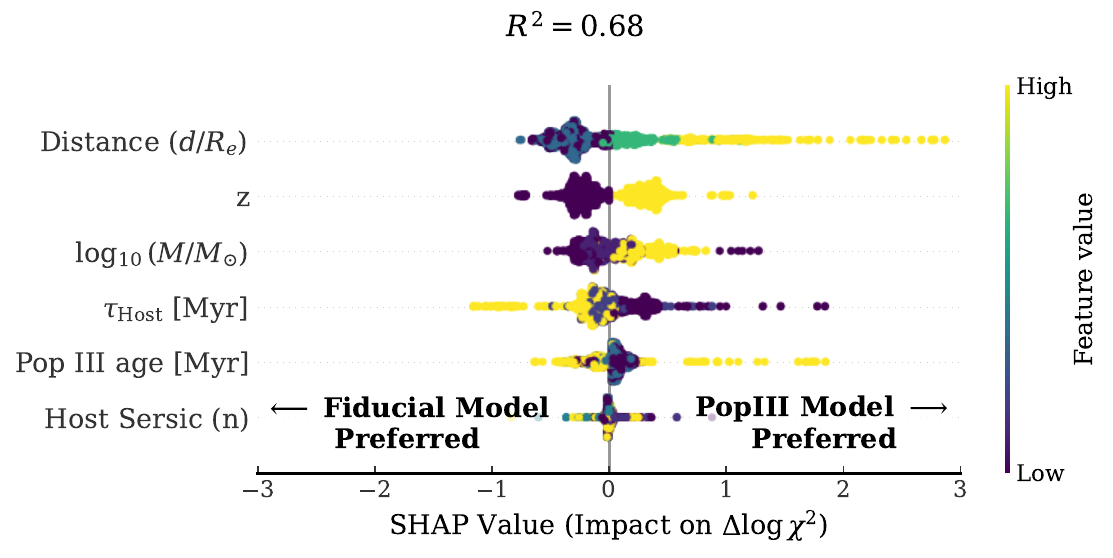}
\caption{SHAP summary for the S\'ersic-overlap experiment, showing the relative contribution of morphology and clump properties to the Pop~III detectability metric. Features are ranked by importance, and SHAP values indicate whether each feature value drives the prediction toward higher or lower recovered Pop~III contribution. This diagnostic highlights which combinations of host structure and Pop~III configuration dominate the recovery performance in the resolved analysis.}
\label{fig:shap_analysis_sersic}
\end{figure}

However, the complexity arises from the high-dimensional parameter space, where SED degeneracies produced by simultaneous changes in multiple simulation properties are difficult to interpret directly. To assess this, we again train a surrogate model to predict $\Delta\log_{10}(\chi^{2})$ from the input properties, obtaining an $R^{2}$ score of 0.68. The lower variance explained by the surrogate model in both the isolated and this resolved-host framework ($R^2 \sim 0.70$), compared to the integrated setup ($R^2 = 0.98$), is fundamentally driven by the emergence of trials where the model preference is positive or marginal ($\Delta \log_{10}\chi^2 \gtrsim 0$). In these resolved regimes, the severe physical degeneracy between Pop~III clump mass and stellar age creates a complex parameter space where subtle variations interact stochastically with the noise floor right at the detection boundary. Conversely, the integrated setup lacks these marginal regimes as the host emission uniformly dominates the signal, rendering the non-detections of Pop~III highly predictable. As shown in Fig.~\ref{fig:shap_analysis_sersic}, the SHAP ranking indicates that projected clump--host separation is the dominant driver, followed by redshift and Pop~III mass. Second-order contributions come from $\tau_{\mathrm{host}}$ and after from Pop~III age, while host morphology (parameterized by the S\'ersic index) is the least influential feature.

The resolved-overlap diagnostics show that morphology in terms of distance controls detectability as strongly as intrinsic stellar-population parameters. This naturally motivates the next step: applying the same pixel-level model-comparison framework to real JWST systems, where local contrast, projected offsets, and host substructure determine whether candidate Pop~III signatures can be separated from enriched Pop~II backgrounds. In this observational context, the resolved $\Delta\log_{10}(\chi^{2})$ maps provide both an interpretive tool and a practical guide for prioritizing the most informative regions for follow-up.

\section{Comparison with an observed Pop~III candidate}
\label{sec:blueberry}

The "blueberry" is a compact companion galaxy at redshift $z=5.124$, located at a projected offset of $\sim 3$ kpc (0.5 arcseconds) from the main nebular-dominated galaxy CAPERS-UDS-32520. They are located in the UDS (Ultra Deep Survey) legacy field. As reported by \cite{Reumert2026}, the observational data for this system consist of JWST/NIRCam imaging from the PRIMER survey and JWST/NIRSpec Prism spectroscopy from the CAPERS survey \citep{austin_inprep}. Photometrically, the companion displays a simple 2D Gaussian-like profile with strong continuum emission in the blue bands. Spectroscopically, it exhibits clear H${\alpha}$ emission and several tentative helium lines (including He~I and He~II), but features a complete absence of detectable metal lines. Because of this pristine chemical signature, the blueberry is theorized to be a surviving low-mass satellite halo dominated by Pop~III stars (or their remnants) that has managed to avoid chemical enrichment from the supernovae of its larger neighbour. 

\begin{figure*}[htbp]
\centering
\includegraphics[width=\textwidth]{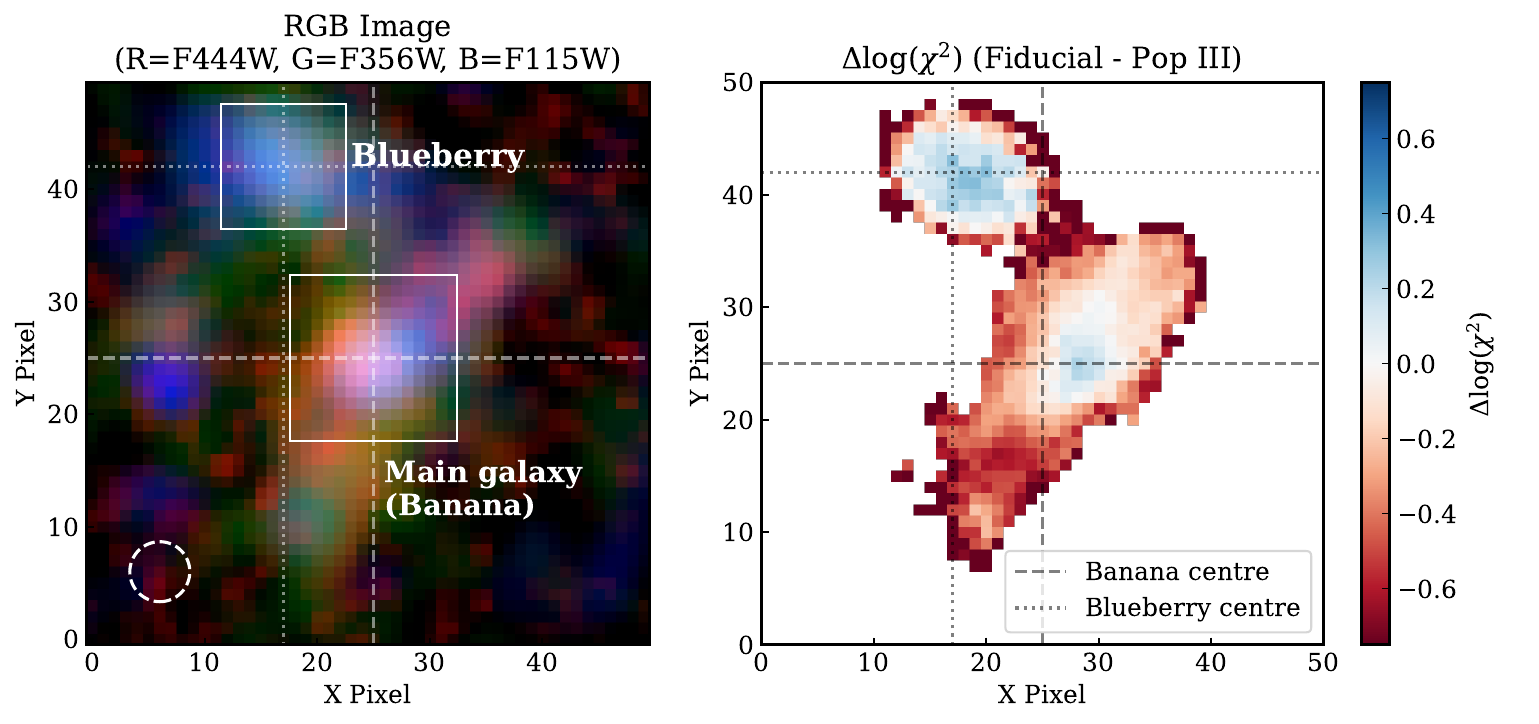}
\caption{Left: RGB
(F444W/F356W/F115W) of the Banana+Blueberry system described in \cite{Reumert2026}, PSF-matched to F444W. We include a white circumference representing the FWHM of the PSF of F444W, as well as squares and lines indicating the two components of the system and their representative pixels. Right: map of model preference $\Delta \log_{10} \chi^2$. Blue regions (positive values) indicate a preference for the Pop~III model over the fiducial PopI/II model. The central Banana pixel, indicated with the dashed lines, is better described by an enriched, nebular-dominated solution, whereas the Blueberry pixel, indicated with the dotted lines, shows a bluer continuum and a stronger preference for the Pop~III model. }
\label{fig:rgb_blueberry}
\end{figure*}

We perform a Bayesian model comparison on a $50\times50$ pixel cutout (of 0.03 arcsec/pixel) centred on CAPERS-UDS-32520. The RGB image of the region fitted, together with the different components are shown in Fig.~\ref{fig:rgb_blueberry}.
The cutout encompasses two morphologically distinct features: (1) the "Banana" --
the main host galaxy expected to harbor standard Pop~I/II stellar populations;
and (2) the "Blueberry" -- the blue compact region and Pop~III candidate \citep{Reumert2026}. An additional
compact feature not mentioned in the article, likely unassociated with
the system based on its photometric properties and redshift, has been excluded from this analysis.

We use the eleven photometric bands available spanning rest-frame UV to optical wavelengths:
HST/ACS (F435W, F606W, F814W) and JWST/NIRCam (F090W, F115W, F150W, F200W, F277W,
F356W, F410M, F444W). Photometric uncertainties and $1\sigma$ detection limits
are adopted from \citet{Donnan2024} for the PRIMER survey depth in the UDS field.
Non-detections are treated as $1\sigma$ upper limits. We retrain the inference with this observational configuration and, following the analysis of \cite{Reumert2026}, perform pixel-by-pixel model comparison between the Pop~III model (Pop~III.2 IMF with $f_{\mathrm{cov}}=1.0$ and $f_{\mathrm{esc,Ly\alpha}}=0.0$) and the fiducial enriched Pop~I/II model (FSPS-based with Dirichlet SFHs). Before proceeding with the SED fitting, we match all images to the F444W PSF using the PSFs from \cite{synference} and homogenization kernels generated with \texttt{PyPHER} \citep{pypher}.

We fit all pixels with $S/N>1$ in the filter redward of Lyman-$\alpha$ (at $z=5.124$, $\lambda_{\mathrm{obs}}\approx0.74\,\mu\mathrm{m}$, corresponding to F814W), and at least six filters above the $1\sigma$ upper limits described above to exclude background regions. We infer posterior distributions for both models, and draw 500 posterior samples to compute posterior-predictive photometry. We then evaluate the residuals against the observed photometry and define the final model-comparison statistic as $\Delta \log \chi^2 = \log \chi^2_{\mathrm{Fiducial}} - \log \chi^2_{\mathrm{PopIII}}$, so positive values favour the Pop~III model and negative values favour the fiducial model.

\begin{figure}[htbp]
\centering
\includegraphics[width=\columnwidth]{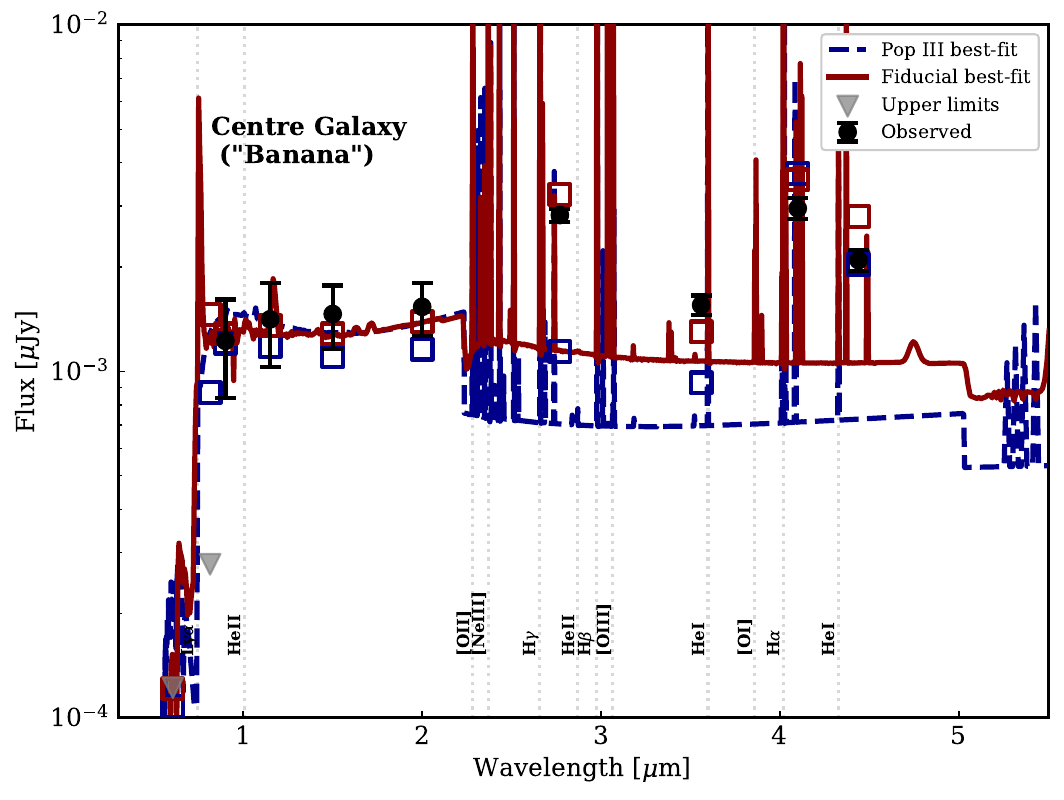}
\includegraphics[width=\columnwidth]{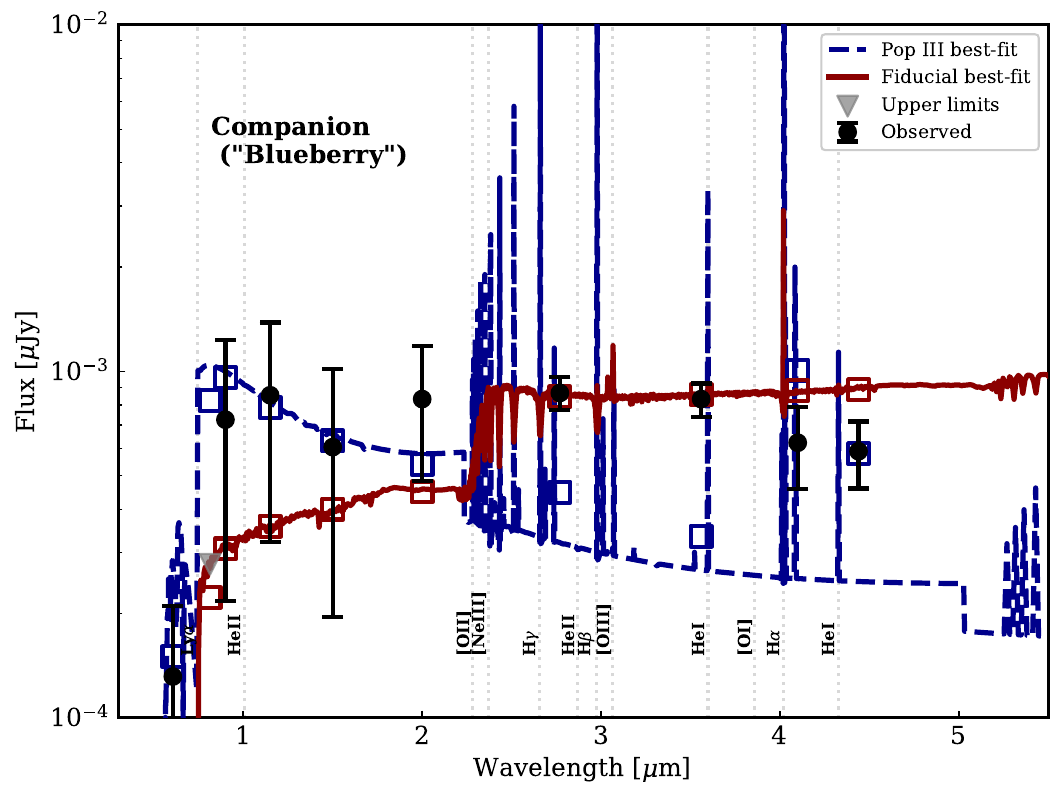}
\caption{Posterior-predictive SED comparison for representative pixels in the Banana+Blueberry system. The upper panel shows a pixel in the central galaxy (``Banana''), while the lower panel shows a pixel in the compact companion (``Blueberry''). Black points indicate the observed broad-band photometry and grey triangles mark upper limits. The blue and red curves show the best-fitting Pop~III and fiducial Pop~I/II models, respectively, with the corresponding model photometry shown as open squares. Vertical dashed lines mark the wavelengths of key rest-frame UV and optical emission lines shifted to $z=5.124$.}
\label{fig:ppc_example_blueberry}
\end{figure}

Figure~\ref{fig:ppc_example_blueberry} shows representative pixels in the Banana and Blueberry, indicated by the dashed and dotted lines in Fig.~\ref{fig:rgb_blueberry}, respectively. The key result is that our photometric model comparison recovers a Pop~III preference in the central Blueberry pixel, coincident with the compact component identified as a Pop~III candidate from spectroscopy. This agreement provides an important validation of the spatially resolved approach: the method highlights the same region as the independent spectroscopic analysis. The fiducial model does not reproduce the blue flux in this pixel, instead favouring a better match at redder wavelengths and missing the blue contribution. The main limitation is that the Pop~III model still predicts significantly lower fluxes than observed in F277W and F356W, possibly indicating an additional contribution from self-enriched material or low-mass stars to the redder emission, and the uncertainties in the bluer part of the SED are considerably large.

For the central pixel of the main galaxy, the best fit is achieved with the fiducial model, which provides successful agreement between the PPC SED and the observed photometry in all bands, whereas the Pop~III model underpredicts the flux in all bands redward of F200W. We also find a weaker Pop~III preference in the central-right region of the main galaxy, with $\Delta \log_{10} \chi^2$ values closer to zero, as shown in Fig.~\ref{fig:rgb_blueberry}. This suggests that the main galaxy itself may contain spatially varying stellar populations. Nevertheless, photometry alone is not sufficient to determine whether this component is genuinely Pop~III-like, although our control analysis of simulated enriched galaxies and real JADES sources, for which no systematic Pop~III preference is found (Appendix~\ref{sec:app_control_galaxies}), suggests it is still statistically significant. In NIRCam imaging, a nebular-dominated enriched galaxy can mimic several of the signatures expected from a primordial system due to the large number of free parameters involved. Specifically, significant systematic uncertainties in modelling nebular emission---such as variations in gas geometry, ionization parameters, and escape fractions---can create complex degeneracies that mimic primordial features, demanding a highly cautious interpretation of these localized regions.

In the case of the centre of the main galaxy, spectroscopy reveals clear metal enrichment through prominent emission lines such as [O\,II] $\lambda\lambda 3726,3729$ (blended with He\,I in the photometry), [O\,III] $\lambda\lambda 4959,5007$ (blended with $\rm{H}_{\beta}$ in the photometry), and other very faint but still detectable metal lines, including [Ne\,III] $\lambda 3869$ and [C\,III] $\lambda\lambda 1907,1909$. These features cannot be robustly identified or excluded from NIRCam photometry alone because their contribution is diluted within the broad filters. Our results therefore reinforce a practical conclusion: photometry can flag promising Pop~III candidates, but spectroscopy remains essential to confirm or reject a primordial interpretation, making IFU follow-up of this system especially interesting. 

The redder preference in the outskirts of both systems is related to the fact that the Pop~III mass prior does not allow such faint pixels to be fit, whereas the fiducial model can fit them by assigning higher stellar masses for the same SED, consistent with the different intrinsic luminosities of Pop~I/II and Pop~III populations.

Our pipeline does confirm that the Blueberry may indeed be a Pop~III dominated region, highlighting its potential use in pre-selecting promising Pop~III candidates. Beyond its value as a single case study, the Blueberry+Banana system illustrates a broader observational paradigm. The scenario it represents --- a compact, blue, low-mass companion offset from a more evolved host --- is precisely the configuration identified by our controlled simulations as the most favourable for Pop~III recovery: high projected separation, young burst age, and a host whose redder stellar populations provide spectral contrast rather than confusion. This alignment between the observational phenomenology and our simulated detectability criteria suggests that the Blueberry is not an isolated curiosity, but a representative of a class of objects that may be findable with a systematic morphology-first search in existing JWST deep fields, such as Hebe \citep{Maiolino2024,Maiolino2026,Rusta2026,bler2026} near GN-z11 \citep{Oesch2016,2023A&A...677A..88B}. A key practical advantage of the pipeline presented here is its computational efficiency: the amortised posteriors can be evaluated in milliseconds per pixel (see Sect.~\ref{sec:isolated}), making it feasible to process large areas of deep imaging without the computational cost that has traditionally limited pixel-by-pixel SED fitting. Such a survey-scale analysis would quantify the rate of photometric Pop~III preference across the full high-redshift population, identify the most promising targets for IFU spectroscopic follow-up, and test whether the Blueberry-type configuration constitutes a statistically significant channel for Pop~III detection at $z \sim 5$--$14$. In that sense, the Blueberry does not only validate our pipeline: it motivates a concrete observational analysis that can be executed with current data.

\section{Limitations}
\label{sec:limitations}

Our framework includes several simplifying assumptions that should be considered when interpreting these results. Specifically, the Pop~III forward model is tied to the \textit{Yggdrasil} templates, which, while providing a well-motivated and widely used baseline, explore a bounded parameter space consisting of discrete IMF families, fixed nebular covering fractions, and simplified Lyman-$\alpha$ transmission scenarios. Moreover, the results depend on the specific stellar-evolution tracks and stellar spectral libraries/properties adopted in the Yggdrasil templates, since other prescriptions exist and may affect the predicted Pop III SEDs \citep{2025A&A...695A..17L}. Consequently, our models do not yet span the full range of spectral behaviour currently discussed for JWST-era primordial candidates---including scenarios involving partial self-enrichment, extended SFHs, complex nebular conditions, or alternative ionizing sources. Future extensions should therefore incorporate a broader suite of primordial and near-primordial models, including the kinds of systems explored in recent observational and theoretical studies of extreme He\,II emitters and chemically evolving Pop~III candidates \citep{Nakajima2022,Rusta2025}. This will be important for testing how strongly our current detectability trends depend on the adopted template family, and for reducing the risk of over-interpreting template-specific signatures as uniquely Pop~III, keeping in mind that these models are also, by necessity, theoretical and may not capture the exact emission properties of real primordial stellar populations.

 Likewise, the fiducial enriched model, although intentionally flexible, cannot span every possible star-formation history, abundance pattern, dust geometry, or ionizing source. In practice, we are not exploring AGN-contaminated systems, that may occupy photometric regions that overlap with the Pop~III solutions. Furthermore, our assumption of fixed metallicity for the host galaxy ($Z=0.001$) simplifies the parameter space but may not fully capture metallicity variations at high-z; sensitivity tests across different metallicityy values would be valuable for understanding how metal abundance affects Pop~III detections. The inferred detectability therefore depends not only on the data quality, but also on how completely the competing models cover the true space of astrophysical alternatives. A related caveat is that our main comparison metric, $\Delta\log_{10}\chi^2$ from the PPCs, does not explicitly penalize model complexity. Since the fiducial family has a larger effective parameter space and broader prior volume than the more restrictive Pop~III family, part of the model preference can reflect this asymmetry rather than only the underlying astrophysics. This makes the comparison conservative in one important sense: when the narrower Pop~III model is nevertheless preferred, such detections are likely to be relatively pure and therefore especially interesting for follow-up.

In addition, our analysis relies on assumptions regarding the observational configuration that are not fully general. We treat the imaging through PSF-matched photometry and empirical noise models; however, residual systematics such as correlated noise, imperfect background subtraction, blending, or PSF mismatches may still affect real data at the pixel level. In the present implementation, the comparison is performed using per-band flux uncertainties without explicitly modelling the full covariance between filters or the spatial covariance between neighbouring pixels introduced by drizzling and PSF matching. A more complete treatment would propagate these covariance terms through the forward model and the posterior-predictive comparison, allowing the model preference maps to account for both inter-band and spatial correlations.

In addition, the framework currently requires redshift information to be fixed or externally constrained. In practice, this will rely on either spectroscopy or photometric-redshift estimates, the latter of which may introduce biases if the template sets do not incorporate primordial stellar populations. This limitation could be partially alleviated in future analyses by including the full redshift posterior distribution as an additional input to the network. A further practical limitation is the observational feasibility: the most favourable Pop~III detections in our grid are typically associated with bright, young, or more massive components, whereas intrinsically faint systems may remain below the detection limit unless enhanced by strong gravitational magnification.

Finally, the application to literature candidates is itself limited by the available data products. A clear example is Hebe, the highly pristine star-forming clump at $z\sim10.6$ near GN-z11 \citep{Oesch2016,2023A&A...677A..88B,Maiolino2024,Maiolino2026,Rusta2026,bler2026}. Spectroscopically, Hebe is extremely interesting because it shows strong He\,II $\lambda1640$ and H$\gamma$ emission while lacking detected metal lines. However, its stellar continuum is not clearly detected in the available deep NIRCam imaging after subtraction of the nearby foreground contamination, leaving only diffuse residual flux at its position. Because our framework is designed to compare spatially resolved broad-band and medium-band photometry, the absence of a robust continuum measurement prevents us from applying the method in the same way as for the Blueberry candidate analysed in Sect.~\ref{sec:blueberry}. This highlights an additional limitation of the present approach: even when spectroscopy strongly motivates a Pop~III interpretation, the photometric component of the analysis can only be robustly carried out when the continuum is detected.


\section{Conclusions, prospects and future work}
\label{sec:prospects}

We present a forward-modelling and simulation-based inference pipeline designed to assess when Pop~III stellar populations can best be distinguished from enriched Pop~I/II systems in JWST imaging. The framework combines Yggdrasil Pop~III templates, flexible fiducial enriched-population models, empirical survey-specific noise modelling, and NPE to generate noise-aware posterior predictive checks at both integrated and pixel-resolved levels. We first validate the approach on isolated Pop~III sources, then test confusion with Pop~II hosts through integrated and spatially resolved mock systems, and finally apply the same methodology to a literature Pop~III candidate. This end-to-end structure allows us to connect physical properties, observational conditions, and morphology to the practical detectability of primordial stellar populations.

Our results indicate that the most effective near-term strategy for discovering Pop~III clumps is to prioritize spatially resolved searches in high-redshift systems where compact blue substructures can be separated from the host light. In practice, this means selecting targets with enough angular resolution and depth to map pixel-scale contrasts around candidate clumps, rather than relying only on integrated photometry. This strategy is computationally feasible at scale because \texttt{Synthesizer} enables efficient mock-observation generation, while \texttt{SBIPIX} uses SBI coupled with normalising flows to provide substantially more efficient posterior estimation than traditional MCMC-based approaches. The resolved $\Delta\chi^{2}$ trends found here show that geometric separation, redshift, and ratio between stellar masses are the main drivers for recovering Pop~III-like signals, and should therefore be treated as first-order criteria for target selection in deep JWST fields.

An important implication is that our sensitivity limits should be interpreted relative to the expected mass scale of Pop~III star-forming events. Current theoretical models and cosmological simulations broadly agree that the stellar mass assembled in a single Pop~III star-forming event is typically in the range $10^3$--$10^5\,M_\odot$, with the upper end requiring unusually massive pristine minihalos or strong Lyman-Werner enhancement of H$_2$ dissociation to delay collapse \citep[e.g.][]{2004ARA&A..42...79B,2010ApJ...716..510G}. The sensitivity limits reached in our controlled experiments --- where isolated sources with $M_*\sim 10^4$--$10^5\,M_\odot$ are recoverable only in the most favourable configurations, combining young ages, high S/N and full nebular coverage --- are therefore physically relevant: they sit near the upper boundary of what is theoretically expected at $z \sim 5$--$14$. More massive cases ($M_* \gtrsim 10^6\,M_\odot$) likely require either unusually massive minihaloes or the co-spatial stacking of multiple Pop~III sub-clusters, and should be treated as optimistic benchmarks rather than typical events. In that sense, our framework is best understood as mapping the observational conditions under which realistic Pop~III masses become detectable, rather than implying that all candidate systems should occupy the upper end of the tested mass grid.

Interestingly, we can also investigate potential contaminants to Pop~III detections --- such as very young dusty galaxies or extremely low-metallicity Pop~II galaxies --- that may occupy photometric regions overlapping with the Pop~III solutions. These alternatives can be explored through the PPCs obtained with the fiducial model.

A natural next step is to combine this morphology-first approach with spectroscopy focused on the highest-probability regions identified by the spatial model-comparison maps. A practical workflow is to use resolved broad-band fitting to flag pixels or clumps with Pop~III-favoured solutions, guide spectroscopic follow-up by placing the slit according to the preference maps, and then test those regions with line diagnostics (e.g., strong He\,II and weak metal lines) to reject enriched alternatives. Applying this pipeline to larger samples will provide population-level constraints on how often Pop~III-like pockets can be isolated in high-redshift galaxies, and will help optimize future JWST programs and upcoming facilities.

This morphology-based criterion is also qualitatively consistent with recent Pop~III formation models, which suggest that observable metal-free or near-primordial episodes at later times are likely to occur in spatially distinct pockets, offset from more chemically enriched regions, rather than as fully isolated pristine galaxies. For example, the MEGATRON simulations find Pop~III formation over a wide range of distances from UV-bright galaxies, while chemically evolving or self-polluted Pop~III models predict hybrid phases in which primordial and enriched components coexist within the same broader system \citep{2026MNRAS.548ag529S,Rusta2025}. These theoretical trends strengthen the case for searches that prioritize spatial offsets and local contrast, particularly toward lower redshifts where enriched host light becomes increasingly important.

Going forward, a promising avenue for refining these observational boundaries is to inject Pop~III components directly into forward-modelled synthetic images derived from cosmological hydrodynamical simulations. A deeper limitation of the toy-model geometries used here—Sérsic profiles plus Gaussian clumps—is that the parameters governing morphology, clump separation, and mass ratio are drawn from broad flat priors that are not anchored to any realistic distribution of galaxy properties. As a result, the derived detectability boundaries are sensitive to the prior volume, and configurations that are rare in nature may be over-represented.

Moving to simulations would replace these flat priors with the covariant structure that cosmological models predict. Several high-resolution simulation suites offer promise in this regard, including the First Light And Reionisation Epoch Simulations (FLARES) suite \citep{2021MNRAS.500.2127L,2021MNRAS.501.3289V} and others providing realistic morphologies at $z\geq5$ \citep{BLUETIDES,SPHINX,astrid,THESAN}. Such an approach would make the resulting sensitivity limits directly comparable to theoretical Pop~III occurrence rates and substantially sharpen the target selection criteria for future JWST programmes.

Looking forward to future developments, we will explore advanced SBI architectures to perform formal model choice. A natural next step is amortised Bayesian model comparison via a marginal classifier trained across the competing model evidence distributions \citep{Cisewski-Kehe2023, 2024MLS&T...5a5008J,  multi-class-extremes2025, Sainsbury-Martinez2026}. This approach requires training a classifier on forward-simulated data marginalized over the parameter priors of both models, enabling the network to approximate marginal likelihood ratios and yield approximate Bayes factors in high-dimensional spaces without relying on PPCs or explicit likelihood evaluations \citep{Sainsbury-Martinez2023}.

\begin{dataavailability}                                                                            
  The JADES data underlying this article are available through the                                        
  project website (\url{https://jades-survey.github.io/}), the MAST                                       
  archive (\url{https://archive.stsci.edu/hlsp/jades}), and the                                           
  database (\url{https://jades.herts.ac.uk/DR4/}).
\end{dataavailability}

\begin{acknowledgements}

PIN thanks the LSST-DA Data Science Fellowship Program, which is funded by LSST-DA, the Brinson Foundation, the WoodNext Foundation, and the Research Corporation for Science Advancement Foundation; her participation in the program has benefited this work. MHC and PIN acknowledge financial support from the State Research Agency of the Spanish Ministry of Science and Innovation (AEI-MCINN) under the grants ``Galaxy Evolution with Artificial Intelligence'' with reference PGC2018-100852-A-I00 and ``BASALT'' with reference PID2021-126838NB-I00. T. Harvey acknowledges a PhD studentship funded by the UK Science Technology and Facilities Council. C. Conselice acknowledges support from the ERC Advanced Investigator Grant EPOCHS (788113). C. Lovell was supported by the research environment and infrastructure of the Handley Lab at the University of Cambridge. Co-funded by the European Union (MSCA Doctoral Network EDUCADO, GA 101119830 and Widening Participation, ExGal-Twin, GA 101158446). JHK acknowledges grant PID2022-136505NB-I00 funded by MCIN/AEI/10.13039/501100011033 and EU, ERDF. BER acknowledges support from the NIRCam Science Team contract to the University of Arizona, NAS5-02105, and JWST Program 3215. A. Bunker acknowledges funding from the “FirstGalaxies” Advanced Grant from the European Research Council (ERC) under the European Union’s Horizon 2020 research and innovation program (Grant agreement No. 789056). H\"U acknowledges support by the Max Planck Society through the Lise Meitner Excellence Program. H\"U acknowledges funding by the European Union (ERC APEX, 101164796). Views and opinions expressed are however those of the authors only and do not necessarily reflect those of the European Union or the European Research Council Executive Agency. Neither the European Union nor the granting authority can be held responsible for them. ZJ acknowledges support from a JWST/NIRCam contract (NAS5-02105) to the University of Arizona. The research of CCW is supported by NOIRLab, which is managed by the Association of Universities for Research in Astronomy (AURA) under a cooperative agreement with the National Science Foundation. 

This work is based on observations made with the NASA/ESA \textit{Hubble Space Telescope} (HST) and NASA/ESA/CSA \textit{James Webb Space Telescope} (JWST) obtained from the \textsc{Mikulski Archive for Space Telescopes} (\textsc{MAST}) at the \textit{Space Telescope Science Institute} (STScI), which is operated by the Association of Universities for Research in Astronomy, Inc., under NASA contract NAS 5-03127 for JWST, and NAS 5–26555 for HST.

\end{acknowledgements}

\bibliographystyle{aa}
\bibliography{references}

\begin{thebibliography}{103}
\expandafter\ifx\csname natexlab\endcsname\relax\def\natexlab#1{#1}\fi

\bibitem[{{Alsing} {et~al.}(2019){Alsing}, {Charnock}, {Feeney}, \& {Wandelt}}]{Alsing2019}
{Alsing}, J., {Charnock}, T., {Feeney}, S., \& {Wandelt}, B. 2019, \mnras, 488, 4440

\bibitem[{{Austin et al.}(in preparation)}]{austin_inprep}
{Austin et al.} in preparation

\bibitem[{{Beers} \& {Christlieb}(2005)}]{2005ARA&A..43..531B}
{Beers}, T.~C. \& {Christlieb}, N. 2005, \araa, 43, 531

\bibitem[{{Bird} {et~al.}(2022){Bird}, {Ni}, {Di Matteo}, {Croft}, {Feng}, \& {Chen}}]{astrid}
{Bird}, S., {Ni}, Y., {Di Matteo}, T., {et~al.} 2022, \mnras, 512, 3703

\bibitem[{{Boucaud} {et~al.}(2016){Boucaud}, {Bocchio}, {Abergel}, {Orieux}, {Dole}, \& {Amine Hadj-Youcef}}]{pypher}
{Boucaud}, A., {Bocchio}, M., {Abergel}, A., {et~al.} 2016, {PyPHER: Python-based PSF Homogenization kERnels}, Astrophysics Source Code Library, record ascl:1609.022

\bibitem[{{Bromm} {et~al.}(1999){Bromm}, {Coppi}, \& {Larson}}]{1999ApJ...527L...5B}
{Bromm}, V., {Coppi}, P.~S., \& {Larson}, R.~B. 1999, \apjl, 527, L5

\bibitem[{{Bromm} \& {Larson}(2004)}]{2004ARA&A..42...79B}
{Bromm}, V. \& {Larson}, R.~B. 2004, \araa, 42, 79

\bibitem[{{Bunker} {et~al.}(2024){Bunker}, {Cameron}, {Curtis-Lake}, {Jakobsen}, {Carniani}, {Curti}, {Witstok}, {Maiolino}, {D'Eugenio}, {Looser}, {Willott}, {Bonaventura}, {Hainline}, {{\"U}bler}, {Willmer}, {Saxena}, {Smit}, {Alberts}, {Arribas}, {Baker}, {Baum}, {Bhatawdekar}, {Bowler}, {Boyett}, {Charlot}, {Chen}, {Chevallard}, {Circosta}, {DeCoursey}, {de Graaff}, {Egami}, {Eisenstein}, {Endsley}, {Ferruit}, {Giardino}, {Hausen}, {Helton}, {Hviding}, {Ji}, {Johnson}, {Jones}, {Kumari}, {Laseter}, {L{\"u}tzgendorf}, {Maseda}, {Nelson}, {Parlanti}, {Perna}, {Rauscher}, {Rawle}, {Rix}, {Rieke}, {Robertson}, {Rodr{\'\i}guez Del Pino}, {Sandles}, {Scholtz}, {Sharpe}, {Skarbinski}, {Stark}, {Sun}, {Tacchella}, {Topping}, {Villanueva}, {Wallace}, {Williams}, \& {Woodrum}}]{Bunker2024}
{Bunker}, A.~J., {Cameron}, A.~J., {Curtis-Lake}, E., {et~al.} 2024, \aap, 690, A288

\bibitem[{{Bunker} {et~al.}(2023){Bunker}, {Saxena}, {Cameron}, {Willott}, {Curtis-Lake}, {Jakobsen}, {Carniani}, {Smit}, {Maiolino}, {Witstok}, {Curti}, {D'Eugenio}, {Jones}, {Ferruit}, {Arribas}, {Charlot}, {Chevallard}, {Giardino}, {de Graaff}, {Looser}, {L{\"u}tzgendorf}, {Maseda}, {Rawle}, {Rix}, {Del Pino}, {Alberts}, {Egami}, {Eisenstein}, {Endsley}, {Hainline}, {Hausen}, {Johnson}, {Rieke}, {Rieke}, {Robertson}, {Shivaei}, {Stark}, {Sun}, {Tacchella}, {Tang}, {Williams}, {Willmer}, {Baker}, {Baum}, {Bhatawdekar}, {Bowler}, {Boyett}, {Chen}, {Circosta}, {Helton}, {Ji}, {Kumari}, {Lyu}, {Nelson}, {Parlanti}, {Perna}, {Sandles}, {Scholtz}, {Suess}, {Topping}, {{\"U}bler}, {Wallace}, \& {Whitler}}]{2023A&A...677A..88B}
{Bunker}, A.~J., {Saxena}, A., {Cameron}, A.~J., {et~al.} 2023, \aap, 677, A88

\bibitem[{Calzetti {et~al.}(2000)Calzetti, Armus, Bohlin, Kinney, Koornneef, \& Storchi{-}Bergmann}]{Calzetti2000}
Calzetti, D., Armus, L., Bohlin, R.~C., {et~al.} 2000, \apj, 533, 682

\bibitem[{{Cameron} {et~al.}(2023){Cameron}, {Katz}, {Rey}, \& {Saxena}}]{Cameron2023}
{Cameron}, A.~J., {Katz}, H., {Rey}, M.~P., \& {Saxena}, A. 2023, \mnras, 523, 3516

\bibitem[{{Castro-Camilo} {et~al.}(2025){Castro-Camilo}, {Sainsbury-Martinez}, \& {Huser}}]{multi-class-extremes2025}
{Castro-Camilo}, D., {Sainsbury-Martinez}, F., \& {Huser}, R. 2025, arXiv e-prints, arXiv:2503.23156

\bibitem[{Cervino \& Valls-Gabaud(2003)}]{cervino2003}
Cervino, M. \& Valls-Gabaud, D. 2003, \mnras, 338, 481

\bibitem[{Choi {et~al.}(2016)Choi, Dotter, Conroy, Cantiello, Paxton, \& Johnson}]{Choi_2016}
Choi, J., Dotter, A., Conroy, C., {et~al.} 2016, \apj, 823, 102

\bibitem[{{Cisewski-Kehe} {et~al.}(2023){Cisewski-Kehe}, {Sainsbury-Martinez}, {Sainsbury-Martinez}, \& {Hagan}}]{Cisewski-Kehe2023}
{Cisewski-Kehe}, J., {Sainsbury-Martinez}, F., {Sainsbury-Martinez}, A., \& {Hagan}, N. 2023, arXiv e-prints, arXiv:2312.05411

\bibitem[{Conroy(2013)}]{Conroy_2013}
Conroy, C. 2013, \araa, 51, 393

\bibitem[{{Conroy} \& {Gunn}(2010)}]{Conroy_2010}
{Conroy}, C. \& {Gunn}, J.~E. 2010, \apj, 712, 833

\bibitem[{{Conroy} {et~al.}(2009){Conroy}, {Gunn}, \& {White}}]{Conroy2009}
{Conroy}, C., {Gunn}, J.~E., \& {White}, M. 2009, \apj, 699, 486

\bibitem[{Cranmer {et~al.}(2020)Cranmer, Brehmer, \& Louppe}]{Cranmer2020}
Cranmer, K., Brehmer, J., \& Louppe, G. 2020, PNAS, 117, 30055

\bibitem[{{Curti} {et~al.}(2025){Curti}, {Witstok}, {Jakobsen}, {Kobayashi}, {Curtis-Lake}, {Hainline}, {Ji}, {D'Eugenio}, {Chevallard}, {Maiolino}, {Scholtz}, {Carniani}, {Arribas}, {Baker}, {Bhatawdekar}, {Boyett}, {Bunker}, {Cameron}, {Cargile}, {Charlot}, {Eisenstein}, {Ji}, {Johnson}, {Kumari}, {Maseda}, {Robertson}, {Silcock}, {Tacchella}, {{\"U}bler}, {Venturi}, {Williams}, {Willmer}, \& {Willott}}]{curti2025}
{Curti}, M., {Witstok}, J., {Jakobsen}, P., {et~al.} 2025, \aap, 697, A89

\bibitem[{{Curtis-Lake} {et~al.}(2026){Curtis-Lake}, {Cameron}, {Bunker}, {Scholtz}, {Carniani}, {Parlanti}, {D'Eugenio}, {Jakobsen}, {Willmer}, {Arribas}, {Baker}, {Charlot}, {Chevallard}, {Circosta}, {Curti}, {Duan}, {Eisenstein}, {Hainline}, {Ji}, {Johnson}, {Jones}, {Maiolino}, {Maseda}, {Perna}, {P{\'e}rez-Gonz{\'a}lez}, {Rawle}, {Rieke}, {Rinaldi}, {Robertson}, {Rodr{\'\i}guez Del Pino}, {Saxena}, {Shivaei}, {Smit}, {Tacchella}, {{\"U}bler}, {Venturi}, {Williams}, \& {Willott}}]{2026MNRAS.tmp..935C}
{Curtis-Lake}, E., {Cameron}, A.~J., {Bunker}, A.~J., {et~al.} 2026, \mnras

\bibitem[{{D'Eugenio} {et~al.}(2025){D'Eugenio}, {Cameron}, {Scholtz}, {Carniani}, {Willott}, {Curtis-Lake}, {Bunker}, {Parlanti}, {Maiolino}, {Willmer}, {Jakobsen}, {Robertson}, {Johnson}, {Tacchella}, {Cargile}, {Rawle}, {Arribas}, {Chevallard}, {Curti}, {Egami}, {Eisenstein}, {Kumari}, {Looser}, {Rieke}, {Rodr{\'\i}guez Del Pino}, {Saxena}, {{\"U}bler}, {Venturi}, {Witstok}, {Baker}, {Bhatawdekar}, {Bonaventura}, {Boyett}, {Charlot}, {Danhaive}, {Hainline}, {Hausen}, {Helton}, {Ji}, {Ji}, {Jones}, {Juod{\v{z}}balis}, {Maseda}, {P{\'e}rez-Gonz{\'a}lez}, {Perna}, {Pusk{\'a}s}, {Shivaei}, {Silcock}, {Simmonds}, {Smit}, {Sun}, {Villanueva}, {Williams}, \& {Zhu}}]{2025ApJS..277....4D}
{D'Eugenio}, F., {Cameron}, A.~J., {Scholtz}, J., {et~al.} 2025, \apjs, 277, 4

\bibitem[{{Donnan} {et~al.}(2024){Donnan}, {McLure}, {Dunlop}, {McLeod}, {Magee}, {Arellano-C{\'o}rdova}, {Barrufet}, {Begley}, {Bowler}, {Carnall}, {Cullen}, {Ellis}, {Fontana}, {Illingworth}, {Grogin}, {Hamadouche}, {Koekemoer}, {Liu}, {Mason}, {Santini}, \& {Stanton}}]{Donnan2024}
{Donnan}, C.~T., {McLure}, R.~J., {Dunlop}, J.~S., {et~al.} 2024, \mnras, 533, 3222

\bibitem[{Draine \& Li(2007)}]{draine2007}
Draine, B.~T. \& Li, A. 2007, \apj, 657, 810

\bibitem[{{Dunlop} {et~al.}(2021){Dunlop}, {Abraham}, {Ashby}, {Bagley}, {Best}, {Bongiorno}, {Bouwens}, {Bowler}, {Brammer}, {Bremer}, {Calabro'}, {Carnall}, {Castellano}, {Cirasuolo}, {Conselice}, {Cullen}, {Dave}, {Dayal}, {Dekel}, {Dickinson}, {Duncan}, {Elbaz}, {Ellis}, {Ferguson}, {Ferrara}, {Finkelstein}, {Fontana}, {Furlanetto}, {Fynbo}, {Gallerani}, {Gardner}, {Giavalisco}, {Grazian}, {Grogin}, {Harikane}, {Hopkins}, {Ilbert}, {Illingworth}, {Juneau}, {Jung}, {Kartaltepe}, {Kassin}, {Kauffmann}, {Khochfar}, {Kirkpatrick}, {Kocevski}, {Koekemoer}, {Labbe}, {Laporte}, {Larson}, {Lucas}, {Magee}, {Mason}, {McCracken}, {McLeod}, {McLure}, {Merlin}, {Mesinger}, {Milvang-Jensen}, {Newman}, {Oesch}, {Ouchi}, {Pacifici}, {Papovich}, {Peacock}, {Peeples}, {Pentericci}, {Perez-Gonzalez}, {Pirzkal}, {Pope}, {Pye}, {Reddy}, {Robertson}, {Salvato}, {Santini}, {Schaerer}, {Shapley}, {Simons}, {Smit}, {Smith}, {Snyder}, {Somerville}, {Stanway}, {Stefanon}, {Tasca}, {Tikkanen}, {Tresse}, {Trump}, {Whitaker},
  {Wilkins}, {Wright}, {Wyithe}, {van Dokkum}, \& {van der Werf}}]{dunlop_primer}
{Dunlop}, J.~S., {Abraham}, R.~G., {Ashby}, M. L.~N., {et~al.} 2021, {PRIMER: Public Release IMaging for Extragalactic Research}, JWST Proposal. Cycle 1, ID. \#1837

\bibitem[{Durkan {et~al.}(2019)Durkan, Bekasov, Murray, \& Papamakarios}]{durkan2019}
Durkan, C., Bekasov, A., Murray, I., \& Papamakarios, G. 2019, Advances in Neural Information Processing Systems, 32

\bibitem[{{Eisenstein} {et~al.}(2026){Eisenstein}, {Willott}, {Alberts}, {Arribas}, {Bonaventura}, {Bunker}, {Cameron}, {Carniani}, {Charlot}, {Curtis-Lake}, {D'Eugenio}, {Ferruit}, {Giardino}, {Hainline}, {Hausen}, {Jakobsen}, {Johnson}, {Maiolino}, {Rauscher}, {Rieke}, {Rieke}, {Rix}, {Robertson}, {Stark}, {Tacchella}, {Williams}, {Willmer}, {Baker}, {Baum}, {Bhatawdekar}, {Boyett}, {Chen}, {Chevallard}, {Circosta}, {Curti}, {Danhaive}, {DeCoursey}, {Endsley}, {de Graaff}, {Dressler}, {Egami}, {Helton}, {Hviding}, {Ji}, {Jones}, {Kumari}, {L{\"u}tzgendorf}, {Laseter}, {Looser}, {Lyu}, {Maseda}, {Nelson}, {Parlanti}, {Perna}, {Pusk{\'a}s}, {Rawle}, {Rodr{\'\i}guez Del Pino}, {Rujopakarn}, {Sandles}, {Saxena}, {Scholtz}, {Sharpe}, {Shivaei}, {Silcock}, {Simmonds}, {Skarbinski}, {Smit}, {Stone}, {Suess}, {Sun}, {Tang}, {Topping}, {{\"U}bler}, {Villanueva}, {Wallace}, {Whitler}, {Witstok}, \& {Woodrum}}]{Eisenstein2026}
{Eisenstein}, D.~J., {Willott}, C., {Alberts}, S., {et~al.} 2026, \apjs, 283, 6

\bibitem[{{Feng} {et~al.}(2016){Feng}, {Di-Matteo}, {Croft}, {Bird}, {Battaglia}, \& {Wilkins}}]{BLUETIDES}
{Feng}, Y., {Di-Matteo}, T., {Croft}, R.~A., {et~al.} 2016, \mnras, 455, 2778

\bibitem[{Ferland {et~al.}(2017)Ferland, Chatzikos, Guzman, {et~al.}}]{Ferland2017}
Ferland, G.~J., Chatzikos, M., Guzman, F., {et~al.} 2017, Revista Mexicana de Astronom\'ia y Astrof\'isica, 53, 385

\bibitem[{{Ferland} {et~al.}(2013){Ferland}, {Porter}, {van Hoof}, {Williams}, {Abel}, {Lykins}, {Shaw}, {Henney}, \& {Stancil}}]{Ferland2013}
{Ferland}, G.~J., {Porter}, R.~L., {van Hoof}, P.~A.~M., {et~al.} 2013, \rmxaa, 49, 137

\bibitem[{Fujimoto {et~al.}(2025)Fujimoto, Naidu, Chisholm, Atek, Endsley, Kokorev, Furtak, Pan, Liu, Bromm, Venditti, Visbal, Sarmento, Weibel, Oesch, Brammer, Schaerer, Adamo, Berg, Bezanson, Bouwens, Chemerynska, Claeyssens, Dessauges-Zavadsky, Frebel, Korber, Labbe, Marques-Chaves, Matthee, McQuinn, Muñoz, Natarajan, Saldana-Lopez, Suess, Volonteri, Zitrin, Fujimoto, Naidu, Chisholm, Atek, Endsley, Kokorev, Furtak, Pan, Liu, Bromm, Venditti, Visbal, Sarmento, Weibel, Oesch, Brammer, Schaerer, Adamo, Berg, Bezanson, Bouwens, Chemerynska, Claeyssens, Dessauges-Zavadsky, Frebel, Korber, Labbe, Marques-Chaves, Matthee, McQuinn, Muñoz, Natarajan, Saldana-Lopez, Suess, Volonteri, \& Zitrin}]{Fujimoto2025}
Fujimoto, S., Naidu, R.~P., Chisholm, J., {et~al.} 2025, \apj, 989, 46

\bibitem[{{Glover} \& {Klessen}(2026)}]{Glover2025}
{Glover}, S. C.~O. \& {Klessen}, R.~S. 2026, in Encyclopedia of Astrophysics, Volume 2, Vol.~2, 211--229

\bibitem[{{Greif} {et~al.}(2010){Greif}, {Glover}, {Bromm}, \& {Klessen}}]{2010ApJ...716..510G}
{Greif}, T.~H., {Glover}, S. C.~O., {Bromm}, V., \& {Klessen}, R.~S. 2010, \apj, 716, 510

\bibitem[{Hahn {et~al.}(2023)Hahn, Kwon, Tojeiro, Siudek, Canning, Mezcua, Tinker, Brooks, Doel, Fanning, Gaztañaga, Kehoe, Landriau, Meisner, Moustakas, Poppett, Tarle, Weiner, \& Zou}]{hahn2023}
Hahn, C., Kwon, K.~J., Tojeiro, R., {et~al.} 2023, \apj, 945, 16

\bibitem[{Hahn \& Melchior(2022)}]{Hahn2022}
Hahn, C. \& Melchior, P. 2022, \apj, 938, 11

\bibitem[{{Hainline} {et~al.}(2026){Hainline}, {Eisenstein}, {Whitler}, {Robertson}, {Johnson}, {Jakobsen}, {Pusk{\'a}s}, {Tacchella}, {Helton}, {Wu}, {Arribas}, {Baker}, {Bunker}, {Cameron}, {Carniani}, {Carreira}, {Charlot}, {Chevallard}, {Curtis-Lake}, {D'Eugenio}, {Duan}, {Egami}, {Hausen}, {Ji}, {Looser}, {Maiolino}, {Mengistu}, {P{\'e}rez-Gonz{\'a}lez}, {Rieke}, {Rinaldi}, {Sun}, {Trussler}, {{\"U}bler}, {Williams}, {Willmer}, {Willott}, \& {Witstok}}]{Hainline2026}
{Hainline}, K.~N., {Eisenstein}, D.~J., {Whitler}, L., {et~al.} 2026, \apj, 1004, 161

\bibitem[{Hainline {et~al.}(2024)Hainline, Johnson, Robertson, Tacchella, Helton, Sun, Eisenstein, Simmonds, Topping, Whitler, Willmer, Rieke, Suess, Hviding, Cameron, Alberts, Baker, Baum, Bhatawdekar, Bonaventura, Boyett, Bunker, Carniani, Charlot, Chevallard, Chen, Curti, Curtis-Lake, D’Eugenio, Egami, Endsley, Hausen, Ji, Looser, Lyu, Maiolino, Nelson, Puskás, Rawle, Sandles, Saxena, Smit, Stark, Williams, Willott, \& Witstok}]{Hainline2024}
Hainline, K.~N., Johnson, B.~D., Robertson, B., {et~al.} 2024, \apj, 964, 71

\bibitem[{{Harvey} {et~al.}(2026){Harvey}, {Lovell}, {Newman}, {Conselice}, {Austin}, {Roper}, {Vijayan}, {Wilkins}, {Iglesias-Navarro}, {Rusakov}, {Li}, {Adams}, {Magdwick}, {Goolsby}, {Huertas-Company}, \& {Ho}}]{synference}
{Harvey}, T., {Lovell}, C.~C., {Newman}, S., {et~al.} 2026, \mnras, 547, stag282

\bibitem[{{Iglesias-Navarro} {et~al.}(2024){Iglesias-Navarro}, {Huertas-Company}, {Mart{\'\i}n-Navarro}, {Knapen}, \& {Pernet}}]{Iglesias-Navarro2024}
{Iglesias-Navarro}, P., {Huertas-Company}, M., {Mart{\'\i}n-Navarro}, I., {Knapen}, J.~H., \& {Pernet}, E. 2024, \aap, 689

\bibitem[{{Iglesias-Navarro} {et~al.}(2025){Iglesias-Navarro}, {Huertas-Company}, {P{\'e}rez-Gonz{\'a}lez}, {Knapen}, {Hahn}, {Koekemoer}, {Finkelstein}, {Villanueva}, \& {Asensio Ramos}}]{iglesias-navarro25}
{Iglesias-Navarro}, P., {Huertas-Company}, M., {P{\'e}rez-Gonz{\'a}lez}, P., {et~al.} 2025, \aap, 703, A229

\bibitem[{{Isobe} {et~al.}(2026){Isobe}, {Curti}, {Maiolino}, {Duan}, {McClymont}, {Pusk{\'a}s}, {D'Eugenio}, {Rinaldi}, {Trussler}, {Scholtz}, {Looser}, {Nelson}, {Ji}, {Langeroodi}, {Tacchella}, {Jones}, {Juod{\v{z}}balis}, {Pascalau}, {Hsiao}, {{\"U}bler}, {Baker}, {Bunker}, {Carniani}, {Charlot}, {Curtis-Lake}, {Geris}, {Koller}, {Lyu}, {Robertson}, {Williams}, \& {Wu}}]{isobe2026}
{Isobe}, Y., {Curti}, M., {Maiolino}, R., {et~al.} 2026, arXiv e-prints, arXiv:2606.11345

\bibitem[{Iyer \& Gawiser(2017)}]{Iyer2017}
Iyer, K. \& Gawiser, E. 2017, \apj, 838, 127

\bibitem[{Iyer {et~al.}(2019)Iyer, Gawiser, Faber, Ferguson, Kartaltepe, Koekemoer, Pacifici, \& Somerville}]{Iyer19}
Iyer, K.~G., Gawiser, E., Faber, S.~M., {et~al.} 2019, \apj, 879, 116

\bibitem[{{Iyer} {et~al.}(2026){Iyer}, {Pacifici}, {Calistro-Rivera}, \& {Lovell}}]{iyer2025spectralenergydistributionsgalaxies}
{Iyer}, K.~G., {Pacifici}, C., {Calistro-Rivera}, G., \& {Lovell}, C.~C. 2026, in Encyclopedia of Astrophysics, Volume 4, Vol.~4, 236--281

\bibitem[{{Jeong} {et~al.}(2026){Jeong}, {Venditti}, {Bromm}, {Jeon}, {Hsiao}, {Finkelstein}, \& {Chisholm}}]{Jeong2026}
{Jeong}, T.~B., {Venditti}, A., {Bromm}, V., {et~al.} 2026, arXiv e-prints, arXiv:2603.23209

\bibitem[{{Jeremiah} {et~al.}(2024){Jeremiah}, {Cole}, \& {Weniger}}]{2024MLS&T...5a5008J}
{Jeremiah}, M., {Cole}, A., \& {Weniger}, C. 2024, Machine Learning: Science and Technology, 5, 015008

\bibitem[{{Johnson} {et~al.}(2026){Johnson}, {Robertson}, {Eisenstein}, {Tacchella}, {Pusk{\'a}s}, {Duan}, {Wu}, {Hainline}, {Rieke}, {Willott}, {Willmer}, {Trussler}, {Alberts}, {Arribas}, {Baker}, {Bunker}, {Cameron}, {Carniani}, {Carreira}, {Cargile}, {Curtis-Lake}, {Egami}, {Hausen}, {Helton}, {Ji}, {Maiolino}, {P{\'e}rez-Gonz{\'a}lez}, {Rinaldi}, {Sun}, {Sun}, {Villanueva}, {Williams}, \& {Zhu}}]{johnson2026}
{Johnson}, B.~D., {Robertson}, B.~E., {Eisenstein}, D.~J., {et~al.} 2026, arXiv e-prints, arXiv:2601.15954

\bibitem[{{Kannan} {et~al.}(2025){Kannan}, {Puchwein}, {Smith}, {Borrow}, {Garaldi}, {Keating}, {Vogelsberger}, {Zier}, {McClymont}, {Shen}, {Popovic}, {Tacchella}, {Hernquist}, \& {Springel}}]{THESAN}
{Kannan}, R., {Puchwein}, E., {Smith}, A., {et~al.} 2025, The Open Journal of Astrophysics, 8, 153

\bibitem[{{Katz} {et~al.}(2023){Katz}, {Kimm}, {Ellis}, {Devriendt}, \& {Slyz}}]{2023MNRAS.524..351K}
{Katz}, H., {Kimm}, T., {Ellis}, R.~S., {Devriendt}, J., \& {Slyz}, A. 2023, \mnras, 524, 351

\bibitem[{{Khullar} {et~al.}(2022){Khullar}, {Nord}, {{\'C}iprijanovi{\'c}}, \& {et al.}}]{Khullar2022}
{Khullar}, G., {Nord}, B., {{\'C}iprijanovi{\'c}}, A., \& {et al.} 2022, Machine Learning: Science and Technology, 3

\bibitem[{{Kinugawa} {et~al.}(2014){Kinugawa}, {Inayoshi}, {Hotokezaka}, {Nakauchi}, \& {Nakamura}}]{2014MNRAS.442.2963K}
{Kinugawa}, T., {Inayoshi}, K., {Hotokezaka}, K., {Nakauchi}, D., \& {Nakamura}, T. 2014, \mnras, 442, 2963

\bibitem[{Kroupa(2001)}]{kroupa2001}
Kroupa, P. 2001, \mnras, 322, 231

\bibitem[{{Lecroq} {et~al.}(2025){Lecroq}, {Charlot}, {Bressan}, {Bruzual}, {Costa}, {Iorio}, {Mapelli}, {Santoliquido}, {Shepherd}, \& {Spera}}]{2025A&A...695A..17L}
{Lecroq}, M., {Charlot}, S., {Bressan}, A., {et~al.} 2025, \aap, 695, A17

\bibitem[{Leja {et~al.}(2019)Leja, Carnall, Johnson, Conroy, \& Speagle}]{Leja2019}
Leja, J., Carnall, A.~C., Johnson, B.~D., Conroy, C., \& Speagle, J.~S. 2019, \apj, 876, 3

\bibitem[{Lemos {et~al.}(2023)Lemos, Coogan, Hezaveh, \& Perreault-Levasseur}]{lemos2023sampling}
Lemos, P., Coogan, A., Hezaveh, Y., \& Perreault-Levasseur, L. 2023, in International Conference on Machine Learning, PMLR, 19256--19273

\bibitem[{{Liu} {et~al.}(2024){Liu}, {Gurian}, {Inayoshi}, {Hirano}, {Hosokawa}, {Bromm}, \& {Yoshida}}]{Liu2024}
{Liu}, B., {Gurian}, J., {Inayoshi}, K., {et~al.} 2024, \mnras, 534, 290

\bibitem[{Lovell {et~al.}(2025)Lovell, Roper, Vijayan, Wilkins, Newman, \& Seeyave}]{Lovell2025Synthesizer}
Lovell, C.~C., Roper, W.~J., Vijayan, A.~P., {et~al.} 2025, The Open Journal of Astrophysics, 8

\bibitem[{{Lovell} {et~al.}(2021)}]{2021MNRAS.500.2127L}
{Lovell}, C.~C. {et~al.} 2021, \mnras, 500, 2127

\bibitem[{Lundberg {et~al.}(2020)Lundberg, Erion, Chen, DeGrave, Prutkin, Nair, Katz, Himmelfarb, Bansal, \& Lee}]{Lundberg2020TreeSHAP}
Lundberg, S.~M., Erion, G., Chen, H., {et~al.} 2020, Nature Machine Intelligence, 2, 56

\bibitem[{Lundberg \& Lee(2017)}]{Lundberg2017SHAP}
Lundberg, S.~M. \& Lee, S.-I. 2017, in Advances in Neural Information Processing Systems 30 (NeurIPS 2017), ed. I.~Guyon, U.~von Luxburg, S.~Bengio, H.~Wallach, R.~Fergus, S.~Vishwanathan, \& R.~Garnett (Curran Associates, Inc.), 4765--4774

\bibitem[{{Maiolino} {et~al.}(2024){Maiolino}, {{\"U}bler}, {Perna}, {Scholtz}, {D'Eugenio}, {Witten}, {Laporte}, {Witstok}, {Carniani}, {Tacchella}, {Baker}, {Arribas}, {Nakajima}, {Eisenstein}, {Bunker}, {Charlot}, {Cresci}, {Curti}, {Curtis-Lake}, {de Graaff}, {Egami}, {Ji}, {Johnson}, {Kumari}, {Looser}, {Maseda}, {Nelson}, {Robertson}, {Rodr{\'\i}guez Del Pino}, {Sandles}, {Simmonds}, {Smit}, {Sun}, {Venturi}, {Williams}, \& {Willmer}}]{Maiolino2024}
{Maiolino}, R., {{\"U}bler}, H., {Perna}, M., {et~al.} 2024, \aap, 687, A67

\bibitem[{{Maiolino} {et~al.}(2026){Maiolino}, {{\"U}bler}, {Perna}, {Witstok}, {Jones}, {Perez-Gonzalez}, {Nakajima}, {Rusta}, {Salvadori}, {Tacchella}, {Madau}, {Trussler}, {D'Eugenio}, {Ji}, {Scholtz}, {Carniani}, {Isobe}, {Katz}, {Arribas}, {Baker}, {B{\"o}ker}, {Bromm}, {Bunker}, {Charlot}, {Chevallard}, {Curti}, {Curtis-Lake}, {Eisenstein}, {Egami}, {Ferrara}, {Graziani}, {Hainline}, {Helton}, {Ivey}, {Jonson}, {Koller}, {Kumari}, {Marconi}, {Mazzolari}, {Laporte}, {Parlanti}, {Pascalau}, {Pentericci}, {Rinaldi}, {Robertson}, {Rodr{\'\i}guez Del Pino}, {Schneider}, {Venditti}, {Venturi}, {Willmer}, {Witten}, \& {Zamora}}]{Maiolino2026}
{Maiolino}, R., {{\"U}bler}, H., {Perna}, M., {et~al.} 2026, arXiv e-prints, arXiv:2603.20362

\bibitem[{Marin {et~al.}(2011)Marin, Pudlo, Robert, \& Ryder}]{marin2011}
Marin, J.~M., Pudlo, P., Robert, C.~P., \& Ryder, R.~J. 2011, Statistics and Computing, 22, 1167

\bibitem[{{Mart{\'\i}n-Navarro} {et~al.}(2026{\natexlab{a}}){Mart{\'\i}n-Navarro}, {Vazdekis}, {Peralta de Arriba}, {Alonso Asensio}, {Angeloudi}, {Iglesias Navarro}, {La Barbera}, {Fahrion}, {Jerabkova}, {Beasley}, {Falc{\'o}n-Barroso}, {Huertas-Company}, {S{\'a}nchez}, \& {Jethwa}}]{martin-navarro26a}
{Mart{\'\i}n-Navarro}, I., {Vazdekis}, A., {Peralta de Arriba}, L., {et~al.} 2026{\natexlab{a}}, arXiv e-prints, arXiv:2605.24093

\bibitem[{{Mart{\'\i}n-Navarro} {et~al.}(2026{\natexlab{b}}){Mart{\'\i}n-Navarro}, {Vazdekis}, {Peralta de Arriba}, {Alonso Asensio}, {Iglesias Navarro}, {Angeloudi}, {La Barbera}, {Cervi{\~n}o}, {Fahrion}, {Jerabkova}, {Beasley}, {Falc{\'o}n-Barroso}, {Huertas-Company}, {S{\'a}nchez}, \& {Jethwa}}]{martin-navarro26b}
{Mart{\'\i}n-Navarro}, I., {Vazdekis}, A., {Peralta de Arriba}, L., {et~al.} 2026{\natexlab{b}}, arXiv e-prints, arXiv:2605.24476

\bibitem[{Nakajima \& Maiolino(2022)}]{Nakajima2022}
Nakajima, K. \& Maiolino, R. 2022, \mnras, 513, 5134

\bibitem[{{Nakajima} {et~al.}(2026){Nakajima}, {Ouchi}, {Harikane}, {Vanzella}, {Ono}, {Isobe}, {Nishigaki}, {Tsujimoto}, {Nakamura}, {Xu}, {Umeda}, \& {Zhang}}]{Nakajima2025}
{Nakajima}, K., {Ouchi}, M., {Harikane}, Y., {et~al.} 2026, \nat, 653, 363

\bibitem[{{Oesch} {et~al.}(2023){Oesch}, {Brammer}, {Naidu}, {Bouwens}, {Chisholm}, {Illingworth}, {Matthee}, {Nelson}, {Qin}, {Reddy}, {Shapley}, {Shivaei}, {van Dokkum}, {Weibel}, {Whitaker}, {Wuyts}, {Covelo-Paz}, {Endsley}, {Fudamoto}, {Giovinazzo}, {Herard-Demanche}, {Kerutt}, {Kramarenko}, {Labbe}, {Leonova}, {Lin}, {Magee}, {Marchesini}, {Maseda}, {Mason}, {Matharu}, {Meyer}, {Neufeld}, {Prieto Lyon}, {Schaerer}, {Sharma}, {Shuntov}, {Smit}, {Stefanon}, {Wyithe}, \& {Xiao}}]{Oesch2023}
{Oesch}, P.~A., {Brammer}, G., {Naidu}, R.~P., {et~al.} 2023, \mnras, 525, 2864

\bibitem[{{Oesch} {et~al.}(2016){Oesch}, {Brammer}, {van Dokkum}, {Illingworth}, {Bouwens}, {Labb{\'e}}, {Franx}, {Momcheva}, {Ashby}, {Fazio}, {Gonzalez}, {Holden}, {Magee}, {Skelton}, {Smit}, {Spitler}, {Trenti}, \& {Willner}}]{Oesch2016}
{Oesch}, P.~A., {Brammer}, G., {van Dokkum}, P.~G., {et~al.} 2016, \apj, 819, 129

\bibitem[{{Pacifici} {et~al.}(2023){Pacifici}, {Iyer}, {Mobasher}, {da Cunha}, {Acquaviva}, {Burgarella}, {Calistro Rivera}, {Carnall}, {Chang}, {Chartab}, {Cooke}, {Fairhurst}, {Kartaltepe}, {Leja}, {Ma{\l}ek}, {Salmon}, {Torelli}, {Vidal-Garc{\'\i}a}, {Boquien}, {Brammer}, {Brown}, {Capak}, {Chevallard}, {Circosta}, {Croton}, {Davidzon}, {Dickinson}, {Duncan}, {Faber}, {Ferguson}, {Fontana}, {Guo}, {Haeussler}, {Hemmati}, {Jafariyazani}, {Kassin}, {Larson}, {Lee}, {Mantha}, {Marchi}, {Nayyeri}, {Newman}, {Pandya}, {Pforr}, {Reddy}, {Sanders}, {Shah}, {Shahidi}, {Stevans}, {Triani}, {Tyler}, {Vanderhoof}, {de la Vega}, {Wang}, \& {Weston}}]{pacifi2023}
{Pacifici}, C., {Iyer}, K.~G., {Mobasher}, B., {et~al.} 2023, \apj, 944, 141

\bibitem[{{Raiter} {et~al.}(2010){Raiter}, {Schaerer}, \& {Fosbury}}]{raiter2010}
{Raiter}, A., {Schaerer}, D., \& {Fosbury}, R.~A.~E. 2010, \aap, 523, A64

\bibitem[{{Reumert} {et~al.}(2026){Reumert}, {Heintz}, {Pollock}, {Cameron}, {Brammer}, {Sneppen}, {Witstok}, {Terp}, \& {Watson}}]{Reumert2026}
{Reumert}, H., {Heintz}, K.~E., {Pollock}, C.~L., {et~al.} 2026, arXiv e-prints, arXiv:2603.13471

\bibitem[{{Rieke} {et~al.}(2015){Rieke}, {Wright}, {B{\"o}ker}, {Bouwman}, {Colina}, {Glasse}, {Gordon}, {Greene}, {G{\"u}del}, {Henning}, {Justtanont}, {Lagage}, {Meixner}, {N{\o}rgaard-Nielsen}, {Ray}, {Ressler}, {van Dishoeck}, \& {Waelkens}}]{2015PASP..127..584R}
{Rieke}, G.~H., {Wright}, G.~S., {B{\"o}ker}, T., {et~al.} 2015, \pasp, 127, 584

\bibitem[{{Rieke} {et~al.}(2023){Rieke}, {Robertson}, {Tacchella}, {Hainline}, {Johnson}, {Hausen}, {Ji}, {Willmer}, {Eisenstein}, {Pusk{\'a}s}, {Alberts}, {Arribas}, {Baker}, {Baum}, {Bhatawdekar}, {Bonaventura}, {Boyett}, {Bunker}, {Cameron}, {Carniani}, {Charlot}, {Chevallard}, {Chen}, {Curti}, {Curtis-Lake}, {Danhaive}, {DeCoursey}, {Dressler}, {Egami}, {Endsley}, {Helton}, {Hviding}, {Kumari}, {Looser}, {Lyu}, {Maiolino}, {Maseda}, {Nelson}, {Rieke}, {Rix}, {Sandles}, {Saxena}, {Sharpe}, {Shivaei}, {Skarbinski}, {Smit}, {Stark}, {Stone}, {Suess}, {Sun}, {Topping}, {{\"U}bler}, {Villanueva}, {Wallace}, {Williams}, {Willott}, {Whitler}, {Witstok}, \& {Woodrum}}]{2023ApJS..269...16R}
{Rieke}, M.~J., {Robertson}, B., {Tacchella}, S., {et~al.} 2023, \apjs, 269, 16

\bibitem[{{Robertson} {et~al.}(2026){Robertson}, {Johnson}, {Tacchella}, {Eisenstein}, {Hainline}, {Alberts}, {Arribas}, {Baker}, {Bunker}, {Cameron}, {Carniani}, {Carreira}, {Chevallard}, {Circosta}, {Curtis-Lake}, {Danhaive}, {Duan}, {Egami}, {Hausen}, {Helton}, {Ji}, {Maiolino}, {P{\'e}rez-Gonz{\'a}lez}, {Pusk{\'a}s}, {Rieke}, {Rinaldi}, {Sun}, {Sun}, {{\"U}bler}, {Trussler}, {Villanueva}, {Whitler}, {Williams}, {Willmer}, {Willott}, {Wu}, \& {Zhu}}]{robertson26}
{Robertson}, B.~E., {Johnson}, B.~D., {Tacchella}, S., {et~al.} 2026, arXiv e-prints, arXiv:2601.15956

\bibitem[{{Roper} {et~al.}(2026){Roper}, {Lovell}, {Vijayan}, {Wilkins}, {Akins}, {Berger}, {Sant Fournier}, {Harvey}, {Iyer}, {Leonardi}, {Newman}, {Pautasso}, {Perry}, {Seeyave}, {Sommovigo}, {Punyasheel}, {d'Hautefort}, \& {Rawlings}}]{Roper2026}
{Roper}, W., {Lovell}, C., {Vijayan}, A., {et~al.} 2026, The Journal of Open Source Software, 11, 9436

\bibitem[{{Rosdahl} {et~al.}(2018){Rosdahl}, {Katz}, {Blaizot}, {Kimm}, {Michel-Dansac}, {Garel}, {Haehnelt}, {Ocvirk}, \& {Teyssier}}]{SPHINX}
{Rosdahl}, J., {Katz}, H., {Blaizot}, J., {et~al.} 2018, \mnras, 479, 994

\bibitem[{{Rusta} {et~al.}(2025){Rusta}, {Salvadori}, {Gelli}, {Schaerer}, {Marconi}, {Koutsouridou}, \& {Carniani}}]{Rusta2025}
{Rusta}, E., {Salvadori}, S., {Gelli}, V., {et~al.} 2025, \apjl, 989, L32

\bibitem[{{Rusta} {et~al.}(2026){Rusta}, {Salvadori}, {Maiolino}, {Gelli}, {Koutsouridou}, {Carniani}, {{\"U}bler}, {Marconi}, \& {Schaerer}}]{Rusta2026}
{Rusta}, E., {Salvadori}, S., {Maiolino}, R., {et~al.} 2026, \apjl, 1003, L14

\bibitem[{{Sainsbury-Martinez} {et~al.}(2026){Sainsbury-Martinez}, {Anbajagane}, {Handley}, {Bevins}, \& {de Lera Acedo}}]{Sainsbury-Martinez2026}
{Sainsbury-Martinez}, F., {Anbajagane}, D., {Handley}, W., {Bevins}, H., \& {de Lera Acedo}, E. 2026, arXiv e-prints, arXiv:2603.26489

\bibitem[{{Sainsbury-Martinez} {et~al.}(2023){Sainsbury-Martinez}, {Cisewski-Kehe}, \& {Hagan}}]{Sainsbury-Martinez2023}
{Sainsbury-Martinez}, F., {Cisewski-Kehe}, J., \& {Hagan}, N. 2023, The American Statistician, 77, 351

\bibitem[{{Schaerer}(2002)}]{2002A&A...382...28S}
{Schaerer}, D. 2002, \aap, 382, 28

\bibitem[{Schaerer(2002)}]{Schaerer2002}
Schaerer, D. 2002, \aap, 382, 28

\bibitem[{{Schaerer}(2003)}]{2003A&A...397..527S}
{Schaerer}, D. 2003, \aap, 397, 527

\bibitem[{{Scholtz} {et~al.}(2026){Scholtz}, {Carniani}, {Parlanti}, {D'Eugenio}, {Curtis-Lake}, {Jakobsen}, {Bunker}, {Cameron}, {Arribas}, {Baker}, {Charlot}, {Chevellard}, {Circosta}, {Curti}, {Duan}, {Eisenstein}, {Hainline}, {Ji}, {Johnson}, {Jones}, {Kumari}, {Maiolino}, {Maseda}, {Perna}, {P{\'e}rez-Gonz{\'a}lez}, {Rawle}, {Rieke}, {Rinaldi}, {Robertson}, {Saxena}, {Shivaei}, {Silcock}, {Sun}, {Rodr{\'\i}guez Del Pino}, {Tacchella}, {{\"U}bler}, {Venturi}, {Williams}, {Willmer}, {Willott}, \& {Witstok}}]{Scholtz26}
{Scholtz}, J., {Carniani}, S., {Parlanti}, E., {et~al.} 2026, \mnras [\eprint[arXiv]{2510.01034}]

\bibitem[{{Stacy} {et~al.}(2010){Stacy}, {Greif}, \& {Bromm}}]{2010MNRAS.403...45S}
{Stacy}, A., {Greif}, T.~H., \& {Bromm}, V. 2010, \mnras, 403, 45

\bibitem[{{Storck} {et~al.}(2026){Storck}, {Katz}, {Devriendt}, {Slyz}, {Cadiou}, {Choustikov}, {Rey}, {Saxena}, {Agertz}, \& {Kimm}}]{2026MNRAS.548ag529S}
{Storck}, A., {Katz}, H., {Devriendt}, J., {et~al.} 2026, \mnras, 548, stag529

\bibitem[{{Tacchella} {et~al.}(2022){Tacchella}, {Conroy}, {Faber}, \& {et al.}}]{Tacchella2022}
{Tacchella}, S., {Conroy}, C., {Faber}, S.~M., \& {et al.} 2022, \apj, 926

\bibitem[{Tejero-Cantero {et~al.}(2020)Tejero-Cantero, Boelts, Deistler, Lueckmann, Durkan, Gonçalves, Greenberg, \& Macke}]{Tejero-Cantero2020}
Tejero-Cantero, A., Boelts, J., Deistler, M., {et~al.} 2020, The Journal of Open Source Software, 5, 2505

\bibitem[{{Trump} {et~al.}(2023){Trump}, {Arrabal Haro}, {Simons}, {Backhaus}, {Amor{\'\i}n}, {Dickinson}, {Fern{\'a}ndez}, {Papovich}, {Nicholls}, {Kewley}, {Brunker}, {Salzer}, {Wilkins}, {Almaini}, {Bagley}, {Berg}, {Bhatawdekar}, {Bisigello}, {Buat}, {Burgarella}, {Calabr{\`o}}, {Casey}, {Ciesla}, {Cleri}, {Cole}, {Cooper}, {Cooray}, {Costantin}, {Croton}, {Ferguson}, {Finkelstein}, {Fujimoto}, {Gardner}, {Gawiser}, {Giavalisco}, {Grazian}, {Grogin}, {Hathi}, {Hirschmann}, {Holwerda}, {Huertas-Company}, {Hutchison}, {Jogee}, {Juneau}, {Jung}, {Kartaltepe}, {Kirkpatrick}, {Kocevski}, {Koekemoer}, {Lotz}, {Lucas}, {Magnelli}, {Matharu}, {P{\'e}rez-Gonz{\'a}lez}, {Pirzkal}, {Rafelski}, {Rose}, {Seill{\'e}}, {Somerville}, {Straughn}, {Tacchella}, {Vanderhoof}, {Weiner}, {Wuyts}, {Yung}, \& {Zavala}}]{trump2023}
{Trump}, J.~R., {Arrabal Haro}, P., {Simons}, R.~C., {et~al.} 2023, \apj, 945, 35

\bibitem[{Trussler {et~al.}(2023)Trussler, Conselice, Adams, Maiolino, Nakajima, Zackrisson, Austin, Ferreira, \& Harvey}]{Trussler2023}
Trussler, J.~A., Conselice, C.~J., Adams, N.~J., {et~al.} 2023, \mnras, 525, 5328

\bibitem[{{{\"U}bler} {et~al.}(2026){{\"U}bler}, {Maiolino}, {P{\'e}rez-Gonz{\'a}lez}, {Isobe}, {Jones}, {Kumari}, {Charlot}, {Rusta}, {Salvadori}, {Nakajima}, {Perna}, {Arribas}, {Bunker}, {Carniani}, {D'Eugenio}, {Rodr{\'\i}guez Del Pino}, {Bertola}, {B{\"o}ker}, {Chevallard}, {Circosta}, {Cresci}, {Curti}, {Curtis-Lake}, {Eisenstein}, {Hainline}, {Johnson}, {Parlanti}, {Rinaldi}, {Robertson}, {Scholtz}, {Tacchella}, {Venturi}, {Witstok}, \& {Zamora}}]{bler2026}
{{\"U}bler}, H., {Maiolino}, R., {P{\'e}rez-Gonz{\'a}lez}, P.~G., {et~al.} 2026, arXiv e-prints, arXiv:2603.20360

\bibitem[{Vazdekis {et~al.}(2010)Vazdekis, Sánchez-Blázquez, Falcón-Barroso, Cenarro, Beasley, Cardiel, Gorgas, \& Peletier}]{vazdekis2010}
Vazdekis, A., Sánchez-Blázquez, P., Falcón-Barroso, J., {et~al.} 2010, \mnras

\bibitem[{{Venditti} {et~al.}(2024){Venditti}, {Bromm}, {Finkelstein}, {Calabr{\`o}}, {Napolitano}, {Graziani}, \& {Schneider}}]{2024ApJ...973L..12V}
{Venditti}, A., {Bromm}, V., {Finkelstein}, S.~L., {et~al.} 2024, \apjl, 973, L12

\bibitem[{{Venditti} {et~al.}(2026){Venditti}, {Graziani}, {Schneider}, {Bromm}, {Munoz}, {Di Cesare}, {Valiante}, {Calabr{\`o}}, {Maiolino}, {Finkelstein}, {Parente}, {Saggini}, \& {Chisholm}}]{venditti2026}
{Venditti}, A., {Graziani}, L., {Schneider}, R., {et~al.} 2026, arXiv e-prints, arXiv:2603.27582

\bibitem[{{Vijayan} {et~al.}(2021)}]{2021MNRAS.501.3289V}
{Vijayan}, A.~P. {et~al.} 2021, \mnras, 501, 3289

\bibitem[{{Villaume} {et~al.}(2015){Villaume}, {Conroy}, \& {Johnson}}]{Villaume2015}
{Villaume}, A., {Conroy}, C., \& {Johnson}, B.~D. 2015, \apj, 806, 82

\bibitem[{{Visbal} {et~al.}(2023){Visbal}, {Bryan}, \& {Haiman}}]{Visbal2023}
{Visbal}, E., {Bryan}, G.~L., \& {Haiman}, Z. 2023, \mnras, 522, 5351

\bibitem[{{Visbal} {et~al.}(2025){Visbal}, {Hazlett}, \& {Bryan}}]{Visbal2025}
{Visbal}, E., {Hazlett}, R., \& {Bryan}, G.~L. 2025, \apjl, 993, L17

\bibitem[{{Wang} {et~al.}(2023){Wang}, {Leja}, {Bezanson}, \& {et al.}}]{Wang2023}
{Wang}, B., {Leja}, J., {Bezanson}, R., \& {et al.} 2023, \apjl, 945

\bibitem[{{Wang} {et~al.}(2024){Wang}, {Cheng}, {Ge}, {Meng}, {Daddi}, {Yan}, {Ji}, {Jin}, {Jones}, {Malkan}, {Arrabal Haro}, {Brammer}, {Oguri}, {Hou}, \& {Zhang}}]{2024ApJ...967L..42W}
{Wang}, X., {Cheng}, C., {Ge}, J., {et~al.} 2024, \apjl, 967, L42

\bibitem[{{Williams} {et~al.}(2023){Williams}, {Tacchella}, {Maseda}, {Robertson}, {Johnson}, {Willott}, {Eisenstein}, {Willmer}, {Ji}, {Hainline}, {Helton}, {Alberts}, {Baum}, {Bhatawdekar}, {Boyett}, {Bunker}, {Carniani}, {Charlot}, {Chevallard}, {Curtis-Lake}, {de Graaff}, {Egami}, {Franx}, {Kumari}, {Maiolino}, {Nelson}, {Rieke}, {Sandles}, {Shivaei}, {Simmonds}, {Smit}, {Suess}, {Sun}, {{\"U}bler}, \& {Witstok}}]{williams2023}
{Williams}, C.~C., {Tacchella}, S., {Maseda}, M.~V., {et~al.} 2023, \apjs, 268, 64

\bibitem[{{Zackrisson} {et~al.}(2011){Zackrisson}, {Rydberg}, {Schaerer}, {{\"O}stlin}, \& {Tuli}}]{Zackrisson2011}
{Zackrisson}, E., {Rydberg}, C.-E., {Schaerer}, D., {{\"O}stlin}, G., \& {Tuli}, M. 2011, \apj, 740, 13

\end{thebibliography}

\begin{appendix}
\section{Detectability as a function of redshift and mass}

\begin{figure}[htbp]
\centering
\includegraphics[width=0.98\columnwidth]{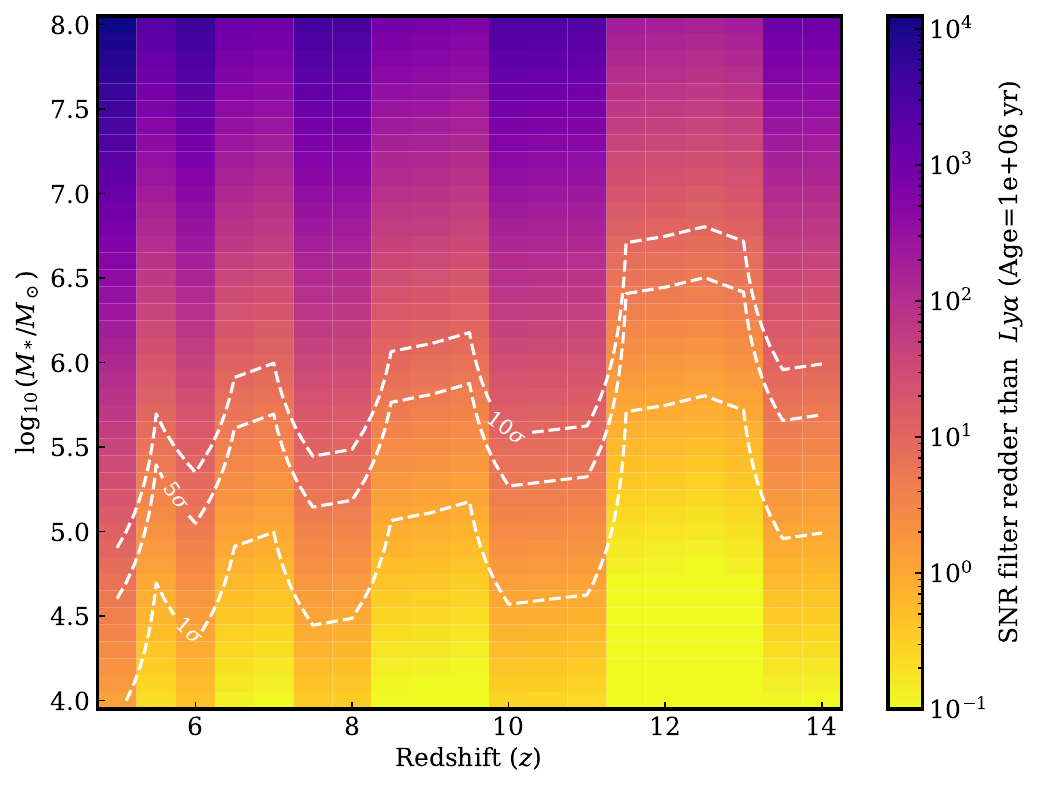}
\caption{S/N in the NIRCam filter immediately redward of Lyman-$\alpha$ as a function of redshift and stellar mass, for a Pop~III.2 IMF with $f_{\mathrm{cov}}=1$ and age $= 10^6$ yr. The dashed contours mark the $1\sigma$, $5\sigma$, and $10\sigma$ detection thresholds for the JADES deep pixel depths. Vertical stripes reflect changes in the filter covering the Lyman-$\alpha$ break across the NIRCam bandpass.}
\label{fig:salpeter_snr_1e6}
\end{figure}

\begin{figure}[htbp]
\centering
\includegraphics[width=0.98\columnwidth]{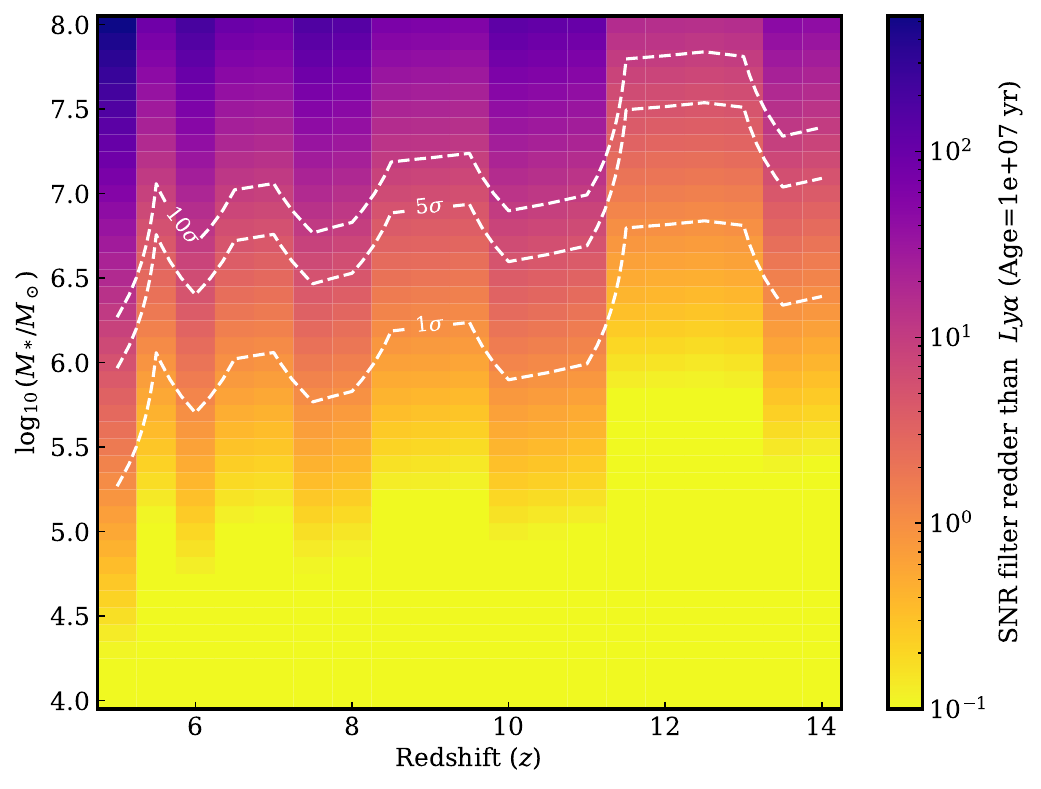}
\caption{Same as Fig.~\ref{fig:salpeter_snr_1e6} but for age $= 10^7$ yr. The detectable region shifts to significantly higher masses, reflecting the decline in ionising luminosity as the stellar population ages. Sources below $\log_{10}(M_*/M_\odot)\sim 6$ fall below $1\sigma$ at most redshifts.}
\label{fig:salpeter_snr_1e7}
\end{figure}

Figures~\ref{fig:salpeter_snr_1e6} and~\ref{fig:salpeter_snr_1e7} show the signal-to-noise ratio (S/N) achieved in the NIRCam filter immediately redward of Lyman-$\alpha$ as a function of redshift ($5\lesssim z \lesssim 14$) and stellar mass ($10^4 \lesssim M_*/M_\odot \lesssim 10^8$) for isolated Pop~III sources with a Pop~III.2 IMF and full nebular coverage ($f_{\mathrm{cov}}=1$). The filter changes with redshift as the Lyman-$\alpha$ break shifts across the NIRCam bandpass, producing the vertical stripes visible in both panels: sharp S/N drops occur when the break falls between filters, while peaks correspond to configurations where the nebular continuum and emission lines land in a sensitive, deep filter. The dashed contours mark the $1\sigma$, $5\sigma$, and $10\sigma$ detection thresholds for the JADES deep pixel depths used throughout this work.

At age $10^6$ yr (Fig.~\ref{fig:salpeter_snr_1e6}), the ionising photon budget is near its peak and the nebular emission contributes maximally to the broadband flux. As a result, even relatively low-mass sources ($\log_{10}(M_*/M_\odot)\sim 4.5$) reach $5\sigma$ detectability over a broad redshift range, and massive sources ($\log_{10}(M_*/M_\odot)\gtrsim 6$) are detected at $>10\sigma$ across most of the explored redshift range. By age $10^7$ yr (Fig.~\ref{fig:salpeter_snr_1e7}), the ionising luminosity has declined significantly as the most massive stars have evolved off the main sequence. The detectable region in the mass--redshift plane shifts upward by roughly $1.5$--$2$ dex: sources below $\log_{10}(M_*/M_\odot)\sim 6$ fall below the $1\sigma$ threshold at most redshifts, and only the most massive end of the grid remains robustly detectable. This strong age dependence is consistent with the results of the main analysis and reinforces that isolated Pop~III detectability is primarily driven by the combination of stellar mass and burst age, with younger and more massive clumps being the most favourable targets for identification with JWST.

\section{Simulation-based calibration}
\label{sec:SBC}

\begin{figure}[htbp]
\centering
\includegraphics[width=0.9\columnwidth]{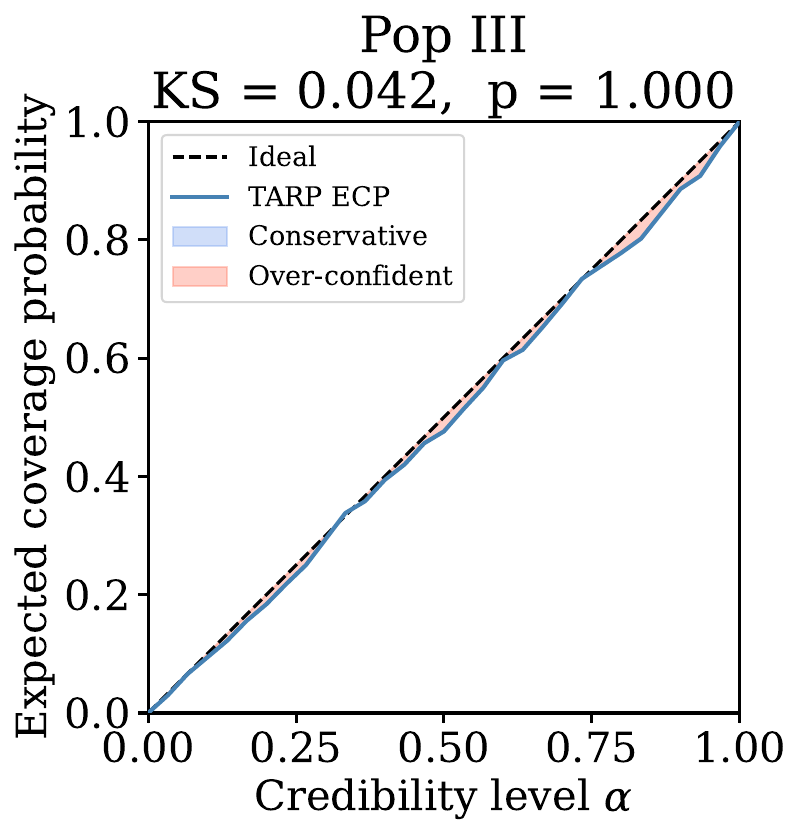}
\caption{TARP calibration test for the Pop~III model. The expected coverage probability (ECP) is shown as a function of credibility level $\alpha$. A well-calibrated posterior lies along the diagonal (dashed line). The blue shaded region indicates conservative coverage (ECP $>$ $\alpha$) and the red shaded region indicates over-confident coverage (ECP $<$ $\alpha$). The KS statistic and $p$-value with respect to the uniform distribution are quoted in the title. The close agreement with the diagonal confirms that the posterior distributions for all model parameters are properly calibrated.}
\label{fig:tarp_popiii}
\end{figure}

\begin{figure}[htbp]
\centering
\includegraphics[width=0.9\columnwidth]{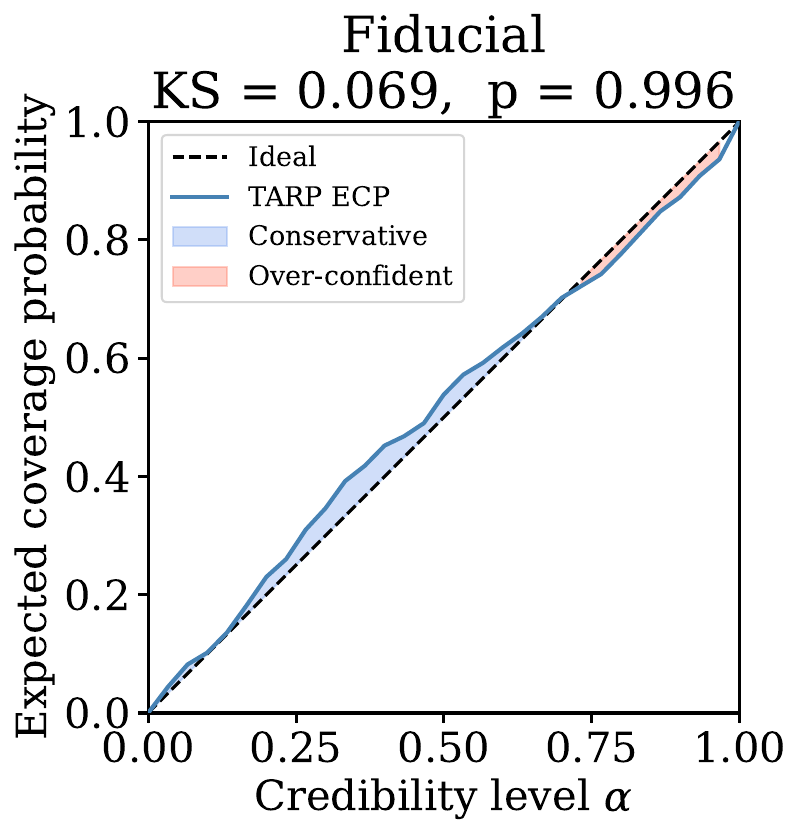}
\caption{Same as Fig.~\ref{fig:tarp_popiii} but for the fiducial model. The posterior distributions for all model parameters are likewise properly calibrated.}
\label{fig:tarp_fiducial}
\end{figure}

To assess the reliability of the trained posteriors, we perform a  test of accuracy with random
points (TARP) calibration test \citep{lemos2023sampling} using the implementation provided by the \texttt{sbi} package
\citep{Cranmer2020}. The TARP diagnostic evaluates whether the posterior credible regions have the
correct frequentist coverage: for a well-calibrated posterior, the fraction of test cases in which the
true parameter lies within the $\alpha$-credibility region should equal $\alpha$ for all $\alpha \in
[0,1]$. Concretely, we draw $N=500$ held-out $(\boldsymbol{\theta}, \boldsymbol{x})$ pairs from the
training simulations and, for each observation $\boldsymbol{x}_i$, sample 1000 posterior draws from the
trained NPE. The resulting expected coverage probability (ECP) curves are shown in
Figs.~\ref{fig:tarp_popiii} and \ref{fig:tarp_fiducial} for the Pop~III and fiducial models
respectively. In both cases the ECP tracks the diagonal closely, and a Kolmogorov--Smirnov test against
the uniform distribution cannot reject calibration (KS $= 0.050$, $p = 1.00$ for Pop~III; KS $= 0.069$,
$p = 1.00$ for the fiducial model), confirming that neither posterior is systematically over- or
under-confident.

\section{No nebular coverage}

The fiducial configuration adopted throughout this work assumes full nebular coverage ($f_{\rm cov}=1$), which maximises the nebular continuum and line contribution to the broadband flux. This is physically motivated: at the very low metallicities characteristic of Pop~III stars, the ionising photon budget is exceptionally large and the surrounding gas is expected to reprocess a significant fraction of the UV radiation into nebular emission \citep{Schaerer2002}. In this regime, the rest-frame UV and optical flux is dominated by nebular continuum and strong emission lines (particularly Ly$\alpha$ and \ion{He}{ii}~$\lambda$1640), which are the primary spectral signatures used to distinguish Pop~III from enriched stellar populations. Models with little or no nebular contribution ($f_{\rm cov}\approx 0$) correspond to an extreme scenario where the ionised gas has been fully disrupted or evacuated, leaving only the stellar continuum. While such configurations may arise in very compact, dense environments or after significant mechanical feedback, they are considered physically unlikely for young Pop~III bursts \citep[e.g.][]{Schaerer2002,raiter2010}.

\begin{figure}[]
\centering
\includegraphics[width=\columnwidth]{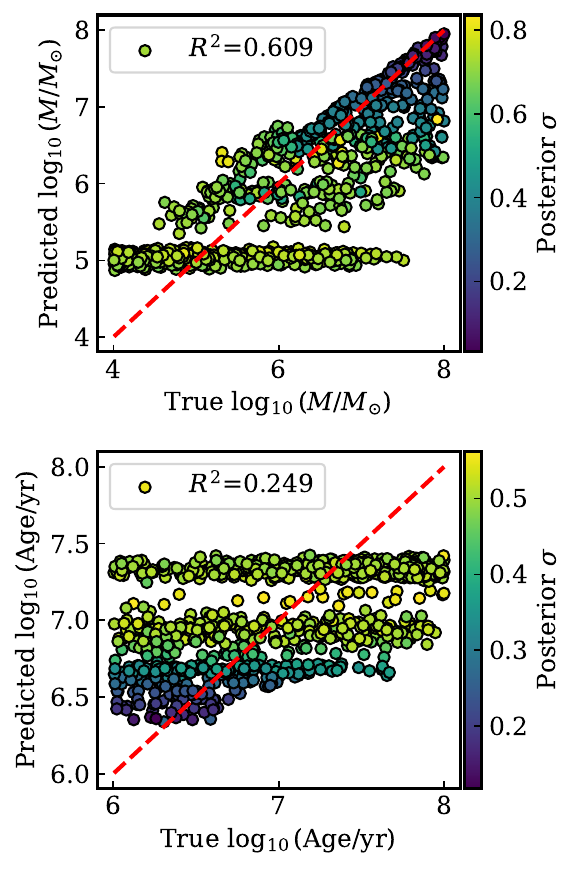}
\caption{As Fig.~\ref{fig:r2} but for a nebular coverage fraction of zero ($f_{\rm cov}=0$). Stellar mass is recovered with $R^2=0.609$, with low-mass sources collapsing to a degenerate band around $\log_{10}(M/M_\odot)\approx 5$. Age recovery is severely degraded ($R^2=0.249$), with predicted values clustering in a narrow range independent of the true age, reflecting the loss of age-sensitive nebular emission features.}
\label{fig:r2_fcov0}
\end{figure}

This appendix presents results for the $f_{\rm cov}=0$ case as a limiting test to quantify how much the detectability degrades when the nebular contribution is absent. Fig.~\ref{fig:r2_fcov0} shows the parameter recovery performance in this configuration. The stellar mass is recovered with $R^2=0.609$, a significant drop compared to the fiducial $f_{\rm cov}=1$ case (Fig.~\ref{fig:r2}). The mass posterior collapses to a narrow band around $\log_{10}(M/M_\odot)\approx 5$ for low-mass sources, indicating that without nebular emission the SBI cannot distinguish low-mass sources from one another — the purely stellar SED provides insufficient contrast to constrain the mass below this threshold. The age recovery is also severely degraded, with $R^2=0.249$: the predicted ages cluster in a narrow range regardless of the true value, reflecting the loss of age-sensitive nebular line ratios. The total broadband flux is substantially lower throughout, since the dominant nebular continuum and line contributions are absent, pushing many sources below the detection threshold and reducing the effective sample size available for inference.

\begin{figure}[htbp]
\centering
\includegraphics[width=0.99\columnwidth]{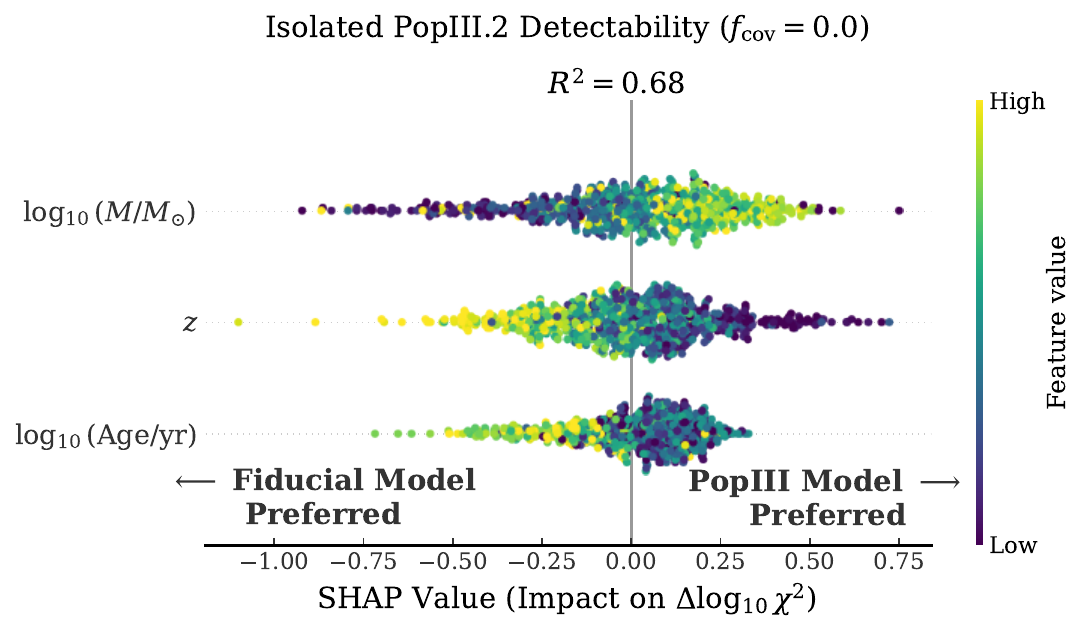}
\caption{As Fig.~\ref{fig:shap_single_clump} but for a nebular coverage fraction of zero ($f_{\rm cov}=0$). The surrogate model achieves $R^2=0.46$, lower than the fiducial case. Redshift becomes the dominant feature, while Pop~III age carries little discriminating power, consistent with the poor age recovery shown in Fig.~\ref{fig:r2_fcov0}.}
\label{fig:shap_single_clump_ly1}
\end{figure}

The SHAP analysis (Fig.~\ref{fig:shap_single_clump_ly1}) reflects this degradation: the surrogate model achieves $R^2=0.46$, compared to $R^2=0.72$ in the fiducial configuration. The ranking of feature importance shifts, with redshift becoming the dominant driver of the $\Delta\log_{10}\chi^2$ statistic, while stellar mass retains moderate importance. Pop~III age becomes less informative as a discriminating feature, consistent with the poor age recovery seen in Fig.~\ref{fig:r2_fcov0}. Overall, the $f_{\rm cov}=0$ scenario represents a substantially more challenging regime for Pop~III identification, underlining the critical role of nebular emission as the primary observational handle on primordial stellar populations with current JWST photometry.

\section{Kroupa IMF}

The Kroupa IMF \citep{kroupa2001} is the standard choice for present-day star formation, with a characteristic slope of $\Gamma \approx -2.3$ above $0.5\,M_\odot$ and a turnover at low masses. However, it is physically disfavoured for metal-free Pop~III environments: in the absence of metal-line and dust cooling, the Jeans mass during cloud fragmentation is expected to be significantly higher than in enriched gas, leading to a top-heavy mass distribution with a characteristic stellar mass of tens to hundreds of solar masses \citep{Schaerer2002,Zackrisson2011}. A Kroupa IMF applied to Pop~III populations therefore underpredicts the number of very massive stars, resulting in substantially fewer ionising photons per unit stellar mass and consequently much weaker nebular emission. The SEDs in this regime resemble an older, less extreme version of the standard Pop~III signature, making them harder to distinguish from enriched stellar populations.

\begin{figure}[]
\centering
\includegraphics[width=\columnwidth]{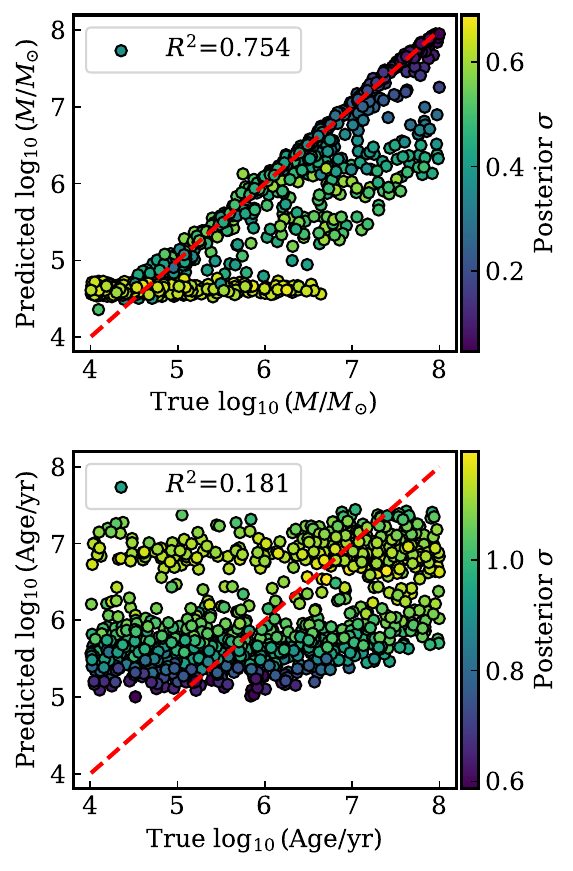}
\caption{As Fig.~\ref{fig:r2} but for a Kroupa IMF. Points are colour-coded by the posterior standard deviation. Stellar mass is recovered with $R^2=0.756$, while age recovery is essentially uninformative ($R^2=0.170$).}
\label{fig:r2_kroupa}
\end{figure}

\begin{figure}[htbp]
\centering
\includegraphics[width=0.99\columnwidth]{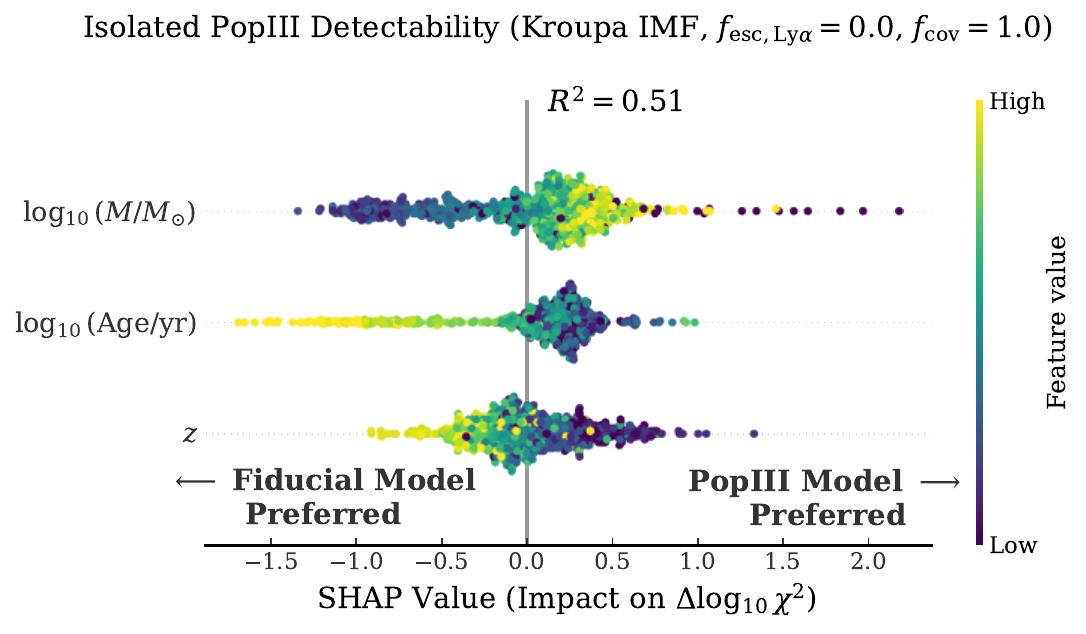}
\caption{As Fig.~\ref{fig:shap_single_clump} but for a Kroupa IMF, achieving an $R^2$ score of 0.51.}
\label{fig:shap_single_clump_kroupa}
\end{figure}

Fig.~\ref{fig:r2_kroupa} shows the parameter recovery performance for a Kroupa IMF model with $f_{\rm cov}=1$ and $f_{\rm esc}^{\rm Ly\alpha}=0$. The stellar mass is recovered with $R^2=0.756$, somewhat worse than the Pop~III.2 fiducial case, reflecting the weaker and less distinctive spectral features available for inference. The age recovery collapses dramatically to $R^2=0.170$, indicating that the posterior is essentially uninformative about the burst age: without the strong age-sensitive nebular line ratios produced by a top-heavy IMF, the age information is largely lost in the broadband photometry. The SHAP analysis (Fig.~\ref{fig:shap_single_clump_kroupa}) confirms the degraded performance, with the surrogate model reaching $R^2=0.51$. Stellar mass remains the dominant driver of $\Delta\log_{10}\chi^2$, but the overall SHAP spread is narrower than in the fiducial case, reflecting the reduced contrast between the Kroupa Pop~III SED and the fiducial model. Age shows negligible discriminating power, consistent with the poor recovery seen in Fig.~\ref{fig:r2_kroupa}. These results reinforce that the detectability of Pop~III sources is strongly coupled to the assumed IMF, and that the Kroupa case represents a conservative lower bound on the expected observational signatures.

\section{Fitting Pop~I/II galaxies with the Pop~III model}
\label{sec:app_control_galaxies}

A necessary sanity check for any model-comparison framework is to verify that the Pop~III model does not spuriously outperform the fiducial model when applied to genuinely enriched stellar populations. If the Pop~III NPE were biased or overfit to a broad range of SEDs, it could produce artificially low $\chi^2$ values even for Pop~I/II galaxies, leading to false positives in the detectability analysis. To test this, we apply the full inference and model-comparison pipeline to a control sample of 2000 simulated enriched galaxies generated with the fiducial model, spanning stellar masses $\log_{10}(M_*/M_\odot) = 5$--$9$, SFH timescales $\tau = 10$--$500$\,Myr, metallicities $Z = 10^{-4}$--$0.025$, and redshifts $5 \leq z \leq 14$. Of these, 1552 pass the detection threshold of at least one band above the noise limit and enter the analysis.

\begin{figure*}[htbp]
\centering
\includegraphics[width=0.98\textwidth]{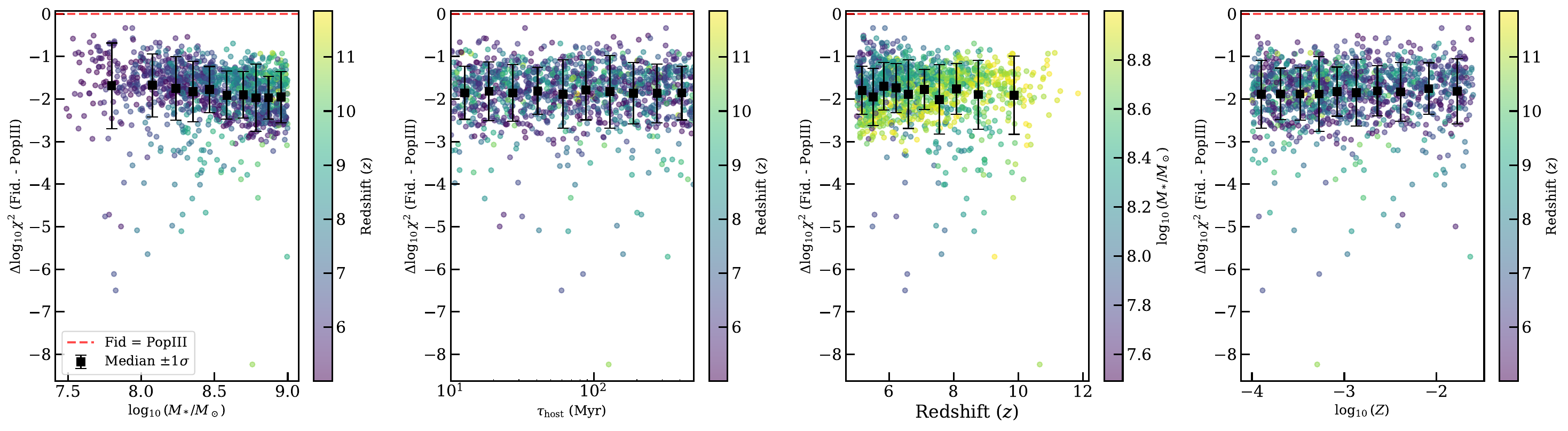}
\caption{$\Delta\log_{10}\chi^2 = \log_{10}\chi^2_{\rm fid} - \log_{10}\chi^2_{\rm PopIII}$ for a control sample of 1235 simulated enriched (Pop~I/II) galaxies with detections in $\geq 2$ bands (from 2000 drawn), fitted with both models, as a function of stellar mass (left), SFH timescale $\tau_{\rm host}$ (centre-left), redshift (centre-right), and metallicity (right). Colours indicate redshift in all panels except centre-right, where they indicate stellar mass. Black squares with error bars show the binned median and $1\sigma$ dispersion. The red dashed line marks $\Delta\log_{10}\chi^2 = 0$; negative values indicate fiducial preference. No case in this subsample shows Pop~III preference ($\Delta\log_{10}\chi^2 > 0$), confirming that the $\geq 2$-band detection threshold effectively eliminates false positives across the full explored parameter space.}
\label{fig:pure_host}
\end{figure*}

Fig.~\ref{fig:pure_host} shows $\Delta\log_{10}\chi^2 = \log_{10}\chi^2_{\rm fid} - \log_{10}\chi^2_{\rm PopIII}$ as a function of stellar mass, SFH timescale $\tau_{\rm host}$, redshift, and metallicity for this control sample. In $93.3\%$ of cases the fiducial model is correctly preferred ($\Delta\log_{10}\chi^2 < 0$), with a median value of $\Delta\log_{10}\chi^2 \approx -1.85$. The success rate rises sharply with the number of detected bands: among the 1235 galaxies with detections in at least two filters, the fiducial model is preferred in $100\%$ of cases. The binned median and $1\sigma$ statistics (black squares) lie consistently below zero across all masses, redshifts, SFH timescales, and metallicities, with no systematic trend suggesting a particular region of parameter space where false positives are more likely. The small fraction of cases where the Pop~III model is nominally preferred tend to occur when only a single band is detected, where the flexibility of both models allows each to produce an acceptable fit to the limited photometric information available. More broadly, the high fiducial preference rate even when only 2--3 filters are detected reflects the greater flexibility of the fiducial model, which has more free parameters and can therefore accommodate a wider range of SEDs, making it systematically harder for the two-parameter Pop~III model to compete. These results confirm that the Pop~III model preference observed for genuine Pop~III mocks in the main analysis is not a generic artefact of the inference framework, but a physically meaningful signal tied to the distinctive spectral features of primordial stellar populations.

\section{Control test on real JADES DR5 galaxies}
\label{sec:app_jades_control}

As a complementary test to the simulated control sample above, we apply the same model-comparison pipeline to real photometry from the JADES Data Release 5 \citep{johnson2026,Eisenstein2026,robertson26}. This test verifies that the Pop~III model does not spuriously outperform the fiducial model on real enriched galaxies, whose photometric properties may differ from our forward-model assumptions in ways not captured by the simulations.

We select all sources in the JADES DR5 photometric catalogue \citep{robertson26} with photometric redshifts from DR5 \citep{Hainline2026} and spectroscopic redshifts from DR1 \citep{Bunker2024}, DR3 \citep{2025ApJS..277....4D} and DR4 \citep{2026MNRAS.tmp..935C,Scholtz26} in the range $5 \leq z \leq 14$. This yields a parent sample of galaxies with 19-band photometry as described in Table~\ref{filters}. We then randomly sample 10000 pixels from the available segmentation maps, applying the same signal-to-noise threshold as in the main analysis, and fit each pixel with both models.

\begin{figure}[htbp]
\centering
\includegraphics[width=0.98\columnwidth]{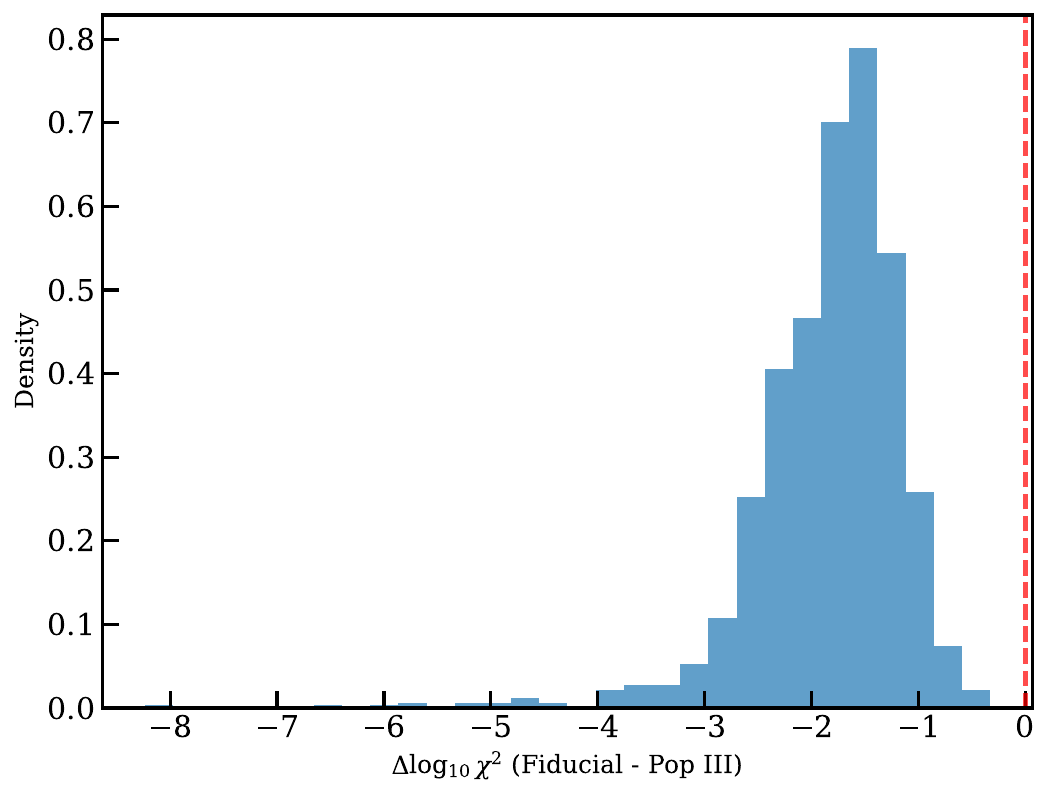}
\caption{$\Delta\log_{10}\chi^2 = \log_{10}\chi^2_{\rm fid} - \log_{10}\chi^2_{\rm PopIII}$ for 10000 pixels drawn from real JADES DR5 high-redshift galaxies ($5 \leq z \leq 14$), fitted with both the fiducial and Pop~III models. Negative values indicate fiducial preference throughout, confirming no systematic Pop~III false-positive signals in real enriched galaxies.}
\label{fig:jades_control}
\end{figure}

The resulting $\Delta\log_{10}\chi^2$ distribution is shown in Fig.~\ref{fig:jades_control}. As expected for a population of enriched galaxies, the fiducial model is systematically preferred, with all the pixels analysed showing $\Delta\log_{10}\chi^2$ lying well below zero. This result is consistent with the simulated control test and confirms that the Pop~III preference identified in the Pop~III candidate is not a generic feature of the pipeline applied to real JWST data, but a physically motivated insight.

\end{appendix}
\end{document}